\documentclass[12pt,twoside]{report}

\usepackage[intlimits]{amsmath}\usepackage{amssymb}\usepackage{xspace}
\usepackage[dvips]{graphicx}
\usepackage[dvips]{epsfig}
\usepackage[francais]{babel}
\def\lb{\label}
\def\be{\begin{equation}}
\def\ee{\end{equation}}
\def\ba{\begin{eqnarray}}
\def\ea{\end{eqnarray}}

\def\bb{\bibitem}
\def\p{\partial}
\def\e{{\rm e}}
\def\td{\tilde{d}}
\def\est{espace-temps }
\def\RG{Relativit\'e G\'en\'erale }

\usepackage{fancyhdr}
\renewcommand{\chaptermark}[1]{
\markboth{\chaptername\ \thechapter.\ #1}{}}

\begin{document}



\thispagestyle{empty}
\begin{tabular}{lr}
N$^\circ$ d'ordre : 073-2004 &
\quad\qquad\qquad\qquad\qquad\qquad\qquad\qquad\qquad\qquad Ann\'ee
2004\\
\\
LAPTH-these-1060/04 &
\end{tabular}


\bigskip
\bigskip
\bigskip
\bigskip
\bigskip
\begin{center}
{ \large THESE}
\bigskip

pr\'esent\'ee devant 
\bigskip

{ \large l'UNIVERSITE CLAUDE BERNARD - LYON 1}
\bigskip

pour l'obtention du
\bigskip

{\large DIPLOME DE DOCTORAT}
\bigskip

Sp\'ecialit\'e : { \large Physique th\'eorique}
\bigskip
\bigskip

pr\'esent\'ee et soutenue publiquement le 14 juin 2004
\bigskip

par
\bigskip

\large {\bf C\'edric Leygnac}
\bigskip
\bigskip
\begin{center}
{\LARGE {\bf Trous noirs non asymptotiquement plats}}
\end{center}

\vspace{2cm}
\hspace*{2cm}
\begin{center}
\begin{tabular}{lll}
JURY:
&G\'erard Cl\'ement , & Directeur de th\`ese \\
&Dmitri Gal'tsov ,& Membre invit\'e\\
&Marc Henneaux ,& Rapporteur\\
&Richard Kerner,& Examinateur\\
&Maurice Kibler ,&Examinateur\\
&David Langlois , &Rapporteur\\
&Paul Sorba ,& Pr\'esident du jury.\\
\end{tabular}
\end{center}
\end{center}

\newpage
\thispagestyle{empty}
\null
\newpage
\thispagestyle{plain}
{\bf {\Huge Remerciements}}

\vspace{2cm}
Je souhaite tout d'abord remercier le professeur G\'erard Cl\'ement
qui m'a offert de travailler en Relativit\'e G\'en\'erale, et plus particuli\`erement
sur les trous noirs, deux domaines qui m'ont toujours int\'eress\'e et
passionn\'e. Je le remercie \'egalement pour m'avoir guid\'e et conseill\'e
dans mes travaux de recherche et pour avoir patiemment r\'epondu \`a
toutes les questions que je lui ai pos\'ees. Gr\^ace \`a lui, j'ai
beaucoup appris au cours de ces trois derni\`eres ann\'ees. Enfin, je le remercie
pour tous les conseils qu'il m'a prodigu\'es au cours de la r\'edaction de ce manuscrit.

\vspace{.2cm}
Toute ma gratitude est aussi destin\'ee au professeur Richard
Kerner. Tout d'abord, je lui suis reconnaissant de m'avoir permis d'effectuer un stage de DEA
en Relativit\'e G\'en\'erale ce qui m'a permis de me forger ma toute premi\`ere
exp\'erience de recherche dans ce domaine. C'est aussi avec son aide
que j'ai pu continuer \`a travailler en Relativit\'e G\'en\'erale avec
le professeur G\'erard Cl\'ement au
Laboratoire d'Annecy-le-Vieux de Physique Th\'eorique (LAPTH). Enfin, je le remercie d'avoir bien
voulu faire partie de mon jury de th\`ese.

\vspace{.2cm}
J'adresse mes remerciements au professeur Paul Sorba  qui a accept\'e de
pr\'esider mon jury de th\`ese. Je lui suis aussi reconnaissant d'avoir facilit\'e ma venue au LAPTH.

\vspace{.2cm}
Je remercie les professeurs Marc Henneaux et David Langlois d'avoir
bien voulu \^etre les rapporteurs de mes travaux de th\`ese. Je les
remercie \'egalement des critiques constructives qu'ils ont \'emises sur ce manuscrit.

\vspace{.2cm}
Enfin, je remercie le professeur Maurice Kibler, pour avoir accept\'e
d'\^etre le repr\'esentant de l'Universit\'e Claude Bernard Lyon I dans mon
jury de th\`ese.

\vspace{.2cm}
Au cours de ma th\`ese, j'ai eu le plaisir de collaborer avec les
professeurs Dmitri Gal'tsov et 
Karim Ait Moussa. Je souhaite les remercier pour les nombreuses
discussions que j'ai pu avoir avec eux. Je les remercie d'avoir
r\'epondu \`a chacune de mes questions et pour leur grande
disponibilit\'e au cours de nos collaborations. J'esp\`ere avoir \`a nouveau l'occasion de travailler avec eux \`a l'avenir.

\newpage
\thispagestyle{empty}
\null
\newpage
\tableofcontents

\chapter*{Introduction}

\addcontentsline{toc}{chapter}{Introduction}

La th\'eorie relativiste de la gravitation, ou Relativit\'e G\'en\'erale, conduit \`a une vision g\'eom\'etrique de l'interaction gravitationnelle. Elle est r\'egie par les \'equations d'Einstein. Parmi les solutions de ces \'equations, il existe des solutions d\'ecrivant des espace-temps contenant un trou noir, c'est-\`a-dire une r\'egion o\`u la gravit\'e est si intense que rien ne peut s'en \'echapper. La fronti\`ere de cette r\'egion est appel\'ee horizon des \'ev\`enements.
Lorsque l'observateur franchit l'horizon, la coordonn\'ee qui \'etait de genre espace \`a l'ext\'erieur de l'horizon devient de genre temps et vice versa. Le c\^one de lumi\`ere de tout observateur franchissant cet horizon bascule et tout retour en arri\`ere est exclu.
De nombreuses solutions trou noir en \RG ont \'et\'e construites et \'etudi\'ees. La premi\`ere fut la solution de Schwarzschild qui d\'ecrit un trou noir statique \`a sym\'etrie sph\'erique. L'espace-temps de Schwarzschild est asymptotiquement Minkowskien. Par la suite, d'autres solutions avec un autre type de comportement asymptotique ont \'et\'e d\'eriv\'ees. Mentionnons la solution de Schwarzschild-anti deSitter (AdS) qui est une solution g\'en\'eralisant la m\'etrique de Schwarzschild au cas des \'equations d'Einstein avec constante cosmologique (n\'egative). Cet espace-temps n'est plus asymptotiquement Minkowskien mais asymptotiquement AdS; \`a l'infini il est domin\'e par la constante cosmologique et tend vers la solution \`a courbure constante de AdS.

Dans cette th\`ese, nous allons nous consacrer \`a la construction et \`a l'\'etude de nouvelles solutions trou noir, dans des th\'eories d\'ecrivant la gravitation coupl\'ee \`a des champs de mati\`ere, avec un comportement asymptotique diff\'erent de ceux que nous venons de mentionner. En effet, ces solutions ne  sont ni asymptotiquement Minkowskiennes ni asymptotiquement AdS. Nous construirons de telles solutions principalement dans des th\'eories \`a quatre dimensions d'espace-temps. Mais nous le ferons aussi \`a $2+1$ dimensions et pour un nombre quelconque de dimensions. 

Pourquoi s'int\'eresser \`a des espace-temps non asymptotiquement plats, alors que ceux-ci semblent a priori exclus par l'observation? Une motivation pourrait \^etre d'essayer d'\'etendre le champ d'application de la conjecture de Maldacena, selon laquelle il y a une correspondance \'etroite (dite correspondance AdS/CFT) entre certaines th\'eories des cordes sur des espace-temps de la forme $AdS_d\times M_{D-d}$, et des th\'eories de champs conformes sur le bord de $AdS_d$. Mais notre motivation principale sera d'utiliser de telles solutions afin d'\'etendre et de v\'erifier la validit\'e de m\'ethodes et d'outils mis au point pour l'\'etude des trous noirs asymptotiquement Minkowskiens ou asymptotiquement AdS.

Un premier probl\`eme est celui du calcul de l'\'energie et du moment angulaire de tels trous noirs. Dans le cas de la solution de Schwarzschild, la composante $g_{00}$ de la m\'etrique est reli\'ee au potentiel gravitationnel, en $1/r$, dont l'``intensit\'e'' donne la masse du trou noir. Une telle approche na\"ive n'est pas possible pour des trous noirs non asymptotiquement plats. Nous utiliserons le formalisme quasilocal qui est l'approche moderne au calcul de l'\'energie en Relativit\'e G\'en\'erale. Intuitivement, cette m\'ethode permet de calculer le flux d'\'energie \`a travers une surface ferm\'ee. L'approche quasilocale, qui permet de calculer la masse et le moment angulaire d'une solution quelque soit son comportement asymptotique, constitue la seule m\'ethode pouvant s'appliquer aux solutions que nous construirons.

\noindent Deuxi\`emement, comme nous le verrons dans le premier chapitre, les trous noirs pr\'esentent des liens intrigants avec la thermodynamique. En effet, quatre lois de la m\'ecanique des trous noirs ont \'et\'e d\'eriv\'ees au d\'ebut des ann\'ees 70. Or, ces quatre lois montrent de fortes ressemblances avec les quatre principes de la thermodynamique. En particulier, la premi\`ere loi relie entre elles les diff\'erentielles de la masse, du moment angulaire, de l'entropie et de la charge \'electrique d'un trou noir. Il est g\'en\'eralement admis que ces quatre lois, qui ont \'et\'e d\'emontr\'ees pour des espace-temps asymptotiquement plats, s'appliquent aussi au cas des espace-temps non asymptotiquement plats. Nous chercherons \`a v\'erifier si les solutions que nous construirons satisfont ou non \`a la premi\`ere loi de la thermodynamique des trous noirs.

La th\`ese est organis\'ee de la mani\`ere suivante.

Dans le premier chapitre, nous ferons une br\`eve introduction historique sur les trous noirs, puis nous introduirons les diagrammes de Penrose et le formalisme quasilocal; deux outils que nous utiliserons tout au long du reste de la th\`ese. Nous conclurons le chapitre en parlant du lien entre trous noirs et thermodynamique.

Dans les chapitres 2 \`a 5, nous allons travailler dans le cadre de th\'eories inspir\'ees des th\'eories des cordes, c'est-\`a-dire, soit des th\'eories obtenues par r\'eduction dimensionnelle des th\'eories des cordes, soit des th\'eories repr\'esentant une partie du secteur bosonique de ces th\'eories.

Dans le deuxi\`eme chapitre, nous introduirons la th\'eorie d'Einstein-Maxwell Dilatonique. Nous rappelerons les solutions trou noir statiques \`a sym\'etrie sph\'erique (asymptotiquement plates et non asymptotiquement plates) de cette th\'eorie. Ensuite, nous nous concentrerons sur les solutions non asymptotiquement plates et nous \'etudierons le mouvement g\'eod\'esique des particules dans le champ de gravitation de ces solutions. Puis nous montrerons comment construire les diagrammes de Penrose de ces solutions. Pour finir, nous adapterons le formalisme quasilocal \`a la th\'eorie d'Einstein-Maxwell dilatonique et nous l'utiliserons pour calculer la masse des solutions non asymptotiquement plates \`a sym\'etrie sph\'erique.

Dans le troisi\`eme chapitre, nous construirons de nouvelles solutions trou noir en rotation de la th\'eorie d'Einstein-Maxwell dilatonique pour une valeur particuli\`ere de la constante de couplage du dilaton: $\alpha^2=3$. Pour ce faire, nous utiliserons les deux ingr\'edients suivants. Premi\`erement, la th\'eorie d'Einstein-Maxwell dilatonique (pour $\alpha^2=3$) est la r\'eduction dimensionnelle de la th\'eorie d'Einstein \`a cinq dimensions sans source par rapport \`a un vecteur de Killing du genre espace. Deuxi\`emement, la th\'eorie d'Einstein \`a cinq dimensions sans source avec un vecteur de Killing du genre temps admet un mod\`ele $\sigma$ invariant sous un groupe de transformations reliant entre elles les diff\'erentes solutions de la th\'eorie. Nous montrerons aussi quels sont les liens entre les nouvelles solutions de la th\'eorie d'Einstein-Maxwell dilatonique (pour $\alpha^2=3$) que nous avons construites et les solutions connues de la th\'eorie d'Einstein \`a cinq dimensions sans source. Ensuite, nous utiliserons le formalisme quasilocal, que nous avons adapt\'e \`a la th\'eorie d'Einstein-Maxwell dilatonique dans le chapitre 2, pour calculer la masse et le moment angulaire des nouvelles solutions.

Dans le quatri\`eme chapitre, nous construirons une nouvelle solution, d\'ecrivant un trou noir en rotation non asymptotiquement plat, de la th\'eorie d'Einstein-Maxwell dilato-axionique. Elle contient la th\'eorie d'Einstein-Maxwell dilatonique lorsque la constante de couplage du dilaton est \'egale \`a un. Nous verrons ensuite quel est le lien entre la nouvelle solution et une certaine solution de la \RG sans source \`a six dimensions, en utilisant le fait que la th\'eorie d'Einstein-Maxwell dilato-axionique s'obtient par r\'eduction dimensionnelle d'un secteur de la \RG sans source \`a six dimensions avec deux vecteurs de Killing. Apr\`es avoir adapt\'e le formalisme quasilocal \`a la th\'eorie d'Einstein-Maxwell dilato-axionique, nous utiliserons ce formalisme pour calculer la masse et le moment angulaire de la nouvelle solution. Enfin, nous conclurons le chapitre en \'etudiant le mouvement g\'eod\'esique des particules, et en d\'eterminant les modes propres d'un champ scalaire dans le champ de gravitation de la nouvelle solution.

Dans le cinqui\`eme chapitre, nous construirons de nouvelles solutions (d\'ecrivant des trous noirs ou des branes noires) non asymptotiquement plates d'une th\'eorie \`a $D$ dimensions d'espace-temps g\'en\'eralisant la th\'eorie d'Einstein-Maxwell dilatonique. Nous r\'esoudrons les \'equations du mouvement de cette th\'eorie en cherchant des solutions statiques. Comme dans les chapitres pr\'ec\'edents, nous calculerons la masse de ces solutions en utilisant le formalisme quasilocal apr\`es l'avoir adapt\'e \`a cette th\'eorie. Puis, pour des valeurs particuli\`eres de la constante de couplage du dilaton, nous montrerons comment g\'en\'eraliser les solutions statiques \`a des solutions en rotation.

Enfin, dans le sixi\`eme chapitre, nous construirons une nouvelle famille de solutions trou noir d'une th\'eorie \`a trois dimensions d'espace-temps, la Gravitation Topologiquement Massive, qui g\'en\'eralise la \RG sans source \`a trois dimensions. Nous construirons les diagrammes de Penrose, et nous calculerons la masse et le moment angulaire de ces solutions. 

Les conventions et les d\'efinitions de termes que nous utiliserons tout au long de cette th\`ese sont regroup\'ees dans l' appendice Conventions et d\'efinitions.

Les travaux regroup\'es dans les chapitres 2 et 3 ont fait l'objet d'un article soumis \`a Physical Review D (voir Appendice C. {\tt ``Non-asymptotically flat, non-AdS dilaton black holes''}). De m\^eme, le travail expos\'e dans le chapitre 4 a \'et\'e publi\'e dans la revue Physical Review D (voir Appendice A. {\tt ``Linear Dilaton Black Holes''}). Enfin, le travail d\'etaill\'e dans le chapitre 6 a \'et\'e publi\'e dans la revue Classical and Quantum Gravity. Cet article est reproduit dans l'appendice B. {\tt ``The Black Holes of Topologically Massive Gravity''}.
\newpage
\thispagestyle{empty}
\null

\chapter{Trous noirs}%
\pagestyle{fancy}
\renewcommand{\chaptermark}[1]{
\markboth{\chaptername\ \thechapter.\ #1}{}}
\fancyhead{}
\renewcommand{\sectionmark}[1]{
\markright{\thesection.\ #1}{}}
\fancyhead[RO]{\rightmark}
\fancyhead[LE]{\leftmark}

Dans ce premier chapitre, nous allons tout d'abord faire une petite introduction historique sur les trous noirs. Puis, dans les sections 2 et 3, nous introduirons deux outils que nous utiliserons tout au long du reste de la th\`ese: les diagrammes de Penrose et le formalisme quasilocal. Pour finir, nous exposerons bri\`evement les liens intrigants existant entre les trous noirs et la thermodynamique.

Dans ce chapitre (sauf la section 1), le syst\`eme d'unit\'e utilis\'e est tel que $G=c=1$.

\section{Introduction}%

Au cours du mois de d\'ecembre 1915, Einstein publia une s\'erie de quatre papiers donnant les grandes lignes de la Relativit\'e G\'en\'erale conduisant \`a une vision g\'eom\'etrique de l'interaction gravitationnelle. Dans cette th\'eorie, la gravitation est r\'egie par les \'equations d'Einstein
\be
R_{\mu\nu}-\frac{1}{2}g_{\mu\nu}R=\frac{8\pi G}{c^2}T_{\mu\nu}.
\ee
 Peu de temps apr\`es, en 1916, Schwarzschild obtint une solution \`a ces \'equations d\'ecrivant le champ de gravitation \`a l'ext\'erieur d'un corps sph\'erique
\be
ds^2=-\left(1-\frac{2GM}{r}\right)dt^2+\left(1-\frac{2GM}{r}\right)^{-1}dr^2+r^2d\Omega^2.
\ee
Cette solution est l'unique solution \`a sym\'etrie sph\'erique des \'equations d'Einstein sans source (th\'eor\`eme de Birkhoff \cite{birkhoff}, 1923). La r\'egion $0<r<2GM$ fut longtemps ignor\'ee et m\^eme consid\'er\'ee comme non physique pour deux raisons: d'une part, l'apparente singularit\'e en $r=2GM$, qui a \'et\'e tout d'abord consid\'er\'ee comme une vraie singularit\'e, et d'autre part le fait que le rayon $r=2GM$ soit en pratique situ\'e bien \`a l'int\'erieur du rayon de l'\'etoile $r_e$, dont la solution de Schwarzschild repr\'esente le champ de gravitation ext\'erieur.
L'extension analytique maximale de la solution de Schwarzschild qui montre, entre autres, que la surface $r=2GM$ n'est pas une vraie singularit\'e, a \'et\'e obtenue par Kruskal \cite{kruskal} et Szekeres \cite{szekeres}.

Plus tard, avec l'\'emergence du concept de trou noir (voir \cite{israelw}), il a \'et\'e r\'ealis\'e que la solution de Schwarzschild pouvait repr\'esenter l'ext\'erieur d'une \'etoile en effondrement ainsi que l'int\'erieur et l'ext\'erieur de l'horizon du trou noir form\'e \`a la fin de cet effondrement.

En 1963, Kerr \cite{Kerr} obtint une solution, g\'en\'eralisant celle de Schwarzschild, d\'ecrivant le champ de gravitation produit par un corps en rotation. Carter \cite{carter71} et Robinson \cite{robinson} ont montr\'e que tout trou noir stationnaire et axisym\'etrique est caract\'eris\'e par seulement deux param\`etres: sa masse et son moment angulaire, et que sa m\'etrique est donn\'ee par la solution de Kerr.
Il est remarquable que l'effondrement gravitationnel de corps diff\'erant par leurs formes et leurs compositions conduise \`a la m\^eme famille de solutions (Kerr) poss\'edant une masse et un moment angulaire. On dit que le trou noir n'a pas de ``cheveux''.

Les solutions de Schwarzschild et de Kerr constituent donc les seules solutions statique \`a sym\'etrie sph\'erique et stationnaire axisym\'etrique, respectivement, de la \RG sans source.

Les solutions trou noir de la \RG avec source ont aussi \'et\'e \'etudi\'ees. Dans la th\'eorie d'Einstein-Maxwell, les solutions de Reissner-Nordstr\"om et de Kerr-Newman sont les g\'en\'eralisations avec charge des solutions de Schwarzschild et de Kerr, respectivement. Il a \'et\'e montr\'e que la solution de Reissner-Norstr\"om est l'unique solution trou noir statique (vide \`a l'ext\'erieur de l'horizon except\'e le champ \'electromagn\'etique produit par la charge du trou noir) \`a sym\'etrie sph\'erique de la th\'eorie d'Einstein-Maxwell \cite{israel}. De m\^eme, la solution de Kerr-Newman est l'unique solution trou noir stationnaire axisym\'etrique vide \`a l'ext\'erieur de l'horizon (except\'e le champ \'electromagn\'etique) \cite{mazur}.

Toutes les solutions pr\'ec\'edemment cit\'ees sont asymptotiquement Minkowskiennes. Il existe cependant des solutions avec un comportement asymptotique diff\'erent. Par exemple, les solutions de de-Sitter (dS) et Anti-de-Sitter (AdS) sont des solutions de la \RG avec constante cosmologique $\Lambda$ 
\be
R_{\mu\nu}-\frac{1}{2}g_{\mu\nu}R=\Lambda g_{\mu\nu}
\ee
($\Lambda>0$ pour dS et $\Lambda<0$ pour AdS). Encore une fois, les g\'en\'eralisations correspondantes des solutions de Schwarzschild (Schwarzschild-dS et Schwarzschild-AdS):
\be
ds^2=-\left(1-\frac{2GM}{r}-\frac{\Lambda r^2}{3}\right)dt^2+\left(1-\frac{2GM}{r}-\frac{\Lambda r^2}{3}\right)^{-1}dr^2+r^2d\Omega^2
\ee
et des autres trous noirs que nous avons mentionn\'es pr\'ec\'edemment: Reissner-Nordstr\"om-(A)dS et Kerr-Newman-(A)dS ont \'et\'e construites. Dans le cas $\Lambda<0$, ces solutions ont asymptotiquement ($r\rightarrow\infty$) la signature $-+++$, avec un comportement domin\'e par le terme proportionnel \`a la constante cosmologique
\be\lb{comp}
ds^2\simeq\frac{\Lambda r^2}{3}dt^2-\frac{3}{\Lambda r^2}dr^2+r^2d\Omega^2.
\ee
On dit que ces solutions sont asymptotiquement AdS.

Tout au long de cette th\`ese nous allons construire des solutions avec un autre type de comportement asymptotique. En effet, ces solutions ne seront ni asymptotiquement Minkowskiennes ni asymptotiquement AdS. Le comportement asymptotique de ces solutions est tel que
\be
ds^2\simeq -r^{1+\gamma}dt^2+r^{-(1+\gamma)}dr^2+r^{1-\gamma}d\Omega^2,\quad -1<\gamma\leq 1.
\ee
Nous voyons que la m\'etrique est asymptotiquement plate uniquement dans le cas limite o\`u $\gamma\rightarrow -1$.

Pour finir, \'evoquons la solution de Taub-NUT (Newman, Tamburini et Unti) \cite{taub,NUT} qui est une solution de la \RG sans source ($G=1$):
\be \lb{taub-NUT}
ds^2=-\frac{r^2-2Mr-N^2}{r^2+N^2}(dt+2N\cos\theta d\varphi)^2+\frac{r^2+N^2}{r^2-2Mr-N^2}dr^2+(r^2+N^2)d\Omega^2.
\ee
Le param\`etre $N$ est appel\'e charge NUT (pour une d\'efinition plus pr\'ecise de la charge NUT voir l'appendice Conventions et d\'efinitions). Le terme crois\'e entre $dt d\varphi$ proportionnel \`a $N$ introduit une singularit\'e appel\'ee corde de Misner \cite{misner63} similaire \`a la corde de Dirac apparaissant en \'electromagn\'etisme dans le cas du monopole magn\'etique. Par cons\'equent, ces trous noirs ne sont pas des trous noirs r\'eguliers, la corde introduisant une singularit\'e \`a l'ext\'erieur de l'horizon.

Les solutions trou noir de th\'eories \`a $2+1$ dimensions \cite{BTZ1,BTZH1} (voir aussi chapitre 6) et de th\'eories \`a plus de 4 dimensions d'espace-temps \cite{GW1,rasheed1,tang1,MP1,RE1} (voir aussi chapitre 5) ont \'et\'e aussi \'etudi\'ees.

Donnons maintenant la d\'efinition d'un trou noir.

$\bullet$
\underline{Trou noir}: On appelle trou noir une r\'egion de l'espace-temps o\`u la gravit\'e est si
 intense que rien ne peut s'en \'echapper y compris la lumi\`ere. Les \'ev\`enements se produisant \`a l'int\'erieur de cette r\'egion sont alors compl\`etement d\'econnect\'es causalement du reste de l'espace-temps. La fronti\`ere de cette r\'egion est appel\'ee \underline{horizon des \'ev\`enements}. Dans le cas des trous noirs statiques, les emplacements des horizons sont ``donn\'es'' par les z\'eros de $g_{tt}$ (par exemple $r=2GM$, dans le cas de la solution de Schwarzschild).

 Plus rigoureusement, une r\'egion $B$ d'un espace-temps $M$ est appel\'ee
 trou noir si cette r\'egion est d\'econnect\'ee causalement de l'infini $I^+$
 c'est-\`a-dire $B=M-J^-(I^+)$ (voir figure 1.1).
L'horizon des \'ev\`enements ou horizon apparent, $H$, est d\'efini comme
 l'intersection entre la fronti\`ere du pass\'e causal de l'infini du genre
 lumi\`ere futur ($\dot{J}^-(I^+)$) et de l'espace-temps $M$: $H=\dot{J}^-(I^+)\cap M$ 
(en se rappelant que les infinis conformes $I^\pm$, $i^\pm$ et $i^0$ ne font pas
 partie de $M$).  

\begin{figure}
\centerline{\epsfxsize=200pt\epsfbox{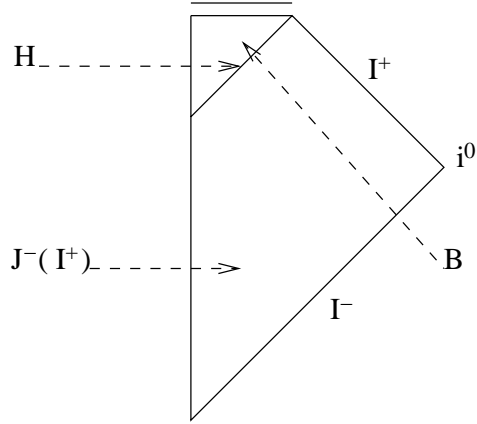}} 
\caption{Diagramme de Penrose (voir section suivante) d'un espace-temps repr\'esentant un trou noir form\'e \`a la suite de l'effondrement gravitationnel d'une \'etoile, par exemple. $I^{\pm}$ sont les infinis conformes de genre lumi\`ere futur et pass\'e. La double barre repr\'esente une singularit\'e cach\'ee derri\`ere l'horizon $H$.}
\end{figure}

Remarquons que la d\'efinition d'un trou noir ne fait pas intervenir la notion de singularit\'e mais uniquement la notion d'horizon (pouvant ou non masquer une singularit\'e). 

D\'efinissons maintenant ce qu'est une singularit\'e.

$\bullet$
\underline{Singularit\'e}: La notion de singularit\'e en \RG est souvent associ\'ee aux singularit\'es de courbure, c'est-\`a-dire aux ``endroits'' o\`u la courbure  de l'espace-temps diverge.

Nous adopterons ici une d\'efinition plus g\'en\'erale. Un espace-temps sera dit singulier si il est \underline{g\'eod\'esiquement incomplet}, c'est-\`a-dire si il poss\`ede au moins une g\'eod\'esique ne pouvant \^etre prolong\'ee au-del\`a d'une valeur finie de son param\`etre affine (la singularit\'e se trouvant au bout de cette g\'eod\'esique).

\section{Diagramme de Penrose}%

Dans cette deuxi\`eme partie, nous allons introduire une repr\'esentation de l'espace-temps propos\'ee par Penrose: le diagramme de Penrose. Il permet de repr\'esenter l'espace-temps en entier sur une feuille de papier, nous donnant ainsi une vision globale de celui-ci.

Pour bien comprendre en quoi consiste ce diagramme et la mani\`ere de le construire, nous allons prendre un exemple: la m\'etrique de Minkowski,
\be \lb{min}
ds^2=-dt^2+dr^2+r^2 d\Omega^2,\quad r>0,\quad t\in]-\infty,\infty[
\ee
o\`u $d\Omega^2=d\theta^2+\sin^2\theta d\varphi^2$.
Nous pouvons aussi param\'etriser la solution de Minkowski en utilisant les coordonn\'ees du c\^one de lumi\`ere : $u=t-r$ et $v=t+r$ ($u$, $v$ $\in]-\infty,+\infty[$) mettant la m\'etrique sous la forme
\be
ds^2=-du\, dv+\frac{(v-u)^2}{4}d\Omega^2,\quad v>u\,(r>0).
\ee
Les points situ\'es \`a l'infini ``$r=\infty$'' et ``$t=\pm\infty$'' ne font bien s\^ur pas partie de l'espace-temps de Minkowski alors que $r=0$ et $\sin\theta=0$ sont des singularit\'es de coordonn\'ees (le syst\`eme de coordonn\'ees est mal adapt\'e pour d\'ecrire l'\est  en ce point; la m\'etrique \'etant cependant r\'eguli\`ere en ces points comme nous le verrons plus loin en utilisant un autre syst\`eme de coordonn\'ees).

Pour construire le diagramme correspondant \`a (\ref{min}), nous allons tout d'abord introduire un nouveau syst\`eme de coordonn\'ees o\`u les coordonn\'ees des points situ\'es \`a l'infini prennent des valeurs finies. Pour cela, nous introduisons deux nouvelles coordonn\'ees du genre lumi\`ere
\be
U=\tan^{-1}u,\qquad V=\tan^{-1}v.
\ee
La m\'etrique devient alors
\be
ds^2=(2\cos U \cos V)^{-2}\left(-4 dU dV+\sin^2(V-U)d\Omega^2\right)
\ee
avec $-\frac{\pi}{2}<U<\frac{\pi}{2}$, $-\frac{\pi}{2}<V<\frac{\pi}{2}$ et $V> U$ ($r>0$).
Les points ``$r=\infty$'' et ``$t=\pm\infty$'' correspondent dans le nouveau syst\`eme de coordonn\'ees \`a $U=\pm\frac{\pi}{2}$ et $V=\pm\frac{\pi}{2}$. Cependant, ces points restent \`a distance g\'eod\'esique infinie \`a cause du terme $(2\cos U \cos V)^{-2}$. Pour les amener \`a distance g\'eod\'esique finie, nous multiplions la m\'etrique par le facteur $\Lambda=(2\cos U \cos V)^{2} $ (transformation conforme). Nous obtenons alors la m\'etrique suivante
\be \lb{minconf2}
d\tilde{s}^2=\Lambda(U,V)ds^2=-4 dU dV+\sin^2(V-U)d\Omega^2
\ee
qui n'est pas solution des \'equations d'Einstein sans source mais qui donne la structure conforme  de la m\'etrique de Minkowski (la transformation conforme n'affecte pas la structure causale de l'espace-temps).
Dans cette m\'etrique (\ref{minconf2}), les points qui \'etaient situ\'es \`a l'infini dans la m\'etrique (\ref{min}) sont maintenant \`a distance g\'eod\'esique finie, ce qui nous permet de les inclure dans l'espace-temps ($-\frac{\pi}{2}\leq U\leq\frac{\pi}{2}$, $-\frac{\pi}{2}\leq V\leq\frac{\pi}{2}$). D'autre part, ils correspondent \`a des coordonn\'ees finies, ce qui va nous permettre de les repr\'esenter sur une feuille de papier. De plus, les singularit\'es de coordonn\'ees ne sont plus pr\'esentes dans le nouveau syst\`eme de coordonn\'ees et nous pouvons consid\'erer aussi bien les valeurs positives de $r$ ($V>U$) que les valeurs n\'egatives de $r$ ($V<U$).

\noindent Finalement, introduisons les coordonn\'ees $T$ et $X$ d\'efinies par 
\be
T=U+V,\quad X=V-U
\ee 
qui nous permettent d'\'ecrire la m\'etrique sous la forme plus famili\`ere 
\be \lb{minconf}
d\tilde{s}^2=-dT^2+dX^2+\sin^2X d\Omega^2
\ee
avec $-\pi\leq T\pm X\leq \pi$.

Nous sommes maintenant en mesure de donner une repr\'esentation de l'espace-temps (\ref{min}) ou plus pr\'ecis\'ement de sa structure conforme (\ref{minconf}). Le diagramme de Penrose ne prend en compte que la partie $(T,X)$ de la m\'etrique et n\'eglige les coordonn\'ees angulaires. Compte-tenu des domaines de variations de $T$ et de $X$, il est facile de voir que l'on obtient la figure 1.2. En se rappelant que l'espace-temps de Minkowski (\ref{min}) ne contient pas les points situ\'es \`a l'infini, nous en d\'eduisons que (\ref{min}) est repr\'esent\'ee par l'int\'erieur de la partie droite du losange ($r>0$) plus la ligne $r=0$.  

\begin{figure}
\centerline{\epsfxsize=200pt\epsfbox{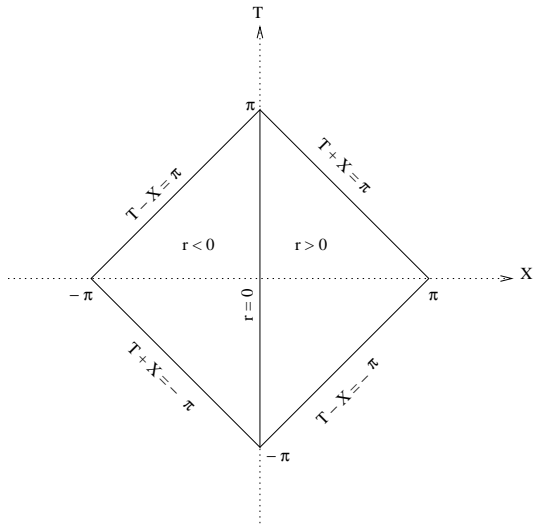}} 
\caption{La repr\'esentation dans le plan (T,X) de (\ref{minconf}) est donn\'ee par l'int\'erieur du losange ainsi que ses c\^ot\'es. (\ref{min}) est repr\'esent\'ee par l'int\'erieur de la partie droite du losange plus la ligne $r=0$ (voir Fig 1.3).}
\end{figure}

Le diagramme de Penrose fait la distinction entre plusieurs types d'infinis appel\'es infinis conformes d\'enot\'es par $i^0$, $i^\pm$ et $I^\pm$ (voir Figure 1.3). $i^0$ est l'infini du genre espace c'est-\`a-dire lorsque $r\rightarrow\infty$ \`a $t$ fini. $i^\pm$ sont les infinis du genre temps futur et pass\'e ($t\rightarrow\pm\infty$ et $r$ fini). Enfin, $I^\pm$ sont les infinis du genre lumi\`ere futur ($t\rightarrow\infty$, $r\rightarrow\infty$ et $r-t$ fini) et pass\'e ($t\rightarrow-\infty$, $r\rightarrow\infty$ et $r+t$ fini). Pour finir, notons que les g\'eod\'esiques du genre lumi\`ere sont repr\'esent\'ees par des lignes inclin\'ees \`a $45^\circ$. Les g\'eod\'esiques du genre temps commencent en $i^-$ et se terminent en $i^+$. Cependant, des courbes du genre temps peuvent, par exemple, commencer en $I^-$ et se terminer en $I^+$. De m\^eme, les g\'eod\'esiques du genre lumi\`ere commencent en $I^-$ et se terminent en $I^+$. Les g\'eod\'esiques du genre espace commencent et se terminent en $i^0$. Les vraies singularit\'es, c'est-\`a-dire n'incluant pas les singularit\'es de coordonn\'ees, seront repr\'esent\'ees dans la suite par une double ligne. 

\begin{figure}
\centerline{\epsfxsize=100pt\epsfbox{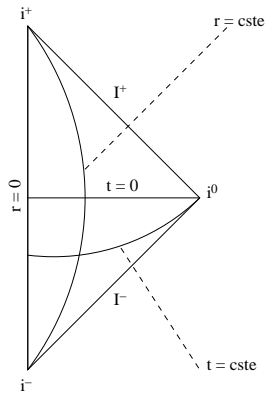}} 
\caption{Le diagramme de Penrose de (\ref{min}) est donn\'e par l'int\'erieur du losange plus la ligne $r=0$. Nous avons repr\'esent\'e une courbe de genre espace $t=\mbox{cste}$ et une courbe de genre temps $r=\mbox{cste}$. Tous les points du diagramme sont des sph\`eres de rayons $\sin X$ sauf $i^0$, $i^\pm$, $I^\pm$ et $r=0$. $i^0$, $i^\pm$ et $r=0$ sont des points alors que $I^\pm$ sont des hypersurfaces de genre lumi\`ere. Il est important aussi de noter que les points $i^\pm$ et $i^0$ sont distincts des lignes $I^\pm$ et $r=0$. }
\end{figure}



\section{Energie quasilocale}%

En Relativit\'e G\'en\'erale, l'existence d'une d\'efinition satisfaisante de l'\'energie est toujours un probl\`eme ouvert. Beaucoup de travaux ont \'et\'e consacr\'es \`a la recherche d'un \'equivalent gravitationnel \`a la densit\'e d'\'energie-impulsion $T^{\mu\nu}$ de la mati\`ere \cite{Trautman, Bergmann,landlif,Moller,weinberg,MTW}. Cependant, les quantit\'es obtenues sont des pseudotenseurs et donc d\'ependent du r\'ef\'erentiel choisi. Ces quantit\'es ne sont donc pas satisfaisantes. Il est maintenant g\'en\'eralement admis qu'il est impossible d'obtenir une d\'efinition locale de l'\'energie en Relativit\'e G\'en\'erale. Nous pouvons le comprendre en se rappelant le postulat du principe d'\'equivalence. En un point, un observateur ne peut faire la diff\'erence entre sa propre acc\'el\'eration et celle du champ de gravitation, rendant impossible la mesure du champ de gravitation en ce point (voir \cite{MTW} pages 466-468 pour une discussion plus approfondie). Autrement dit, nous pouvons toujours choisir, en un point, un rep\`ere o\`u le champ de gravitation est nul alors que la gravitation est pourtant pr\'esente.

Il existe des d\'efinitions permettant de calculer la masse et le moment angulaire \`a l'infini. Dans le cas d'espace-temps asymptotiquement plats, la d\'efinition la plus utilis\'ee est celle de Arnowitt, Deser et Misner (ADM) \cite{ADM}. Pour des espace-temps asymptotiquement AdS, mentionnons la d\'efinition de Abott et Deser \cite{AD} qui adapte la m\'ethode utilis\'ee par ADM au cas d'espace-temps asymptotiquement AdS. Cependant, nous construirons, dans les chapitres suivants, des solutions qui ne seront ni asymptotiquement Minkowskiennes ni asymptotiquement AdS. Nous ne pourrons donc pas utiliser les formules pr\'ec\'edentes pour calculer la masse et le moment angulaire de nos solutions.

\noindent L'impossibilit\'e de d\'efinir une \'energie locale en Relativit\'e G\'en\'erale a conduit les physiciens (Penrose ayant semble-t-il \'et\'e le premier \cite{Pen}) \`a introduire le concept d'\'energie quasilocale, qui comme nous le verrons plus loin, est associ\'ee \`a une surface \`a deux dimensions de genre espace. De nombreuses d\'efinitions ont \'et\'e propos\'ees pour l'\'energie quasilocale (voir \cite{CN2} pour un r\'esum\'e des approches utilis\'ees et les r\'ef\'erences associ\'ees). Nous utiliserons dans cette th\`ese la d\'efinition propos\'ee par Hawking et Horowitz \cite{HH} qui utilise l'approche Hamiltonienne (voir aussi \cite{BY,CN}). Rappelons bri\`evement en quoi consiste cette d\'efinition et les principales \'etapes du calcul. 

Pour cela, prenons l'exemple de la gravitation sans source donn\'ee par l'action d'Einstein-Hilbert
\be \lb{actioneinst}
S=\frac{1}{16\pi}\int_M R\sqrt{|g|} \,d^4x+\frac{1}{8\pi}\int_{\partial M} K \sqrt{h}\,d^3x
\ee
\`a laquelle nous avons ajout\'e un terme de surface \cite{RegTe} ($\partial_M$ d\'esignant la fronti\`ere de l'\est $M$). L'ajout de ce terme de surface est n\'ecessaire. En effet, l'action d'Einstein-Hilbert est g\'en\'eralement pr\'esent\'ee comme l'action dont la variation par rapport \`a la m\'etrique $g_{\mu\nu}$ donne les \'equations d'Einstein. Or, la densit\'e Lagrangienne $\sqrt g R$ d\'ependant de la m\'etrique $g_{\mu\nu}$ ainsi que de ses d\'eriv\'ees premi\`ere et seconde, la variation de l'action d'Einstein-Hilbert est
\be
\delta S=\int (\mbox{\'eqs du mouvement})\delta g^{\mu\nu}\sqrt{|g|}\,d^4x+ \mbox{terme de surface}(\delta g^{\mu\nu},\delta \dot{g}^{\mu\nu}).
\ee

\noindent Pour avoir un principe variationnel bien d\'efini le terme de surface doit s'annuler, ce qui implique d'imposer \`a la fois $\delta g^{\mu\nu}=0$ (conditions aux limites de type Dirichlet) et $\delta \dot{g}^{\mu\nu}=0$ (conditions aux limites de type Neumann) ce qui est impossible en g\'en\'eral. Choisissons, par exemple, d'imposer des conditions aux limites de type Dirichlet. Le terme de surface ne contient plus alors qu'un terme proportionnel \`a $\delta\dot{g}^{\mu\nu}$. Or, nous pouvons montrer (voir \cite{Wald} page 458) que ce terme n'est autre que la variation de la courbure extrins\`eque $K$ de $\partial_M$. C'est la raison pour laquelle nous ajoutons le terme $\int_{\partial M} K d^3x$ \`a l'action d'Einstein-Hilbert, ce terme permettant d'\'eliminer la contribution proportionnelle \`a $\delta\dot{g}^{\mu\nu}$ lors de la variation. 

Revenons \`a la d\'efinition de Hawking et Horowitz de l'\'energie quasilocale. Tout d'abord nous devons calculer le Hamiltonien de la th\'eorie d'Einstein. Pour cela, nous effectuons une d\'ecomposition \`a la ADM (Arnowitt, Deser et Misner) \cite{ADM} de la m\'etrique
\be \lb{ADM}
ds^2=-N^2 dt^2+h_{ij}(dx^i+N^i dt)(dx^j+N^j dt)
\ee
qui correspond au d\'ecoupage de l'espace-temps (voir Figure 1.4) en une famille d'hypersurfaces de genre espace $\Sigma_t$ de m\'etrique $h_{ij}$. La fronti\`ere de l'espace-temps consiste en une hypersurface de genre temps $\Sigma^\infty$ et deux hypersurfaces de genre espace $\Sigma_{t_1}$ et $\Sigma_{t_2}$. Pour la simplicit\'e du calcul les hypersurfaces $\Sigma_t$ sont prises orthogonales \`a $\Sigma^\infty$ (voir \cite{HHu} pour le calcul dans le cas non orthogonal). Les intersections entre les $\Sigma_t$ et $\Sigma^\infty$ sont appel\'ees $S_t^\infty$. Ces surfaces ont pour m\'etrique $\sigma_{ab}$. Lorsque nous avons donn\'e la d\'efinition d'un trou noir dans la premi\`ere section, nous avons indiqu\'e que, pour les trous noirs statiques, les emplacements des horizons sont donn\'es par les z\'eros de $g_{tt}$. Dans le cas d'un trou noir stationnaire, les emplacements des horizons sont donn\'es par les z\'eros de $N$. Le z\'ero, $r_s$ (situ\'e avant l'horizon des \'ev\`enements $r_s>r_h$), de $g_{tt}$ est appel\'e limite statique car il ne peut exister d'observateur statique dans la zone $r_h<r<r_s$ (appel\'ee ergosph\`ere).

\begin{figure}
\centerline{\epsfxsize=210pt\epsfbox{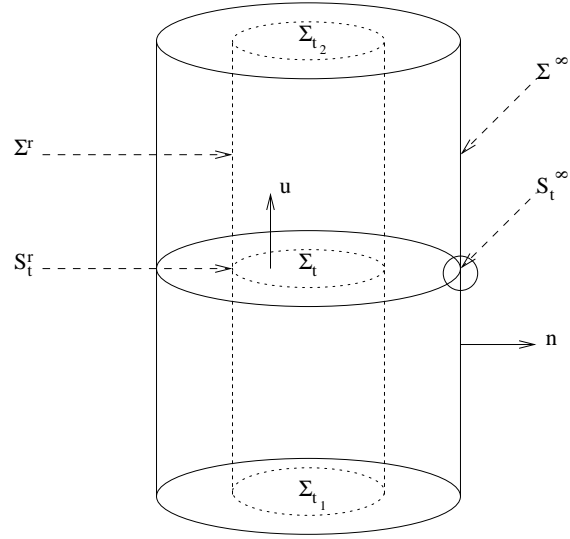}}
\caption{Repr\'esentation de l'espace-temps (\ref{ADM}). On appelle $u$ la normale (de genre temps) aux hypersufaces $\Sigma_t$ et $n$ la normale (de genre espace) aux hypersurfaces $\Sigma^r$. La m\'etrique de l'espace-temps $g_{\mu\nu}$ induit une m\'etrique $h_{ij}=g_{ij}+u_iu_j$ sur les hypersurfaces $\Sigma_t$ et une m\'etrique $\sigma_{ab}=g_{ab}+u_au_b-n_an_b$ sur les hypersurfaces $\Sigma^r$.}
\end{figure}

En utilisant (\ref{ADM}), nous pouvons r\'e\'ecrire l'action (\ref{actioneinst}) en faisant appara\^itre le Hamiltonien 
\be
S=\int dt\left[\int_{\Sigma_t}\left(p^{ij}\dot{h}_{ij}-N {\cal H}-N^i {\cal H}_i\right)d^3x-\oint_{S^r_t}\left(N \epsilon+2N^i\pi_{ij}n^j\right)d^2x\right]
\ee
o\`u
\be
p^{ij}=\frac{1}{16\pi}\sqrt{h}(h^{ij}K-K^{ij})
\ee
est le moment conjugu\'e de $h_{ij}$, $\pi^{ij}=\frac{\sqrt{\sigma}}{\sqrt h}p_{ij}$ et $\epsilon=\frac{1}{8\pi}\sqrt \sigma \,k$.

\noindent Les quantit\'es $k$ et $K$ sont, respectivement, la trace de la courbure extrins\`eque de $S^\infty_t$ et $\Sigma_t$ donn\'ees par
\ba
K_{\mu\nu}&=&-h_\mu^{\alpha}\nabla_\alpha n_\nu=-\frac{1}{2N}\left(\dot{h}_{\mu\nu}-2D_{(\mu}N_{\nu)}\right)\\
k_{\mu\nu}&=&-\sigma_\mu^{\alpha}D_\alpha n_\nu
\ea
o\`u $\nabla$ est la d\'eriv\'ee covariante associ\'ee \`a la m\'etrique $g_{\mu\nu}$ et $D$ celle associ\'ee \`a la m\'etrique $h_{ij}$.

\noindent Les quantit\'es ${\cal H}$ et ${\cal H}_i$ sont des contraintes associ\'ees aux multiplicateurs de Lagrange  $N$ et $N^i$
\be
{\cal H}=-R_3+\frac{1}{\sqrt h}\left(p^{\mu\nu}p_{\mu\nu}-\frac{1}{2}p^2\right),\quad {\cal H}_i=-2D_\mu\left(\frac{p^\mu_i}{\sqrt h}\right).
\ee
Maintenant que nous avons obtenu le Hamiltonien de la Relativit\'e G\'en\'erale, nous d\'efinissons l'\'energie comme \'etant la valeur de ce Hamiltonien ``sur couche'' (c'est-\`a-dire lorsque les \'equations d'Einstein sont satisfaites) . Etant donn\'e que les contraintes ${\cal H}$ et ${\cal H}_i$ s'annulent pour une solution de la th\'eorie, l'\'energie est simplement donn\'ee par le terme de surface du Hamiltonien (qui provient en partie de l'action d'Einstein-Hilbert et en partie du terme de surface ajout\'e \`a celle-ci d\`es le d\'epart)
\be \lb{equas}
E=\oint_{S^r_t}\left(N \epsilon+2N^i\pi_{ij}n^j\right)d^2x.
\ee
En particulier, nous voyons que l'\'energie ne d\'epend que de quantit\'es \'evalu\'ees sur la surface \`a deux dimensions du genre espace $S^r_t$.

Cependant, l'\'energie quasilocale est g\'en\'eralement divergente. Pour obtenir une \'energie finie, il faut retrancher la contribution $E^0$ d'une solution de fond. Pour une solution ($g_{\mu\nu}$, $\phi$) ($\phi$ repr\'esentant d'\'eventuels champs de mati\`ere), la solution de fond ($g_{\mu\nu}^0$, $\phi^0$) est choisie de telle mani\`ere que les ($g_{\mu\nu}$, $\phi$) et les ($g^0_{\mu\nu}$, $\phi^0$) co\"incident\footnote{ou au moins sont \'egaux \`a un ordre d'approximation suffisant.} sur l'hypersurface $S^r_t$. En supposant que la solution de fond est statique ($N_0^i=0$) (ce qui sera le cas dans cette th\`ese), l'\'energie est alors donn\'ee par la formule suivante:
\be \lb{equas2}
E=\oint_{S^r_t}\left(N (\epsilon-\epsilon_0)+2N^i\pi_{ij}n^j\right)d^2x.
\ee  

Pour finir, en examinant la variation du Hamiltonien sous une rotation infinit\'esimale $x^\mu\rightarrow x^\mu+\delta x^\mu$,
\be
\delta {\cal H}=J_\mu \delta x^\mu,
\ee
nous en d\'eduisons l'expression suivante pour le moment angulaire ($J^0=0$ pour une solution de fond statique)
\be \lb{jquas}
J_i=-2\oint_{S^r_t}n_\mu\pi^{\mu}{}_id^2x.
\ee

Nous pouvons v\'erifier que ces d\'efinitions co\"incident avec les pr\'ec\'edentes d\'efinitions de l'\'energie en Relativit\'e G\'en\'erale \cite{HH}.

\section{La thermodynamique des trous noirs}%

Au d\'ebut des ann\'ees 1970, Bekenstein \cite{Beckenphd,Becken73} obtint la formule suivante
\be \lb{loib}
dM={\cal T}_h dA+\Omega_h dJ+\Phi_h dQ
\ee
reliant entre elles les diff\'erentielles de la masse $M$, de l'aire $A$ de l'horizon, du moment angulaire $J$ et de la charge \'electrique $Q$ de la solution de Kerr-Newman. Les quantit\'es ${\cal T}_h$, $\Omega_h$ et $\Phi_h$ sont la tension effective de surface de l'horizon, la vitesse angulaire de l'horizon et le potentiel \'electrostatique sur l'horizon, respectivement. Peu de temps apr\`es, toujours pour la solution de Kerr-Newman, Smarr \cite{Smarr} d\'eriva \`a partir de (\ref{loib}), la formule suivante
\be \lb{lois}
M={\cal T}_h A+2\,\Omega_h J+\Phi_h Q.
\ee

En 1971, Hawking d\'emontre un th\'eor\`eme sur l'aire d'un trou noir. Ce th\'eor\`eme dit que l'aire d'un trou noir ne diminue jamais
\be
\delta A\geq 0.
\ee
Ce th\'eor\`eme fait fortement penser au deuxi\`eme principe de la thermodynamique qui dit que pour tout processus physique l'entropie d'un syt\`eme ne peut diminuer. Cependant, le th\'eor\`eme de Hawking est plus fort encore car pour un syst\`eme de trous noirs, c'est l'aire de chaque trou noir qui ne peut diminuer. De plus, lorque deux trous noirs fusionnent et qu'un nouveau trou noir en r\'esulte, l'aire du trou noir final ne peut \^etre inf\'erieure \`a la somme des aires initiales.
L'aire d'un trou noir semble donc reli\'ee \`a l'entropie de celui-ci.
Dans l'article \cite{BCH}, Bardeen, Carter et Hawking ont g\'en\'eralis\'e les r\'esultats obtenus par Bekenstein (\ref{loib}) et Smarr (\ref{lois}) au cas d'une solution asymptotiquement plate quelconque et, en poursuivant l'analogie avec la thermodynamique, ont formul\'e les quatre lois de la m\'ecanique des trous noirs.

Rappelons ici leur expression:

$\centerdot$ La ``loi z\'ero'': La gravit\'e de surface est constante sur l'horizon pour un trou noir stationnaire.

\noindent \`a comparer avec le principe z\'ero de la thermodynamique: la temp\'erature $T$ d'un corps en \'equilibre est constante.

$\centerdot$ La ``premi\`ere loi'':
\be
dM=\frac{\kappa}{8\pi}dA+\Omega_h dJ+\Phi_h dQ
\ee
o\`u $\kappa$ est la gravit\'e de surface de l'horizon, \`a comparer avec le premier principe de la thermodynamique: $dE=T dS-p dV$, les termes $\Omega dJ$ et $\Phi dQ$ \'etant des termes de travail.

$\centerdot$ La ``deuxi\`eme loi'':
\be
\delta A\geq 0 \ \mbox{pour tout processus physique}
\ee
\`a comparer avec le deuxi\`eme principe de la thermodynamique: $\delta S\geq 0$ pour tout processus physique .

$\centerdot$ La ``troisi\`eme loi'':
\be
\mbox{On ne peut atteindre $\kappa=0$ par aucun processus physique}
\ee
\`a comparer avec le troisi\`eme principe de la thermodynamique: On ne peut atteindre $T=0$ par aucun processus physique.

Nous avons donc quatre lois de la m\'ecanique des trous noirs qui montrent une tr\`es grande ressemblance avec les quatre principes de la thermodynamique. En particulier, il semble y avoir un lien entre la temp\'erature et la gravit\'e de surface d'une part et l'entropie et l'aire du trou noir d'autre part. Cependant, en \RG classique les trous noirs, comme leur nom l'indique, ne font qu'absorber et n'\'emettent rien. Leur temp\'erature est nulle. Apparemment, il ne peut donc y avoir de lien entre la temp\'erature $T$ et la gravit\'e de surface $\kappa$, et l'analogie entre les lois de la m\'ecanique des trous noirs et les quatre principes de la thermodynamique ne semble \^etre rien d'autre qu'une analogie.

En fait, il existe bel et bien un lien entre les quatre lois et les quatre principes. En effet, Hawking a montr\'e qu'en incluant les effets quantiques, les trous noirs rayonnent. Au voisinage de l'horizon, il y a cr\'eation de paires particules-antiparticules. Les antiparticules tombent dans le trou noir et les particules partent vers l'infini. Tout se passe donc comme si le trou noir \'emettait des particules en se comportant comme un corps noir \`a la temp\'erature
\be
T=\frac{\kappa}{2\pi}.
\ee
Ce ph\'enom\`ene a pour nom le rayonnement de Hawking \cite{H75}. Il s'agit donc d'une \'emission spontan\'ee, ph\'enom\`ene purement quantique qui n'a pas d'\'equivalent classique.
Il y a donc bien un lien entre la temp\'erature et la gravit\'e de surface. En poursuivant l'analogie, nous voyons en comparant deuxi\`eme loi et deuxi\`eme principe que
\be
S=\frac{A}{4}.
\ee
Nous pouvons montrer que l'une des cons\'equences du rayonnement de Hawking est que le trou noir peut s'\'evaporer totalement et donc que son aire tend vers z\'ero. En tenant compte des effets quantiques la deuxi\`eme loi est donc viol\'ee. Mais nous pouvons introduire une deuxi\`eme loi g\'en\'eralis\'ee \cite{beken73c}.

$\centerdot$ La ``deuxi\`eme loi g\'en\'eralis\'ee'':
\be
\delta S_T\geq 0 \ \mbox{pour tout processus physique}
\ee
o\`u $S_T$ est la somme de l'entropie du trou noir $S_{tn}$ et de l'entropie $S_{ext}$ de tout ce qui se trouve \`a l'ext\'erieur de celui-ci.

Ces quatre lois montrent qu'il existe un lien entre gravitation, th\'eorie quantique et physique statistique (voir \cite{Waldthermo} pour une revue relativement r\'ecente de la thermodynamique des trous noirs).

Aucune th\'eorie satisfaisante de la gravitation quantique n'a pour l'instant \'et\'e obtenue \cite{Wald}. Cependant de nombreux r\'esultats (dont le rayonnement de Hawking) ont pu \^etre obtenus en faisant les calculs dans l'euclidien \cite{H79}, c'est-\`a-dire en effectuant la continuation analytique 
\be \lb{wrot}
\tau=i t,
\ee
appel\'ee rotation de Wick, dans la m\'etrique.
Prenons l'exemple de la m\'etrique de Schwarzschild dans laquelle nous effectuons la rotation de Wick (\ref{wrot})
\be
ds^2=\left(1-\frac{2M}{r}\right)d\tau^2+\left(1-\frac{2M}{r}\right)^{-1}dr^2+r^2d\Omega^2.
\ee
Au voisinage de l'horizon $r=2M$, en posant $x^2=r-2M$, nous avons
\be
ds^2=8M\left(dx^2+\frac{x^2}{16M^2}d\tau^2\right)+r^2d\Omega^2.
\ee
La m\'etrique de Schwarzschild euclidienne est r\'eguli\`ere si $\tau$ est une variable angulaire de p\'eriode $\beta=8\pi M$. Or, la r\'egularit\'e de la m\'etrique euclidienne est n\'ecesssaire pour qu'une th\'eorie quantique des champs \`a la temp\'erature $T=\beta^{-1}$ soit en \'equilibre avec un trou noir. D'autre part, nous pouvons v\'erifier que la temp\'erature $\beta^{-1}$ est bien \'egale \`a la temp\'erature de Hawking ($\kappa=1/4M$ pour Schwarzschild).

De mani\`ere g\'en\'erale pour une m\'etrique stationnaire \'ecrite sous la forme ADM \cite{ADM}
\be
ds^2=-N^2dt^2+n_r^2 dr^2+h_{\theta\theta} d\theta^2 +h_{\varphi\varphi}(d\varphi+N^\varphi dt)^2
\ee
nous avons, en adoptant un rep\`ere tournant avec l'horizon $r_h$ ($d\varphi\rightarrow d\varphi-N^\varphi(r_h) dt$) et en effectuant une rotation de Wick,
\be
ds^2=N^2d\tau^2+n_r^2 dr^2
\ee
o\`u nous avons n\'eglig\'e la partie angulaire ($\theta,\varphi$).
Au voisinage de l'horizon, $r=r_h+x$ et $N=xN_h^{'}+ \ldots$, la m\'etrique devient
\be
ds^2\simeq x^2N^{'}|_h d\tau^2+n^2_r|_h dx^2\simeq n^2_r|_h(dx^2+(n^i\p_iN)|_hd\tau^2)
\ee
qui est r\'eguli\`ere si $\tau$ est une variable angulaire de p\'eriode $\beta=2\pi/(n^i\p_iN)|_h$. Nous en d\'eduisons alors la formule g\'en\'erale de la temp\'erature d'un trou noir
\be \lb{fT}
T=\beta^{-1}=\frac{1}{2\pi}(n^i\p_iN)|_h
\ee
que nous utiliserons r\'eguli\`erement dans cette th\`ese.

Dans la suite, nous nous int\'eresserons uniquement \`a la premi\`ere loi. Nous v\'erifierons si les nouvelles solutions trou noir que nous d\'eriverons satisfont ou non \`a cette premi\`ere loi.

\newpage
\thispagestyle{empty}
\null
\chapter{La th\'eorie d'Einstein-Maxwell dilatonique}%

Dans ce chapitre, nous allons dans une premi\`ere partie pr\'esenter la th\'eorie d'Einstein-Maxwell dilatonique (EMD). Puis nous rappelerons les solutions \`a sym\'etrie sph\'erique (d\'ecrivant des trous noirs asymptotiquement plats et non asymptotiquement plats) de cette th\'eorie. Dans les troisi\`eme et quatri\`eme parties, nous \'etudierons le mouvement g\'eod\'esique et nous construirons les diagrammes de Penrose des solutions non asymptotiquement plates. Enfin, dans les cinqui\`eme et sixi\`eme parties, nous adapterons le formalisme quasilocal \`a EMD puis nous l'appliquerons aux solutions non asymptotiquement plates.

Dans ce chapitre, le syst\`eme d'unit\'e utilis\'e est tel que $G=1$.

\section{La th\'eorie d'Einstein-Maxwell dilatonique}%
La th\'eorie d'Einstein-Maxwell dilatonique (EMD), donn\'ee par l'action

\be \lb{actionEMD}
S = \frac{1}{16\pi}\int d^4x\sqrt{|g|} \left\{R-2\partial_\mu\phi\partial^\mu\phi-\e^{-2\alpha\phi} F_{\mu\nu} F^{\mu\nu}\right\}
\ee
contient un champ scalaire, le dilaton $\phi$, ainsi qu'un vecteur $A$ ab\'elien, qui sont coupl\'es \`a la gravit\'e. $F_{\mu\nu}=\p_\mu A_\nu-\p_\nu A_\mu$ est le tenseur \'electromagn\'etique. La force du couplage entre le dilaton et le champ \'electromagn\'etique est ``mesur\'ee'' par $\alpha$, appel\'ee constante de couplage du dilaton . Nous pouvons nous restreindre, sans perte de g\'en\'eralit\'e, au cas $\alpha \geq 0$, le cas $\alpha<0$ \'etant ais\'ement obtenu par red\'efinition du dilaton ($\phi\rightarrow -\phi$).

Remarquons que l'action (\ref{actionEMD}) est invariante sous la transformation de dualit\'e
\be \lb{transdual}
(g_{\mu\nu},\phi,F_{\mu\nu})\rightarrow (\hat{g}_{\mu\nu}=g_{\mu\nu},\hat{\phi}=-\phi,\hat{F}_{\mu\nu}=\pm\e^{-2\alpha\phi}\tilde{F}_{\mu\nu})
\ee
avec 
\be \lb{Fdual}
\tilde{F}^{\mu\nu}=\frac{1}{2\sqrt g}\varepsilon^{\mu\nu\rho\sigma}F_{\rho\sigma}
\ee
o\`u $\varepsilon^{\mu\nu\rho\sigma}$ est le symbole totalement antisym\'etrique\footnote{Nous adopterons la convention $\varepsilon^{1230}=1$ o\`u $x^0=t$.}. Cette transformation peut \^etre utilis\'ee pour passer d'une solution charg\'ee \'electriquement $(g_{\mu\nu},\phi,F_{\mu\nu})$ \`a sa version charg\'ee magn\'etiquement $(g_{\mu\nu},\hat{\phi},\hat{F}_{\mu\nu})$ et vice versa. En effet, prenons l'exemple d'une solution statique charg\'ee \'electriquement. Les seules composantes non nulles du tenseur \'electromagn\'etique sont alors $F_{k0}$. En utilisant (\ref{transdual}), nous avons:
\be
\hat{F}^{ij}=\frac{1}{\sqrt{g}}\e^{-2\alpha\phi}\varepsilon^{ijk0}F_{k0}.
\ee
La transformation (\ref{transdual}) n'affectant pas la m\'etrique, nous avons donc la m\^eme solution mais avec cette fois-ci un tenseur \'electromagn\'etique dont les seules composantes non nulles sont $\hat{F}_{ij}$, donc une solution charg\'ee magn\'etiquement.

Trois cas particuliers, d\'ependant de la valeur de $\alpha$, sont \`a signaler. Le cas $\alpha=0$ se r\'eduit \`a la th\'eorie d'Einstein-Maxwell coupl\'ee \`a un champ scalaire de masse nulle. 

\noindent Le cas $\alpha=1$ provient de la limite \`a basse \'energie (compactification des dimensions suppl\'ementaires sur un tore) de la th\'eorie des cordes h\'et\'erotiques. Il peut aussi \^etre vu comme la troncation du secteur bosonique de la supergravit\'e $D=4$, $N=4$. 

\noindent Enfin, le cas $\alpha=\sqrt 3$ peut \^etre obtenu par r\'eduction dimensionnelle de la th\'eorie d'Einstein \`a 5 dimensions par rapport \`a un vecteur de Killing de genre espace.

En variant l'action par rapport \`a $\phi$, $A_\mu$ et $g_{\mu\nu}$, nous obtenons les \'equations du mouvement suivantes :

\ba \lb{eqmvtemd1}
\frac{1}{\sqrt{|g|}}\partial_\mu(\sqrt{|g|}\partial^\mu\phi)=-\frac{\alpha}{2} \e^{-2\alpha\phi}F^2\\
 \lb{eqmvtemd2}
\frac{1}{\sqrt{|g|}}\partial_\mu(\sqrt{|g|} \e^{-2\alpha\phi}F^{\nu\mu})=0\\
 \lb{eqmvtemd3}
R_{\mu\nu}=2\,\partial_\mu\phi\partial_\nu\phi-\frac{1}{2}g_{\mu\nu}\e^{-2\alpha\phi}F^2+2\,\e^{-2\alpha\phi}F_\mu{}^\lambda F_{\nu\lambda}.
\ea
Notons que, lorsque $\phi=0$, EMD ne se r\'eduit pas \`a la th\'eorie d'Einstein-Maxwell puisque l'\'equation du dilaton (\ref{eqmvtemd1}) impose la contrainte $F^2=0$ sur le champ \'electromagn\'etique. La th\'eorie d'Einstein-Maxwell n'est retrouv\'ee qu'en imposant $\phi=0$ et $\alpha=0$. Enfin, dans la limite $\alpha\rightarrow\infty$, nous retrouvons la th\'eorie d'Einstein coupl\'ee \`a un champ scalaire de masse nulle et la th\'eorie d'Einstein sans source en imposant en plus que le dilaton $\phi$ soit constant.

\section{Solutions \`a sym\'etrie sph\'erique de EMD}%

Dans l'article \cite{MS}, les auteurs ont d\'emontr\'e que les solutions obtenues par Gibbons et Maeda \cite{GM,GHS} sont les seules solutions asymptotiquement plates d\'ecrivant des trous noirs r\'eguliers (c'est-\`a-dire poss\`edant un horizon non singulier)  \'electrostatiques et magn\'etostatiques de EMD. La solution \'electrostatique est donn\'ee par : 
\ba \lb{solGHS}
ds^2&=&-\frac{(r-r_+)(r-r_-)^\gamma}{r^{1+\gamma}} dt^2 
+\frac{r^{1+\gamma}\,dr^2}{(r-r_+)(r-r_-)^\gamma}+r^2 \left(1-\frac{r_-}{r}\right)^{1-\gamma}d\Omega^2\\
F&=&\frac{Q \e^{\alpha\phi_\infty}}{r^2}dr\wedge dt,\quad \e^{2\alpha(\phi-\phi_\infty)}=\left(1-\frac{r_-}{r}\right)^{1-\gamma}\lb{solGHS2}
\ea
o\`u
\be
\gamma\equiv \frac{1-\alpha^2}{1+\alpha^2},\qquad -1<\gamma\leq 1,
\ee
$r_+>r_->0$ et $\phi_\infty$ est la valeur asymptotique du dilaton. Dans le cas g\'en\'eral ($\gamma\neq 1$), cette solution d\'ecrit un trou noir avec un horizon en $r_+$ et une singularit\'e du genre espace en $r_-$.
Lorsque $\gamma=1$, la solution se r\'eduit \`a la solution de Reissner-Nordstr\"om et poss\`ede deux horizons en $r_\pm$ et une singularit\'e du genre temps en $r=0$.

La masse et la charge \'electrique sont :
\ba
{\cal M}=\frac{r_+}{2}+\frac{\gamma}{2}r_-,\qquad Q=\e^{-\alpha\phi_\infty}\sqrt{\frac{1+\gamma}{2}}\sqrt{r_+r_-}\lb{GHSQM}
\ea
La version magn\'etostatique s'obtient ais\'ement en utilisant la transformation de dualit\'e (\ref{transdual}).

Etudions, maintenant, la limite pr\`es de l'horizon (on fait tendre la coordonn\'ee radiale $r$ vers $r_h$, le rayon de l'horizon) et pr\`es de l'extr\'emalit\'e (on fait tendre les deux horizons $r_\pm$ l'un vers l'autre) de ces solutions. Cependant, il ne s'agit pas de faire brutalement $r \rightarrow r_-$ et $r_+\rightarrow r_-$ dans la m\'etrique. D\'etaillons le calcul.
Tout d'abord posons
\be
r_+=r_-+\epsilon \,b\quad\mbox{ et }\quad r=r_-+\epsilon\, \bar{r}\lb{limitph1},
\ee
o\`u $\epsilon$ est un param\`etre infinit\'esimal (que nous ferons tendre vers z\'ero plus loin). La solution  (\ref{solGHS})-(\ref{solGHS2}) devient
\ba \lb{limitg}
ds^2&=&-\frac{\epsilon^{1+\gamma}\,\bar{r}^\gamma\,(\bar{r}-b)}{(r_-+\epsilon \,\bar{r})^{1+\gamma}} dt^2 
+\frac{\epsilon^2(r_-+\epsilon \,\bar{r})^{1+\gamma}}{\epsilon^{1+\gamma}\,\bar{r}^\gamma\,(\bar{r}-b)}d\bar{r}^2+(\epsilon\,\bar{r})^{1-\gamma}(r_-+\epsilon \,\bar{r})^{1+\gamma}d\Omega^2\\
F&=&\e^{-\alpha\phi_\infty}\sqrt{\frac{1+\gamma}{2}}\sqrt{r_-(r_-+\epsilon b)}\,\epsilon\,d\bar{r}\wedge dt,\quad
 \e^{2\alpha\phi}=\e^{2\alpha\phi_\infty}\left(\frac{\epsilon\, \bar{r}}{r_-+\epsilon \,\bar{r}}\right)^{1-\gamma} \lb{limitFphi}.
\ea
Il est clair en examinant (\ref{limitg}) et (\ref{limitFphi}) que nous ne pouvons pas prendre $\epsilon\rightarrow 0$ brutalement.
Pour pouvoir prendre la limite $\epsilon\rightarrow 0$ ``proprement'', il est n\'ecessaire de changer l'\'echelle de $r_-$ et de $t$ ainsi que de fixer la valeur de $\phi_\infty$
\be \lb{limitph2}
t=\epsilon^{-1}\,\bar{t},\quad r_-=\epsilon^{-\alpha^2}r_0,\quad \phi_\infty=-\alpha\ln\epsilon.
\ee
La solution s'\'ecrit alors 
\be
ds^2=-\frac{\bar{r}^\gamma\,(\bar{r}-b)}{(r_0+\epsilon^{1+\alpha^2} \,\bar{r})^{1+\gamma}} d\bar{t}^2 
+\frac{(r_0+\epsilon^{1+\alpha^2} \,\bar{r})^{1+\gamma}}{\bar{r}^\gamma\,(\bar{r}-b)}d\bar{r}^2+\bar{r}^{1-\gamma}(r_0+\epsilon^{1+\alpha^2} \,\bar{r})^{1+\gamma}d\Omega^2
\ee
\be
F=\sqrt{\frac{1+\gamma}{2}}\sqrt{r_0(r_0+\epsilon^{1+\alpha^2} b)}d\bar{r}\wedge d\bar{t},\quad
 \e^{2\alpha\phi}=\left(\frac{\bar{r}}{r_0+\epsilon^{1+\alpha^2} \,\bar{r}}\right)^{1-\gamma}.
\ee
Nous pouvons maintenant prendre la limite $\epsilon\rightarrow 0$ sans probl\`eme et nous obtenons ainsi une solution \`a deux param\`etres de EMD, solution pr\'ec\'edemment obtenue par r\'esolution directe des \'equations du mouvement dans l'article \cite{CHM},

\ba \lb{solnas}
ds^2=-\frac{r^\gamma(r-b)}{r_0^{1+\gamma}}dt^2+\frac{r_0^{1+\gamma}}{r^\gamma(r-b)}[dr^2+r(r-b)d\Omega^2]\\
F=\sqrt{\frac{1+\gamma}{2}}\frac{1}{r_0} dr\wedge dt,\quad \e^{2\alpha\phi}=\left(\frac{r}{r_0}\right)^{1-\gamma} \lb{solnaselec},
\ea
o\`u nous avons renomm\'e $\bar{r}$ en $r$ et $\bar{t}$ en $t$. 

Le param\`etre $r_0$ est reli\'e \`a la charge \'electrique 
\be \lb{defQ}
Q=\frac{1}{4\pi}\int\e^{-2\alpha\phi}F^{0r}\sqrt{|g|}d\theta d\varphi=\sqrt{\frac{1+\gamma}{2}}r_0
\ee
et le param\`etre $b$ \`a la masse (voir section 6 pour le calcul)
\be \lb{mas}
{\cal M}=\frac{1-\gamma}{4}b.
\ee
La version magn\'etique est facilement obtenue en utilisant (\ref{transdual}),
\be \lb{solnasmag}
F=\sqrt{\frac{1+\gamma}{2}}\,r_0 \sin\theta\, d\theta\wedge d\varphi,\qquad \e^{2\alpha\phi}=\left(\frac{r}{r_0}\right)^{\gamma-1},
\ee
la m\'etrique \'etant toujours donn\'ee par (\ref{solnas}). Le param\`etre $r_0$ est cette fois-ci reli\'e \`a la charge magn\'etique
\be \lb{defP}
P=\frac{1}{4\pi}\int F_{\theta\varphi}d\theta d\varphi =\sqrt{\frac{1+\gamma}{2}}r_0.
\ee

Lorsque $\alpha\rightarrow\infty$ ($\gamma\rightarrow -1$), le champ scalaire se d\'ecouple alors que le champ \'electromagn\'etique s'annule. La solution (\ref{solnas}) se r\'eduit alors \`a la solution de Schwarzschild ($b=2M$) ce qui est en accord avec la valeur de la masse donn\'ee par la formule (\ref{mas}) ${\cal M}=M$. Lorsque $\alpha=0$ ($\gamma=1$), le dilaton s'annule et la m\'etrique (\ref{solnas}) peut se ramener \`a 
\be
ds^2=r_0^2\left(-(x^2-c^2)d\tau^2+\frac{dx^2}{x^2-c^2}+d\Omega^2\right)
\ee
en posant $x=r-b/2$, $c=b/2$ et $t=r_0^2\tau$. Cette m\'etrique est la solution de Bertotti-Robinson ($AdS_2\times S^2$) \cite{bertotti,robinson2}:
\be
ds^2=x^2dt^2+\frac{dx^2}{x^2}+d\Omega^2
\ee
\'ecrite dans un syst\`eme de coordonn\'ees couvrant l'espace-temps entier \cite{CG01}.
La solution de Bertotti-Robinson n'est pas un trou noir, ce qui est en accord avec le fait que la masse (\ref{mas}) s'annule pour $\gamma=1$.
Donc, la solution (\ref{solnas}) interpole contin\^ument entre la solution de Bertotti-Robinson ($\alpha=0$) et la solution de Schwarzschild ($\alpha\rightarrow\infty$).

Les solutions (\ref{solnas}) poss\`edent plusieurs caract\'eristiques inhabituelles. Premi\`erement, contrairement aux solutions (\ref{solGHS}), elles sont non asymptotiquement Minkowskiennes et non asymptotiquement AdS. Cependant, nous pouvons v\'erifier que la courbure scalaire
\be
R=\frac{1-\gamma^2}{2}r_0^{-1-\gamma}r^{\gamma-2}(r-b)
\ee
ainsi que les autres invariants de courbure, qui tendent vers les expressions suivantes \`a l'infini
\ba
R^{\mu\nu}R_{\mu\nu}&\simeq& \frac{(1+\gamma)^2(3+\gamma^2)}{4r_0^{2(1+\gamma)}}r^{2(\gamma-1)},\\ R^{\mu\nu\lambda\sigma}R_{\mu\nu\lambda\sigma}&\simeq& \frac{(1+\gamma)^2(11-10\gamma+7\gamma^2)}{4r_0^{2(1+\gamma)}}r^{2(\gamma-1)},
\ea
s'annulent \`a l'infini ($\gamma-1=-2\alpha^2/(1+\alpha^2)\leq 0$). De m\^eme, ces invariants restent finis dans tout l'espace except\'e bien s\^ur en $r=0$ o\`u se situe la singularit\'e de courbure.

Deuxi\`emement, contrairement au cas des solutions de Reissner-Nordstr\"om et de Kerr-Newman, le fond ($b=0$) sur lequel les trous noirs ($b>0$) se forment n'est pas Minkowski. De plus, ce fond n'est pas neutre et le param\`etre $r_0$ n'est donc pas un param\`etre associ\'e au trou noir mais un param\`etre associ\'e au fond. Nous verrons que ceci aura des cons\'equences lorsque nous calculerons la masse des solutions (\ref{solnas}) et lorsque nous examinerons la thermodynamique de ces solutions. En fait, il existe une famille de fonds charg\'es (\'electriquement ou magn\'etiquement) et \`a chaque membre de cette famille est associ\'ee une famille de trous noirs ($b>0$) (voir fig. 2.1)

Enfin, terminons cette section en remarquant que nous pouvons, sans modifier l'action (\ref{actionEMD}), effectuer la transformation suivante:
\be \lb{tphi}
\phi\rightarrow\phi+\phi_0,\qquad F\rightarrow \e^{\alpha\phi_0}F.
\ee
Or, les d\'efinitions des charges (\ref{defQ}) et (\ref{defP}) ne sont pas invariantes sous cette transformation. Par exemple, en appliquant cette transformation avec $\phi_0=-\alpha^{-1}\ln r_0$ pour la version \'electrique et $\phi_0=\alpha^{-1}\ln r_0$ pour la version magn\'etique, nous obtenons:
\be \lb{r0elec2}
F=\sqrt{\frac{1+\gamma}{2}}dr\wedge dt,
\ee
\be \lb{r0mag2}
F=\sqrt{\frac{1+\gamma}{2}}\sin\theta d\theta\wedge d\varphi.
\ee
Les valeurs des charges correspondantes sont:
\ba
Q=\frac{1}{4\pi}\int\e^{-2\alpha\phi}F^{0r}\sqrt{|g|}d\theta d\varphi=\sqrt{\frac{1+\gamma}{2}}\\
P=\frac{1}{4\pi}\int F_{\theta\varphi}d\theta d\varphi=\sqrt{\frac{1+\gamma}{2}}.
\ea
La transformation (\ref{tphi}) permet de s\'electionner un fond particulier ($r_0=1$ dans ce cas) et de passer d'un fond \`a un autre.
\begin{figure}
\centerline{\epsfxsize=250pt\epsfbox{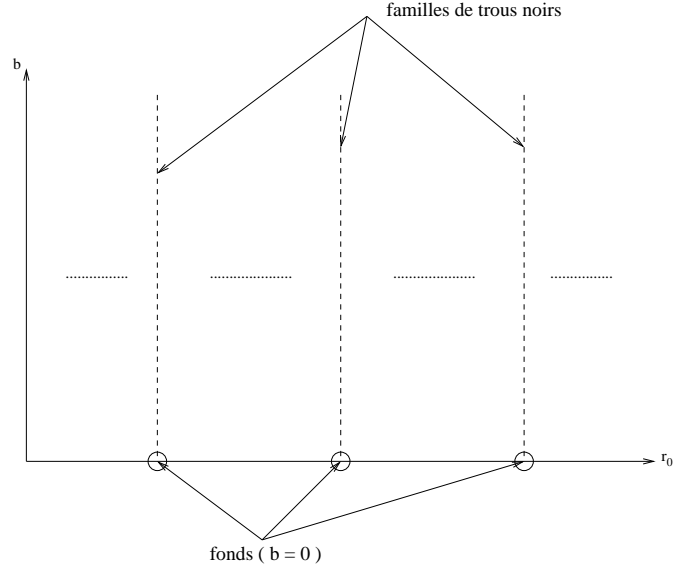}} 
\caption{A chaque valeur de $r_0$ correspond un fond ($b=0$) sur lequel une famille de trous noirs ($b> 0$) peut se former.}
\end{figure}

\section{Mouvement g\'eod\'esique dans la m\'etrique (\ref{solnas})}%

Examinons maintenant le mouvement g\'eod\'esique des particules dans le champ de gravitation donn\'e par la m\'etrique (\ref{solnas}). 
La trajectoire d'une particule se d\'epla\c{c}ant dans le champ de gravitation de (\ref{solnas}) est donn\'ee par l'\'equation des g\'eod\'esiques
\be \lb{eqdgeo2}
\frac{d^2x^\mu}{d\lambda^2}+\Gamma^\mu_{\nu\sigma}\frac{dx^\nu}{d\lambda}\frac{dx^\sigma}{d\lambda}=0
\ee
qui d\'erive du Lagrangien:
\be \lb{lag2}
{\cal L}=g_{\mu\nu}\dot{x}^\mu\dot{x}^\nu=-\frac{r^\gamma(r-b)}{r_0^{1+\gamma}}\dot{t}^2+\frac{r_0^{1+\gamma}}{r^\gamma(r-b)}\dot{r}^2+r_0^{-1-\gamma}r^{\gamma-1}(\dot{\theta}^2+\sin^2\theta\,\dot{\varphi}^2 )
\ee
o\`u $\dot{}=\p_\lambda$; $\lambda$ \'etant le param\`etre affine param\'etrisant la g\'eod\'esique. Or, $\lambda$ doit \^etre reli\'e au temps propre $\tau$ pour les g\'eod\'esiques du genre temps. D'autre part, $\tau$ est d\'efini par
\be
\tau=\int\sqrt{-g_{\mu\nu}dx^\mu dx^\nu}=\int_0^\lambda\sqrt{-g_{\mu\nu}\dot{x}^\mu \dot{x}^\nu}d\lambda.
\ee
Nous obtenons $\tau=\lambda$ si nous imposons la contrainte:
\be 
g_{\mu\nu}\dot{x}^\mu \dot{x}^\nu=-1
\ee
pour les g\'eod\'esiques du genre temps.
Nous pouvons \'etendre la validit\'e de cette \'equation au cas des g\'eod\'esiques du genre lumi\`ere et du genre espace en rempla\c{c}ant $1$ par $\varepsilon$,
\be\lb{cont2}
g_{\mu\nu}\dot{x}^\mu \dot{x}^\nu=-\varepsilon.
\ee
Nous allons utiliser cette contrainte pour \'etudier les g\'eod\'esiques de genre temps ($\varepsilon=1$, $ds^2<0$), du genre lumi\`ere ($\varepsilon=0$, $ds^2=0$) et du genre espace ($\varepsilon=-1$, $ds^2>0$).

Le Lagrangien (\ref{lag2}) \'etant ind\'ependant de $t$ et $\varphi$, nous en d\'eduisons les deux constantes du mouvement $E$ (\'energie) et $L$ (moment angulaire):
\ba
\frac{\p {\cal L}}{\p \dot{t}}&=&\frac{r^\gamma(r-b)}{r_0^{1+\gamma}}\dot{t}=E\\
\frac{\p {\cal L}}{\p \dot{\varphi}}&=&r_0^{1+\gamma}r^{1-\gamma}\sin^2\theta\,\dot{\varphi}=L
\ea
ce qui donne pour $\dot{t}$ et $\dot{\varphi}$
\ba
\dot{t}&=&\frac{r_0^{1+\gamma}}{r^\gamma(r-b)}E\\
\dot{\varphi}&=&\frac{L}{r_0^{1+\gamma}r^{1-\gamma}\sin^2\theta}.
\ea
En rempla\c{c}ant $\dot{t}$ et $\dot{\varphi}$ par leurs expressions, la contrainte (\ref{cont2}) devient
\be \lb{eqgeo2}
\dot{r}^2=E^2-r(r-b)\dot{\theta}^2-\frac{r-b}{r_0^{2(1+\gamma)}r^{1-2\gamma}}L^2-\frac{r^\gamma(r-b)}{r_0^{1+\gamma}}\varepsilon .
\ee

Etudions le mouvement g\'eod\'esique dans le plan \'equatorial ($\theta=\pi/2$). L'\'equation (\ref{eqgeo2}) devient:
\be
\dot{r}^2+V=E^2
\ee
avec
\be
V=\frac{r-b}{r_0^{2(1+\gamma)}r^{1-2\gamma}}L^2+\frac{r^\gamma(r-b)}{r_0^{1+\gamma}}\varepsilon.
\ee
Examinons tout d'abord le comportement des g\'eod\'esiques au voisinage de l'infini du genre espace. Lorsque $r$ tend vers l'infini le potentiel effectif est donn\'e par:
\be
V\rightarrow\left\{\begin{array}{ll}\frac{r^{1+\gamma}}{r_0^{1+\gamma}}\varepsilon,&\quad\varepsilon\neq 0\\\frac{r^{2\gamma}}{r_0^{2(1+\gamma)}}L^2,&\quad \varepsilon=0,\, L\neq 0\\0,&\quad\varepsilon=0,\, L=0.\end{array}\right.
\ee
Nous voyons que les g\'eod\'esiques du genre espace ($\varepsilon=-1$) peuvent s'\'etendre jusqu'\`a $r\rightarrow\infty$ ($1+\gamma>0$). Au contraire, les g\'eod\'esiques du genre temps ($\varepsilon=1$) ne peuvent pas s'\'etendre jusqu'\`a l'infini. En effet, le potentiel effectif $V$ tend vers l'infini ($1+\gamma>0$) et les g\'eod\'esiques du genre temps sont r\'eflechies par la barri\`ere de potentiel. Dans le cas des g\'eod\'esiques du genre lumi\`ere, deux cas sont \`a distinguer. Premi\`erement, lorsque $L=0$ ou $\gamma\leq 0$ ($\alpha\geq 1$), les g\'eod\'esiques du genre lumi\`ere peuvent se propager jusqu'\`a l'infini. Deuxi\`emement, lorsque $L\neq 0$ et $\gamma>0$ ($\alpha<1$), les g\'eod\'esiques du genre lumi\`ere sont r\'eflechies par la barri\`ere de potentiel.

Au voisinage de l'horizon, le potentiel effectif s'annule et l'\'equation de contrainte devient
\be
\dot{r}^2=E^2
\ee
qui s'int\`egre par
\be
r=E\lambda+\mbox{cste}.
\ee
Nous voyons alors que l'horizon ($r=b$) se situe \`a distance g\'eod\'esique finie ($\lambda$ fini) et qu'il est traversable.

Finalement, au voisinage de la singularit\'e, le potentiel devient
\be
V\rightarrow\left\{\begin{array}{cl}-\frac{b\,L^2}{r_0^{2(1+\gamma)}}r^{2\gamma-1},&\quad L\neq 0\\-\frac{b\,\varepsilon}{r_0^{1+\gamma}}r^\gamma,&\quad L=0, \varepsilon\neq 0 \\ 0,&\quad L=\varepsilon=0.\end{array}\right. 
\ee
Nous voyons que la singularit\'e est \`a distance g\'eod\'esique finie. Lorsque $L\neq 0$, toutes les g\'eod\'esiques atteignent la singularit\'e ($2\gamma-1<0$, $-1<\gamma<1$). Lorsque $L=0$, les g\'eod\'esiques du genre lumi\`ere atteignent la singularit\'e ($V=0$) alors que plusieurs cas, en fonction de la valeur de $\gamma$, sont \`a distinguer pour les g\'eod\'esiques des genres temps et espace ($\varepsilon\neq 0$). Lorsque $\gamma>0$, le potentiel effectif tend vers z\'ero et toutes les g\'eod\'esiques atteignent la singularit\'e. Lorsque $\gamma=0$, les g\'eod\'esisques du genre temps et les g\'eod\'esiques du genre espace (telle que $E+b\,\varepsilon/r_0^{1+\gamma}>0$) atteignent la singularit\'e. Enfin, lorsque $\gamma<0$, les g\'eod\'esiques du genre temps atteignent la singularit\'e ($\varepsilon=1$, $V\rightarrow -\infty$) alors que les g\'eod\'esiques du genre espace sont r\'eflechies par la singularit\'e.

\section{Diagrammes de Penrose de (\ref{solnas})}%

Dans cette section, nous allons construire les diagrammes de Penrose de la solution (\ref{solnas}).
Dans la section 2 du chapitre 1, nous avons obtenu l'extension analytique maximale de l'espace-temps de Minkowski en introduisant un syst\`eme de coordon\'ees appropri\'e puis nous avons construit le diagramme de Penrose correspondant. Cependant, il n'est pas toujours ais\'e de trouver un tel syst\`eme de coordonn\'ees dans le cas d'espace-temps plus complexes. Il est alors plus commode de construire le diagramme de Penrose morceau par morceau comme nous allons le voir maintenant.

En n\'egligeant les coordonn\'ees angulaires, la solution (\ref{solnas}) peut s'\'ecrire
\be
ds^2=\frac{r^\gamma(r-b)}{r_0^{1+\gamma}}\left(-dt^2+dx^2\right)
\ee
o\`u nous avons introduit la nouvelle coordonn\'ee $x$ d\'efinie par 
\be
dx=\frac{r_0^{1+\gamma}}{r^\gamma(r-b)}dr.
\ee
La partie $(t,r)$ de la m\'etrique (\ref{solnas}) est donc conforme \`a celle de la solution de Minkowski. Par cons\'equent, sa structure conforme est identique \`a celle de Minkowski et les diagrammes de Penrose correspondants sont constitu\'es des m\^emes morceaux (voir fig. 2.2) que ceux composant le diagramme de Penrose de Minkowski (fig. 1.2).

\begin{figure}
\centerline{\epsfxsize=300pt\epsfbox{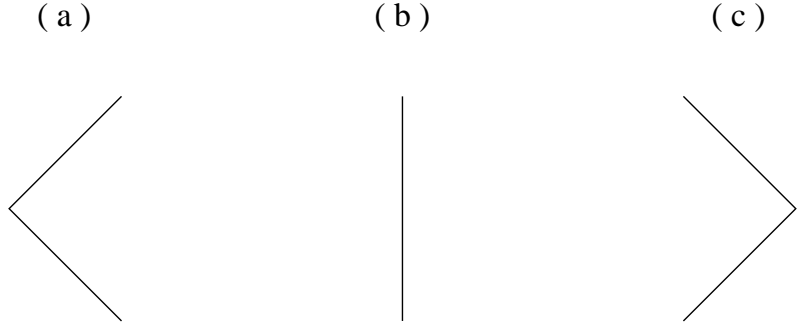}} 
\caption{Les diff\'erents ``morceaux'' composant le diagramme de Penrose de la structure conforme de la solution de Minkowski.}
\end{figure}

Pour construire les diagrammes de Penrose associ\'es \`a la solution (\ref{solnas}), il faut distinguer plusieurs cas en fonction des valeurs de la constante de couplage du dilaton $\alpha$ ($\alpha=0$, $0<\alpha<1$, $\alpha=1$ et $\alpha>1$) et du param\`etre $b$ ($b<0$, $b=0$ et $b>0$).

Le cas $\alpha=0$ correspond, comme nous l'avons signal\'e pr\'ec\'edemment, \`a la solution de Bertotti-Robinson. 

D\'etaillons la construction des diagrammes de Penrose (qui sont donn\'es dans la fig. 2.3) dans le cas $\alpha=1$ ($\gamma=0$).
\begin{figure}
\centerline{\epsfxsize=400pt\epsfbox{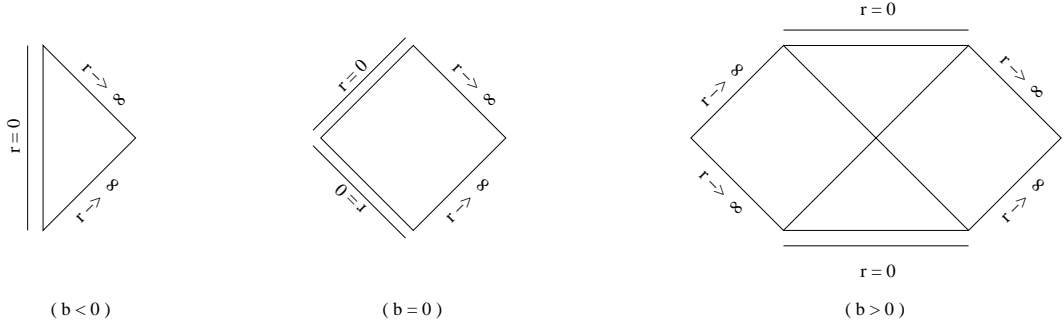}} 
\caption{Diagrammes de Penrose de (\ref{solnas}) lorsque $\alpha= 1$}
\end{figure}
Lorsque $b<0$, il n'y a pas d'horizon et $r$ varie de z\'ero (o\`u se trouve la singularit\'e de courbure) \`a l'infini. Lorsque $r$ tend vers z\'ero, la variable $x$
\be
x=r_0\ln(r-b)+\mbox{cste}
\ee
tend vers une valeur finie que nous pouvons fixer \`a z\'ero en posant $\mbox{cste}=-r_0\ln(-b)$. Par cons\'equent, lorsque $r=0$, la partie du diagramme de Penrose correspondante est identique \`a celle de Minkowski lorsque $r=0$. Il s'agit donc d'une ligne de genre temps (verticale)(voir. 2.2 (b)).
De m\^eme, lorsque $r$ tend vers l'infini la variable $x$ tend vers l'infini et la partie du diagramme de Penrose correspondante est alors identique \`a celle de Minkowski lorsque $r$ tend vers l'infini (voir fig. 2.2 (c)). En r\'eunissant les deux parties (b) et (c), nous voyons que le diagramme de Penrose est identique \`a celui de Minkowski (voir fig. 1.3) avec une double ligne du genre temps en $r=0$ pour signaler la pr\'esence d'une singularit\'e de courbure.

Lorsque $b=0$, il n'y a pas d'horizon et $r$ varie toujours de z\'ero \`a l'infini. Cependant, cette fois-ci la variable $x$ est
\be
x=r_0\ln(r)\in]-\infty,\infty[.
\ee
Lorsque $r=0$, la partie du diagramme de Penrose est identique \`a celle de Minkowski lorsque $x\rightarrow -\infty$ (fig 2.2 (a)). Lorsque $r\rightarrow\infty$, la partie du diagramme de Penrose est identique \`a celle de Minkowski pour $x\rightarrow\infty$ (fig. 2.2 (c)). En r\'eunissant les deux parties, nous obtenons le deuxi\`eme diagramme de la figure 2.3 o\`u la singularit\'e de courbure est signal\'ee par une double ligne de genre lumi\`ere (inclin\'ee \`a $45^\circ$).

Le cas $b>0$ est un peu plus complexe. Il y a un horizon en $r=b$ et il faut alors distinguer deux domaines de variation de $r$. Lorsque $r\in[b,\infty[$, la variable $x$ 
\be
x=r_0\ln(r-b)
\ee
varie de $-\infty$ \`a $+\infty$. Lorsque $r=b$, la variable $x$ tend vers $-\infty$ et la partie du diagramme de Penrose est donn\'ee par la figure 2.2(a). Lorsque $r\rightarrow\infty$, la variable $x$ tend vers  $\infty$ et la partie du diagramme de Penrose correspondante est repr\'esent\'ee dans la figure 2.2(c). En regroupant les parties (a) et (c), nous obtenons le morceau de diagramme repr\'esent\'e dans la figure figure 2.4(b).
\begin{figure}
\centerline{\epsfxsize=250pt\epsfbox{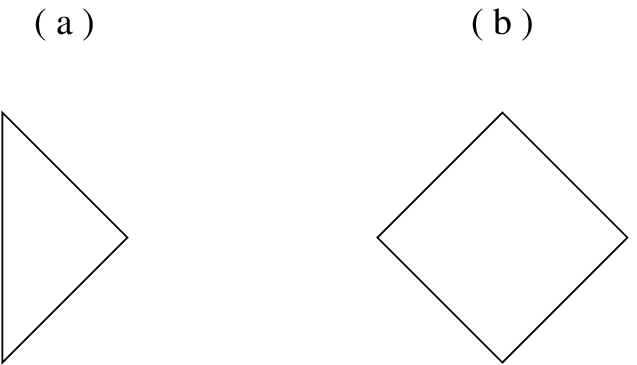}} 
\caption{Les deux parties constituant le diagramme de Penrose de la solution (\ref{solnas}) lorsque $\alpha=1$ et $b>0$.}
\end{figure}
Lorsque $r\in[0,b[$, la variable $x$ est 
\be
x=r_0\ln(b-r)+\mbox{cste}
\ee
qui varie de $-\infty$ \`a $0$ ($\mbox{cste}=-r_0\ln(b)$). Lorsque $r=b$, la variable $x$ tend vers $-\infty$ et la partie du diagramme de Penrose est donn\'ee dans la figure 2.2(a). Lorsque $r=0$, $x=0$ et la partie est fig. 2.2(b). Le morceau du diagramme de Penrose pour $r\in[0,b[$ est donn\'e par la figure 2.4(a). Le diagramme de Penrose lorsque $b>0$ est un assemblage des parties (a) et (b) de la figure 2.4 en prenant garde au changement de signature de la m\'etrique, qui se traduit par un basculement (rotation de $90^\circ$) du morceau du diagramme de Penrose correspondant (fig. 2.4 (a) dans notre cas), lors du franchissement de l'horizon ($r=b$ ici).

Les diagrammes de Penrose dans les cas $0<\alpha<1$ et $\alpha>1$ se construisent de mani\`ere similaire et sont repr\'esent\'es dans les figures 2.5 et 2.6.

\begin{figure}
\centerline{\epsfxsize=400pt\epsfbox{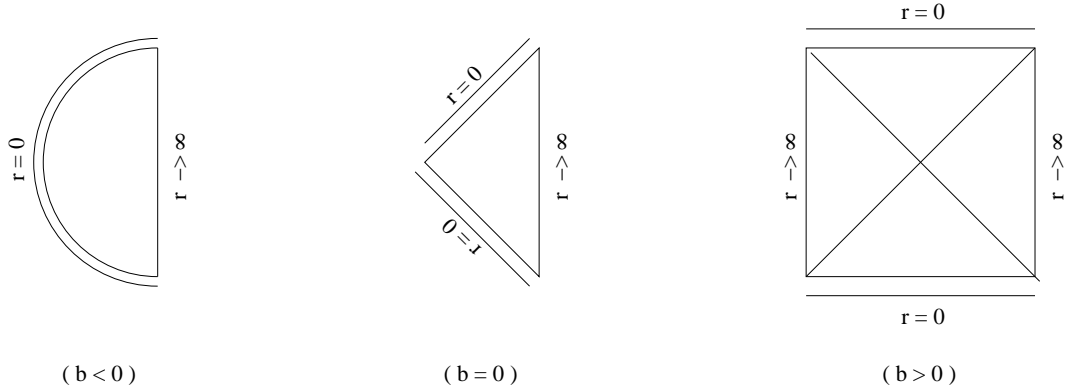}} 
\caption{Diagrammes de Penrose de (\ref{solnas}) lorsque $0<\alpha< 1$}
\end{figure}

\begin{figure}
\centerline{\epsfxsize=400pt\epsfbox{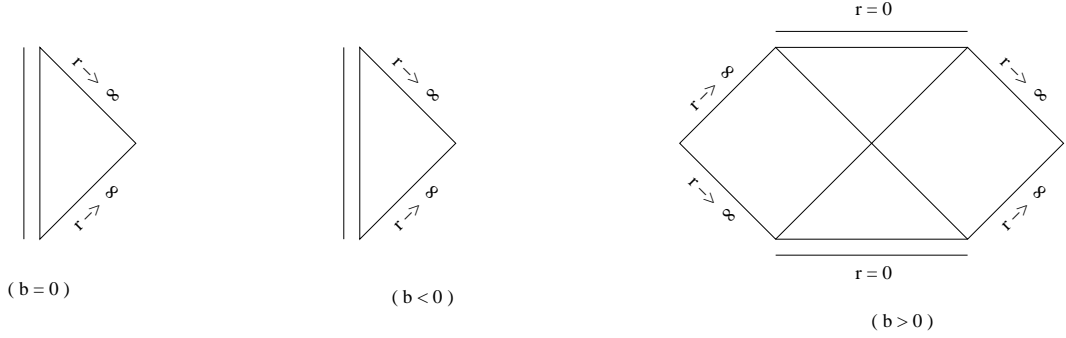}} 
\caption{Diagrammes de Penrose de (\ref{solnas}) lorsque $\alpha> 1$}
\end{figure}

\section{Le formalisme quasilocal}%

Le formalisme quasilocal peut bien s\^ur \^etre adapt\'e au cas de la gravitation en pr\'esence de sources. Le cas de la th\'eorie d'Einstein-Maxwell a \'et\'e examin\'e dans les r\'ef\'erences \cite{HH2,HR}. Pour EMD, le calcul a \'et\'e conduit de mani\`ere d\'etaill\'ee dans \cite{booth}. Nous allons maintenant rappeler les grandes lignes du calcul afin de mieux comprendre l'origine de la contribution de la mati\`ere \`a l'\'energie quasilocale et au moment angulaire quasilocal. Consid\'erons la partie mat\'erielle de l'action de EMD (\ref{actionEMD}) 

\be \lb{actionEMDmat}
S_m =- \frac{1}{16\pi}\int dt\int_{\Sigma_t} d^3x \sqrt{g} \left\{2\partial_\mu\phi\partial^\mu\phi+\e^{-2\alpha\phi} F_{\mu\nu} F^{\mu\nu}\right\}.
\ee
Les moments conjugu\'es de $\phi$ et de $A_i$ sont  
\be
p_\phi=-\frac{\sqrt g}{4\pi}\p^0\phi,\quad \Pi^i=\frac{\sqrt g}{4\pi}\e^{-2\alpha\phi}F^{i0}\equiv\frac{\sqrt g}{4\pi}E^i
\ee
o\`u $E^i$ est le champ \'electrique modifi\'e, par rapport au cas de la th\'eorie d'Einstein-Maxwell, par la pr\'esence du dilaton.

\noindent En utilisant les relations
\be \lb{relphi}
\p_0\phi=\frac{4\pi N^2}{\sqrt g}p_\phi+N^i\p_i\phi,\quad \p^i\phi=h^{ij}\p_j\phi+\frac{4\pi N^i}{\sqrt g}p_\phi,
\ee
\be\lb{relF}
F_{0i}=N^j\bar{F}_{ji}+\frac{4\pi N}{\sqrt h}\e^{2\alpha \phi}\Pi_i,\quad F^{ij}=\bar{F}^{ij}+\frac{4\pi}{\sqrt g}(N^i\Pi^j-N^j\Pi^i)\e^{2\alpha\phi}
\ee
(o\`u $\bar{F}_{ij}=\p_iA_j-\p_jA_i$ est un tenseur \`a trois dimensions dont les indices sont \'elev\'es et abaiss\'es par la m\'etrique $h_{ij}$) obtenues en utilisant la m\'etrique (\ref{ADM}) et son inverse
\be \lb{ADMinv}
g^{\mu\nu}\p_\mu\p_\nu=-N^{-2}\p_t\p_t+2N^{-2}N^i\p_i\p_t+(h^{ij}-N^{-2}N^iN^j)\p_i\p_j,
\ee
nous pouvons r\'e\'ecrire l'action (\ref{actionEMDmat}) sous la forme suivante
\be \lb{act}
S =\int dt\left[\int_{\Sigma_t} d^3x \left(p_\phi\p_0\phi+\Pi^i\p_0A_i-N{\cal H}-N^i{\cal H}_i\right)-\int_{\Sigma_t} d^3x \,\Pi^i\p_iA_0\right]
\ee
o\`u
\ba
{\cal H}&=&\frac{2\pi}{\sqrt h}p_\phi^2+\frac{\sqrt h}{8\pi}h^{ij}\p_i\phi\p_j\phi+\frac{2\pi}{\sqrt h}\e^{2\alpha\phi}\Pi_i\Pi^i+\frac{\sqrt h}{16\pi}\e^{-2\alpha\phi}\bar{F}_{ij}\bar{F}^{ij}\\
{\cal H}_i&=&p_\phi\p_i\phi+\bar{F}_{ij}\Pi^j
\ea
sont les contraintes Hamiltoniennes associ\'ees aux multiplicateurs de Lagrange $N$ et $N^i$.
Le dernier terme dans l'action (\ref{act}) s'int\`egre par partie:
\be
\int_{\Sigma_t} d^3x \,\Pi^i\p_iA_0=\int_{\Sigma_t} d^3x\, \p_i(A_0\Pi^i)-\int_{\Sigma_t} d^3x \,A_0\p_i\Pi^i
\ee
puis, en utilisant le th\'eor\`eme de Gauss, nous pouvons int\'egrer le premier terme
\be
\int_{\Sigma_t} d^3x\, \p_i(A_0\Pi^i)=\frac{1}{4\pi}\int_{\Sigma_t}\sqrt hD_i(NA_0E^i)d^3x=\frac{1}{4\pi}\int_{S^r_t}\sqrt \sigma d^2x NA_0E^in_i
\ee
ce qui donne finalement pour l'action (\ref{actionEMDmat})
\be \lb{act2}
S_m =\int dt\left[\int_{\Sigma_t} d^3x \left(p_\phi\p_0\phi+\Pi^i\p_0A_i-N{\cal H}-N^i{\cal H}_i-A_0 {\cal H}_A\right)-\frac{1}{4\pi}\int_{S^r_t}\sqrt \sigma d^2x NA_0E^in_i\right]
\ee
o\`u
\be
{\cal H}_A=-\p_i\Pi^i=-\frac{1}{4\pi}\p_i(\sqrt g E^i)
\ee
est la contrainte, qui impose que la divergence du champ \'electrique soit nulle, associ\'ee au multiplicateur de Lagrange $A_0$. En conclusion, nous voyons que le dilaton ne contribue pas directement \`a l'\'energie quasilocale. Il n'appara\^it qu'indirectement, dans la contribution du champ \'electromagn\'etique (dans le $E^i$). 

Les contributions \`a l'\'energie quasilocale et au moment angulaire quasilocal de la partie mat\'erielle (\ref{actionEMDmat}) de EMD sont donc:
\be \lb{equasdil}
E_m=\int_{S^r_t}A_0\bar{\Pi}^in_id^2x,\quad J_m=-\int_{S^r_t}A_\varphi\bar{\Pi}^in_id^2x
\ee
o\`u nous avons introduit $\bar{\Pi}^i=(\sqrt \sigma/\sqrt h) \Pi^i$.

\section{Masse et thermodynamique des solutions non asymptotiquement plates}%

Nous sommes maintenant en mesure de calculer la masse des solutions (\ref{solnas}). En regroupant la contribution gravitationnelle (\ref{equas}) et la contribution mat\'erielle (\ref{equasdil}), nous obtenons la formule suivante pour l'\'energie quasilocale
\be \lb{equasb}
E=\oint_{S^r_t}\left(N( \epsilon-\epsilon_0)+2N^i\pi_{ij}n^j+A_0(\bar{\Pi}^i-\bar{\Pi}_0^i)n_i\right)d^2x
\ee
o\`u nous avons retranch\'e la contribution d'une solution de fond statique ($N_0^i=0$). Cette solution de fond est choisie de telle sorte que les $N$, $N^i$, $\sqrt \sigma$ et $A_0$ de la solution dont nous voulons calculer l'\'energie quasilocale et ceux de la solution de fond co\"incident sur la surface d'int\'egration $S_t^r$, afin de pouvoir imposer les m\^emes conditions aux limites de type Dirichlet. La masse est donn\'ee par l'\'energie quasilocale (\ref{equasb}) calcul\'ee sur une surface \`a l'infini $S^\infty_t$.

Commen\c{c}ons par calculer la masse de la version \'electrique (\ref{solnaselec}) de la solution (\ref{solnas}). En comparant (\ref{ADM}) et (\ref{solnas}), nous en d\'eduisons 
\be
N=\sqrt{\frac{r^\gamma(r-b)}{r_0^{1+\gamma}}},\quad N^i=0, \quad h_{ij}dx^idx^j=\frac{r_0^{1+\gamma}}{r^\gamma(r-b)}\left[dr^2+r(r-b)d\Omega^2\right].
\ee
De plus, la m\'etrique de $S^r_t=\Sigma_t\cap \Sigma^r$ (voir section 1.3 et fig. 1.4) est donn\'ee par
\be
\sigma_{ab}dx^a dx^b=(h_{ab}-n_an_b)dx^a dx^b=r_0^{1+\gamma} r^{1-\gamma}d\Omega^2
\ee
qui est la m\'etrique d'une sph\`ere o\`u 
\be
n^r=\sqrt{\frac{r^\gamma(r-b)}{r_0^{1+\gamma}}}
\ee
est la normale aux hypersurfaces $\Sigma^r$.
La solution (\ref{solnas}) \'etant statique ($N^i=0$), le deuxi\`eme terme dans la formule (\ref{equasb}) ne contribue pas. Calculons les deux autres termes. Le premier terme est proportionnel \`a la courbure extrins\`eque de $S^r_t$:
\be
k=\sigma^{\mu\alpha}D_\alpha n_\mu=-\left(\sigma^{\theta\theta}\,{}^3\Gamma_{\theta\theta}^r+\sigma^{\varphi\varphi}\,{}^3\Gamma_{\varphi\varphi}^r\right)n_r=\frac{\gamma-1}{\sqrt{r_0^{1+\gamma}}}r^{\frac{\gamma-1}{2}}\sqrt{1-\frac{b}{r}}
\ee
o\`u $D$ est la d\'eriv\'ee covariante et ${}^3\Gamma$ sont les symboles de Christoffel associ\'es \`a la m\'etrique $h_{ij}$ 
\be
{}^3\Gamma_{\varphi\varphi}^r=\sin^2\theta\, {}^3\Gamma_{\theta\theta}^r=\frac{1-\gamma}{2}(r-b)\sin^2\theta.
\ee
Le deuxi\`eme terme est proportionnel au moment conjugu\'e de $A_r$
\be
\Pi^r=\bar{\Pi}^rn_r=-\frac{1}{4\pi}\sqrt{\frac{1+\gamma}{2}}r_0\sin\theta.
\ee
Les deux termes contribuant \`a l'\'energie quasilocale sont donc
\be
N\epsilon=\frac{\gamma-1}{8\pi}r\sin\theta,\quad A_0\Pi^r=-\frac{1+\gamma}{8\pi}r\sin\theta.
\ee
Or, nous voyons que ces deux termes divergent lorsque $r$ tend vers l'infini ce qui n'est pas surprenant puisque, comme nous l'avons fait remarqu\'e dans le chapitre 1 (section 3), ceci est g\'en\'eralement le cas. Il est n\'ecessaire de retrancher la contribution d'une solution de fond. Le choix le plus logique semble \^etre de choisir le fond charg\'e ($b=0$) sur lequel les trous noirs existent. Nous pouvons v\'erifier que ce choix implique bien que les ($N$, $N^i$, $\sqrt{\sigma}$ et $A_t$) co\"incident avec les ($N_0$, $N_0^i$, $\sqrt{\sigma_0}$ et $A^0_t$) sur la surface d'int\'egration $S^\infty_t$.
La contribution de la solution de fond \'etant ajout\'ee pour annuler la contribution du terme dominant dans l'\'energie, il faut tenir compte de l'ordre d'approximation suivant en $1/r$ pour $\epsilon$ et $\Pi^r$. Pour le terme gravitationnel, nous avons
\be
\epsilon=\frac{\gamma-1}{8\pi}\sqrt{r_0^{1+\gamma}}r^{\frac{1-\gamma}{2}}\left(1-\frac{b}{2r}+\cdots\right)\sin\theta
\ee
et la contribution du terme gravitationnel \`a l'\'energie devient
\be
N(\epsilon-\epsilon_0)=\frac{1-\gamma}{16\pi}b\sin\theta.
\ee
Dans le cas de la contribution \'electromagn\'etique, $\Pi^r$ est constant et ne d\'epend pas de $b$. Il s'ensuit que $\Pi_0^r=\Pi^r$ et que la contribution du terme \'electromagn\'etique est nulle
\be
A_0(\Pi^r-\Pi_0^r)=0.
\ee
En regroupant les deux contributions, nous obtenons pour la masse
\be \lb{massolnas}
{\cal M}=\frac{1-\gamma}{4}b.
\ee

Nous pouvons aussi calculer la temp\'erature (en utilisant la formule (\ref{fT})) et l'entropie du trou noir
\ba
T&=&\frac{1}{2\pi}(n^i\p_iN)|_h=\frac{b^\gamma}{4\pi r_0^{1+\gamma}},\\
S&=&\frac{A}{4}=\frac{1}{4}\frac{1}{4\pi}\int\sqrt{\sigma}d\theta d\varphi=\pi r_0^{1+\gamma}b^{1-\gamma}.
\ea

Nous pouvons maintenant v\'erifier si la version \'electrique de la solution (\ref{solnas}) v\'erifie la premi\`ere loi de la thermodynamique des trous noirs. Nous avons: 
\be
d{\cal M}=\frac{1-\gamma}{4}db,\quad TdS-A_0 dQ=\frac{1-\gamma}{4}db-\frac{1+\gamma}{4r_0}b \,dr_0.
\ee
La premi\`ere loi est donc satisfaite uniquement si $dr_0=0$, c'est-\`a-dire si nous fixons la valeur de la charge \'electrique. Ce r\'esultat peut se comprendre en se rappelant que la charge \'electrique n'est pas associ\'ee au trou noir mais au fond sur lequel le trou noir se forme. Il est donc logique de ne pas varier la charge puisque ce n'est pas un param\`etre du trou noir. De plus, en calculant la masse avec le formalisme quasilocal nous avons d\'ej\`a retranch\'e la contribution du fond. En conclusion, la version \'electrique de la solution (\ref{solnas}) v\'erifie la premi\`ere loi si nous ne varions pas la charge (ce qui semble logique au regard de ce que nous avons dit plus haut). Cependant, seule une \'etude approfondie de la thermodynamique de ces trous noirs au comportement asymptotique atypique pourrait ent\'eriner ce r\'esultat.

Remarquons que la premi\`ere loi d\'epend de la jauge dans laquelle nous \'ecrivons $A_0$. Dans le cas d'un trou noir asymptotiquement plat, il y a une jauge naturelle qui est celle dans laquelle $A_0$ est fini sur l'horizon et nul \`a l'infini. Dans le cas d'un trou noir non asymptotiquement plat il n'y a pas de choix de jauge pr\'eferentiel car $A_0$ diverge \`a l'infini.

Nous pourrions bien s\^ur refaire le m\^eme genre de calcul dans le cas de la version magn\'etique (\ref{solnasmag}), mais nous pouvons obtenir le r\'esultat sans calcul en faisant les remarques suivantes. Tout d'abord $A_0=0$ et donc le troisi\`eme terme ne contribue pas \`a l'\'energie quasilocale. Donc, seul le premier terme contribue, comme pour la version \'electrique. Or, ce terme ne d\'epend que de quantit\'es calcul\'ees \`a partir de la m\'etrique. La m\'etrique \'etant la m\^eme pour la version \'electrique et pour la version magn\'etique, la contribution de ce terme \`a l'\'energie quasilocale sera la m\^eme que pour la version \'electrique. Autrement dit la masse de la version magn\'etique est la m\^eme (\ref{massolnas}) que celle de la version \'electrique. Concernant la premi\`ere loi, nous avons
\be
d{\cal M}=\frac{1-\gamma}{4}db,\quad TdS=\frac{1-\gamma}{4}db+\frac{1+\gamma}{4r_0}b\, dr_0.
\ee
et donc la m\^eme conclusion que pour la version \'electrique.

\newpage
\thispagestyle{empty}
\null
\chapter{Nouvelles solutions trou noir de EMD pour $\alpha^2=3$}

Dans ce chapitre nous allons construire de nouvelles solutions de la th\'eorie d'Einstein-Maxwell dilatonique pour la valeur particuli\`ere de la constante de couplage du dilaton $\alpha^2=3$. Pour cela, nous allons tout d'abord introduire le lien unissant EMD3 (EMD avec $\alpha^2=3$) et la th\'eorie d'Einstein \`a cinq dimensions (E5) ainsi que le mod\`ele $\sigma$ de E5. Puis nous utiliserons ces deux ingr\'edients pour g\'en\'erer de nouvelles solutions de EMD3 et pour montrer quels sont les liens unissant ces nouvelles solutions et les solutions de E5. Pour finir, en utilisant le formalisme quasilocal, que nous avons adapt\'e \`a EMD dans la section 5 du chapitre pr\'ec\'edent, nous calculerons les masses et moments angulaires des nouvelles solutions. 

Dans ce chapitre, le syst\`eme d'unit\'e utilis\'e est tel que $G=1$.

\section{Lien entre EMD3 et la th\'eorie d'Einstein \`a cinq dimensions}%

Avant d'aborder le lien existant entre EMD3 et E5, nous allons tout d'abord introduire la notion de mod\`ele $\sigma$ \cite{NK,BM84,SKMHH}. Soit ${\cal M}$ un espace Riemannien de dimension $m$ de m\'etrique $g_{\mu\nu}(x)$ et ${\cal N}$ un espace de dimension $n$ (appel\'e espace des potentiels ou espace cible) de m\'etrique ${\cal G}_{AB}(\Phi)$. On dit que l'espace ${\cal M}$ admet un mod\`ele $\sigma$ si il existe une fonction $\Phi^A(x^\mu)$ de ${\cal M}$ dans ${\cal N}$ dite harmonique, c'est-\`a-dire satisfaisant aux \'equations d'Euler-Lagrange
 d\'erivant de l'action
\be \lb{actionsigma}
S_\sigma=\int \sqrt g \,{\cal G}_{AB}\, g^{\mu\nu}\Phi^A_{,\mu}\Phi^B_{,\nu}\,.
\ee
Nous voyons que (\ref{actionsigma}) est invariante sous les transformations infinit\'esimales
\be \lb{transinf}
\Phi^A\rightarrow\Phi^A+\epsilon \,X^A(\Phi)
\ee
si $X^A$ est un vecteur de Killing de ${\cal N}$, c'est-\`a-dire si $X_{(A;B)}=0$. L'un des int\'er\^ets du mod\`ele $\sigma$ est que les transformations finies (r\'esultant des transformations infinit\'esimales (\ref{transinf})) peuvent \^etre utilis\'ees pour g\'en\'erer une nouvelle solution $(\Phi^{'}, g^{'})$ \`a partir d'une solution connue $(\Phi,g)$. Les th\'eories d'Einstein \cite{ernst} et d'Einstein-Maxwell \cite{IW,K} admettent un tel mod\`ele. Le mod\`ele $\sigma$ est coupl\'e \`a la gravitation
\be \lb{actionsemd}
S_\sigma=\frac{1}{2}\int \left({\cal R}_{ij} -{\cal G}_{AB} \partial_i \Phi^A\partial_j \Phi^B\right) h^{ij}\sqrt h \,d^3x.
\ee
Les \'equations du mouvement de ce mod\`ele sont l'\'equation d'Einstein \`a trois dimensions et les \'equations du mouvement des $\Phi^A$ 
\ba \lb{eqe}
{\cal R}_{ij}={\cal G}_{AB}\partial_i \Phi^A \partial_j\Phi^B\\
\frac{1}{\sqrt h}\partial_i(\sqrt h\, h^{ij}{\cal G}_{AB}\partial_j \Phi^B)=0. \lb{eqmvtPhi}
\ea
Les mod\`eles $\sigma$ de nombreuses th\'eories \`a quatre dimensions, d\'ecrivant la \RG coupl\'ee \`a des champs de jauge ab\'eliens et \`a des champs scalaires, sont d\'eriv\'es dans la r\'ef\'erence \cite{BGM}. Mentionnons aussi les travaux de Geroch \cite{Geroch} et Cremmer et al. \cite{CreJu}.
 
Dans l'article \cite{GGK}, les auteurs ont montr\'e que, dans le cas g\'en\'eral ($\alpha\neq 0,\sqrt 3$), les g\'en\'erateurs du groupe laissant invariant le mod\`ele $\sigma$ de EMD sont au nombre de cinq et que trois de ces g\'en\'erateurs conduisent \`a des transformations de jauge. Les deux autres transformations sont des combinaisons de translations du dilaton et de transformations d'\'echelle. Le faible nombre de transformations non triviales du groupe dans le cas $\alpha\neq 0,\sqrt 3$ rend d\'elicate la g\'en\'eration de nouvelles solutions, bien que cela soit possible \cite{GGK}. Au contraire, dans les cas particuliers $\alpha=0$ et $\alpha^2=3$, le nombre de g\'en\'erateurs est de huit. Parmi les transformations associ\'ees se trouvent, entre autres, les transformations de Ehlers et Harrison \cite{ehlers,harr} qui sont absentes dans le cas g\'en\'eral ($\alpha\neq 0,\sqrt 3$). Les transformations du groupe sont alors suffisamment g\'en\'erales pour permettre de g\'en\'erer plus facilement de nouvelles solutions que dans le cas g\'en\'eral ($\alpha\neq 0,\sqrt 3$). 

Dans le cas $\alpha=0$, le dilaton se d\'ecouple du champ \'electromagn\'etique. Ce cas est donc peu int\'eressant \`a \'etudier. Nous allons nous concentrer, dans le reste du chapitre, sur le cas $\alpha^2=3$. Or, dans ce cas, EMD peut \^etre obtenue par r\'eduction dimensionnelle de la th\'eorie d'Einstein \`a cinq dimensions (E5). En effet, en supposant l'existence d'un vecteur de Killing du genre espace ($\p_5$), toute solution de E5
\be \lb{actionE5}
S_5=\frac{1}{16\pi}\int \sqrt G R_5d^5x
\ee
peut \^etre \'ecrite sous la forme
\be \lb{gE5}
ds_5^2=\e^{2\phi/\sqrt 3}ds_4^2+\e^{-4\phi/\sqrt 3}(dx^5+2A_\mu dx^\mu)^2
\ee
o\`u $ds_4^2$, $\phi$ et $A_\mu$ sont ind\'ependants de $x^5$. En utilisant cette param\'etrisation de la m\'etrique, la courbure scalaire \`a cinq dimensions se d\'ecompose de la fa\c{c}on suivante
\be
\sqrt{G}R_5=\sqrt{g_4}\left[R_4-2(\p_\phi)^2-\frac{1}{4}\,\e^{-2\sqrt 3\phi}F^2-\frac{2}{\sqrt 3}\nabla^2\phi\right]
\ee
et nous retrouvons l'action de EMD3 (en n\'egligeant la divergence).

\noindent Donc, \`a toute solution $ds_4^2$ de EMD3 correspond une solution de E5.

Les solutions d\'ecrivant des trous noirs de E5 sont bien connues. Il existe trois familles de solutions d\'ecrivant des trous noirs asymptotiquement plats de E5. Gibbons et Wiltshire \cite{GW5} ont obtenu les solutions d\'ecrivant des trous noirs \`a sym\'etrie sph\'erique dans trois des quatre dimensions d'espace. Cette solution fut ensuite g\'en\'eralis\'ee au cas stationnaire par Rasheed \cite{rasheed}. Les solutions d\'ecrivant des trous noirs \`a sym\'etrie sph\'erique dans les quatre dimensions d'espace ont \'et\'e obtenues par Tangherlini \cite{tang}. Elles ont ensuite \'et\'e g\'en\'eralis\'ees au cas stationnaire par Myers et Perry \cite{MP}. Enfin, r\'ecemment, Reall et Emparan ont obtenu des solutions d\'ecrivant des anneaux noirs \cite{RE}.

D\'erivons maintenant le mod\`ele $\sigma$ de E5. Pour cela, nous allons utiliser le formalisme de Maison \cite{Maison} pour r\'eduire la th\'eorie de cinq \`a trois dimensions. La m\'etrique \`a cinq dimensions doit poss\'eder deux vecteurs de Killing, l'un de genre espace $\p_5$ et l'autre de genre temps $\p_t$. Elle peut alors se mettre sous la forme
\be \lb{mai}
ds_5^2=G_{AB}dx^Adx^B=\lambda_{ab}(dx^a+\mbox{\LARGE $a$}^a_i dx^i)(dx^b+\mbox{\LARGE $a$}^b_j dx^j)+\tau^{-1}h_{ij}dx^idx^j
\ee
o\`u $\lambda_{ab}$, $\tau=|det(\lambda)|$, $\mbox{\LARGE $a$}^a_i$ et $h_{ij}$ ne d\'ependent pas des coordonn\'ees $x^4=t$ et $x^5$. Les indices $i,j\ldots$ varient de $1$ \`a $3$ alors que les indices $a,b\ldots$ prennent les valeurs $4$ et $5$.

Les composantes ${}^5\mbox{R}_{ai}=0$ de l'\'equation d'Einstein ${}^5\mbox{R}_{AB}=0$ sont trivialement satifaites en introduisant le $2$-vecteur $\omega_a$ tel que 
\be\lb{dual}
\omega_{a,i} \equiv |h|^{-1/2}\tau\lambda_{ab}h_{il}\epsilon^{jkl}
\mbox{\LARGE $a$}^b_{j,k}
\ee
($d\omega$ est le dual de $d\mbox{\LARGE $a$}$).
En utilisant (\ref{mai}) et (\ref{dual}), les deux autres composantes ${}^5\mbox{R}_{ab}$ et ${}^5\mbox{R}_{ij}$ deviennent
\ba \lb{sig1}
(\chi^{-1}\chi^{,i})_{;i} =  0,\\
{\cal R}_{ij} =  \frac{1}{4}{\rm
Tr}(\chi^{-1}\chi_{,i}\chi^{-1}\chi_{,j})
\lb{sig2}
\ea
o\`u $\chi$ est une matrice anti-unimodulaire
\be\lb{chi}
\chi = \left( \begin{array}{ccc}
            \lambda_{ab} - \tau^{-1}\omega_a\omega_b & -\tau^{-1}\omega_a\\
            -\tau^{-1}\omega_b & -\tau^{-1} \end{array} \right)\,.
\ee 
Ces deux \'equations d\'erivent de l'action
\be \lb{msemd}
S_\sigma=\int \sqrt h\, d^3x\left({\cal R}-\frac{1}{4}Tr\left(\chi^{-1}\chi_{,i}\chi^{-1}\chi^{,i}\right)\right).
\ee
Nous avons ainsi obtenu le mod\`ele $\sigma$ de E5 qui est param\'etris\'e par la matrice $\chi\in SL(3,R)/SO(3)$, la m\'etrique de l'espace cible \'etant 
\be \lb{msm}
dl^2=\frac{1}{4}Tr\left(d\chi d\chi^{-1}\right).
\ee
En particulier, nous voyons imm\'ediatement que l'action  (\ref{msemd}) est invariante sous la transformation
\be \lb{transchi}
\chi\rightarrow \bar{\chi}=U^T \chi U, \qquad U\in SL(3,R).
\ee
Nous utiliserons dans les sections suivantes cette transformation pour g\'en\'erer de nouvelles solutions $(\bar{\chi},\bar{G})$ de E5 conduisant \`a de nouvelles solutions de EMD en utilisant (\ref{gE5}). De m\^eme, remarquons que les \'equations du mouvement (\ref{sig1}) et (\ref{sig2}) sont elles aussi invariantes sous la transformation (\ref{transchi}). De (\ref{sig1}), nous d\'eduisons que $\chi^{-1}\chi^{,i}=\bar{\chi}^{-1}\bar{\chi}^{,i}=cste$. En reportant ce r\'esultat dans (\ref{sig2}), nous voyons que ${\cal R}_{ij}=\bar{{\cal R}}_{ij}$ et donc que $h_{ij}=\bar{h}_{ij}$. Donc, toutes solutions $\chi$ et $\bar{\chi}$ reli\'ees par la transformation (\ref{transchi}) poss\`edent forc\'ement la m\^eme m\'etrique r\'eduite $h_{ij}$. 

Consid\'erons maintenant le cas particulier o\`u la matrice $\chi$ ne d\'epend des coordonn\'ees $x^i$ que par l'interm\'ediaire d'un potentiel $\sigma$, $\chi(\sigma(x^i))$. Nous pouvons montrer que le potentiel $\sigma$ peut toujours \^etre choisi harmonique ($\Delta\sigma=0$).
Pour cela, utilisons la param\'etrisation du mod\`ele $\sigma$ en fonction des $\Phi^A$. Si nous faisons l'hypoth\`ese que les $\Phi^A$ ne d\'ependent des $x^i$ uniquement par l'interm\'ediaire d'un potentiel $\sigma$, c'est-\`a-dire que $\Phi^A=\Phi^A(\sigma(x^i))$, alors l'\'equation pour les potentiels $\Phi^A$ (\ref{eqmvtPhi}) devient \cite{NK,SKMHH}:
\be
\frac{d\Phi^A}{d\sigma} D_i\partial^i\sigma+\left(\frac{d^2\Phi^A}{d\sigma^2}+\Gamma^A_{BC}\frac{d \Phi^B}{d\sigma}\frac{d \Phi^C}{d\sigma}\right)\partial_i\sigma\partial^i\sigma=0. \lb{eqmvtPhi2}
\ee
Les $\Phi^A$ \'etant des fonctions de $\sigma$ seulement, nous pouvons r\'e\'ecrire (\ref{eqmvtPhi2}) de la fa\c{c}on suivante
\be
 D_i\partial^i\sigma=-\frac{\frac{d^2\Phi^A}{d\sigma^2}+\Gamma^A_{BC}\frac{d \Phi^B}{d\sigma}\frac{d \Phi^C}{d\sigma}}{\frac{d\Phi^A}{d\sigma}}\partial_i\sigma\partial^i\sigma=g(\sigma)\partial_i\sigma\partial^i\sigma. \lb{eqmvtPhi3}
\ee
Or, comme les $x^i$ sont d\'efinis \`a une transformation de coordonn\'ees pr\`es, $x^i\rightarrow x'^i(x^i)$, nous avons la m\^eme libert\'e de choix pour le potentiel $\sigma$, $\sigma\rightarrow\sigma'(\sigma)$. Cette libert\'e de choix du $\sigma$ peut \^etre utilis\'ee pour mettre (\ref{eqmvtPhi3}) sous la forme ($\partial_i\sigma\partial^i\sigma\neq 0$)
\be
 D_i\partial^i\sigma=0.
\ee 
En utilisant cette relation, (\ref{eqmvtPhi}) devient 
\be
\frac{d^2\Phi^A}{d\sigma^2}+\Gamma^A_{BC}\frac{d \Phi^B}{d\sigma}\frac{d \Phi^C}{d\sigma}=0
\ee
qui est simplement l'\'equation des g\'eod\'esiques sur l'espace cible.

En utilisant \`a nouveau la param\'etrisation du mod\`ele $\sigma$ en fonction de $\chi$, nous avons pour l'\'equation des g\'eod\'esiques de l'espace cible (\ref{sig1}) et l'\'equation d'Einstein \`a trois dimensions (\ref{sig2})
\ba \lb{sig3}
\frac{d}{d\sigma}\left(\chi^{-1}\frac{d\chi}{d\sigma}\right)=0\\ \lb{sig4}
{\cal R}_{ij}=\frac{1}{4}Tr\left(\chi^{-1}\frac{d\chi}{d\sigma}\right)^2\p_i\sigma\p_j\sigma.
\ea
L'\'equation des g\'eod\'esiques (\ref{sig3}) admet pour solution 
\be \lb{solgeo}
\chi=\eta\e^{A\sigma}
\ee
o\`u $\eta$ et $A$ sont deux matrices constantes. Donc, toute solution de EMD dont la matrice $\chi$ peut se mettre sous la forme $\chi(\sigma(x^i))$ est repr\'esent\'ee par une g\'eod\'esique dans l'espace cible. Les \'equations (\ref{sig4}) et (\ref{msm}) deviennent
\ba
{\cal R}_{ij}=\frac{1}{4}Tr(A^2)\p_i\sigma\p_j\sigma\\
dl^2=\frac{1}{4}Tr(A^2)d\sigma^2.
\ea
Donc, en particulier, nous voyons que les solutions ayant une matrice $A$ telle que $Tr(A^2)=0$, sont des g\'eod\'esiques du genre lumi\`ere de l'espace cible. De plus, les composantes du tenseur de Ricci sont nulles. Or, \`a trois dimensions, le tenseur de Riemann \'etant une combinaison lin\'eaire du tenseur de Ricci et de sa trace, nous en d\'eduisons que les composantes du tenseur de Riemann sont elles aussi nulles. Donc, les solutions avec $Tr(A^2)=0$ sont des g\'eod\'esiques du genre lumi\`ere de l'espace cible repr\'esentant des solutions de EMD ayant des sections spatiales $h_{ij}$ plates.

Dans le cas o\`u les sections spatiales $h_{ij}$ sont plates, l'\'equation de Laplace, $\Delta\sigma=0$, admet comme solution $\sigma=1/\vec{r}$. A partir de cette solution, nous pouvons construire une solution multi-centres
\be
\sigma=\sum_i\sigma_i=\sum_i\frac{c_i}{|\vec{r}-\vec{r_i}|},\quad \chi=\eta\,\e^{A\sum_i\sigma_i}.
\ee

Pour finir signalons que nous pouvons aussi consid\'erer le cas o\`u $\chi$ d\'epend de plusieurs potentiels \cite{G86,GC96}
\be
\chi=\eta\e^{\sum_a A_a\sigma_a}
\ee
et d\'eriver les conditions n\'ecessaires pour pouvoir construire les solutions multi-centres correspondantes.

\section{G\'en\'eration de solutions en rotation - Secteur magn\'etique}%

Dans cette section, nous allons consid\'erer uniquement la version magn\'etique (\ref{solnasmag}) de la solution (\ref{solnas}). Dans le cas $\alpha^2= 3$, la solution s'\'ecrit:
\ba \lb{tn3}
ds^2=-\frac{r^{-1/2}(r-b)}{\sqrt{r_0}}dt^2+\frac{\sqrt{r_0}}{r^{-1/2}(r-b)}[dr^2+r(r-b)d\Omega^2],\\
F=\frac{r_0}{2} \sin\theta\, d\theta\wedge d\varphi,\qquad \e^{2\sqrt 3\phi}=\left(\frac{r}{r_0}\right)^{-\frac{3}{2}}. \lb{tn3m}
\ea
Son partenaire \`a cinq dimensions s'obtient facilement en utilisant la formule (\ref{gE5}): 
\be\lb{5m1}
ds_5^2 = -\frac{r-b}{r}\,dt^2 + \frac{r_0}{r-b}\left(dr^2 +
r(r-b)d\Omega^2\right) + \frac{r}{r_0}\left(dx^5 -
r_0\cos\theta\,d\varphi\right)^2\,.
\ee
Deux points sont \`a souligner concernant cette solution.
Premi\`erement, cette solution poss\`ede la m\^eme m\'etrique r\'eduite $h_{ij}$ que la m\'etrique de Schwarzschild plong\'ee dans cinq dimensions ($b=2M$)
\be \lb{schw5}
ds_5^2 = \left(dx^5\right)^2-\frac{r-b}{r}dt^2 + \frac{r}{r-b}\left(dr^2 +
r(r-b)\right)d\Omega^2.
\ee
Par cons\'equent, comme nous l'avons vu dans la section pr\'ec\'edente, ces deux m\'etriques sont li\'ees par la transformation (\ref{transchi}),
\be \lb{transm}
\chi_m=U_{m}^T\chi_SU_{m}
\ee
o\`u $\chi_m$ et $\chi_S$ 
\be\lb{chim}
\chi_m = \left( \begin{array}{ccc} -\frac{r-b}{r} & 0 & 0
\\0 & -\frac{br}{r_0(r-b)} & \frac{r}{r-b} \\ 0 & \frac{r}{r-b} &
-\frac{r_0}{r-b}
\end{array}\right),\quad
\chi_S = \left( \begin{array}{ccc} -\frac{r-b}{r} & 0 & 0 \\ 0 & 1 & 0
\\ 0 & 0 & -\frac{r}{r-b}  \end{array}
\right)\,,
\ee
sont les matrices associ\'ees aux solutions (\ref{5m1}) et (\ref{schw5}), respectivement. La matrice $U_m$ est
\be
 U_{m} =\left(
\begin{array}{ccc}
1&0&0\\0&0&\sqrt{\frac{r_0}{b}}
\\ 0&-\sqrt{\frac{b}{r_0}}&\sqrt{\frac{r_0}{b}}
\end{array}
\right).
\ee

Deuxi\`emement, la m\'etrique (\ref{5m1}) peut-\^etre r\'e\'ecrite sous la forme (en posant $r=x^2/4r_0$ et $x^5=r_0\eta$)
\be\lb{5m2}
ds_5^2 = -\bigg(1 - \frac{\mu}{x^2}\bigg)\,dt^2 + \bigg(1 -
\frac{\mu}{x^2}\bigg)^{-1}\,dx^2 + x^2\,d\Omega_3^2\,,
\ee
o\`u $d\Omega_3^2$ est la m\'etrique d'une  $3$-sph\`ere de rayon unit\'e,
\be
d\Omega_3^2 = \frac14\left(d\theta^2 + \sin^2\theta\,d\varphi^2  +
(d\eta - \cos\theta\,d\varphi)^2\right).
\ee
Or, nous reconnaissons dans cette m\'etrique la solution de Tangherlini \cite{tang} qui est contenue dans la solution plus g\'en\'erale de Myers et Perry \cite{MP} (la solution de Tangherlini s'obtient en prenant les deux moments angulaires $a_\pm=0$):
\ba \lb{MP}
ds_5^2 & = & -dt^2 + \frac{\mu}{\rho^2}\bigg[dt+a_+\sin^2\bar{\theta}\,d\varphi_+ 
+a_-\cos^2\bar{\theta}\,d\varphi_-\bigg]^2 + \rho^2\,d\bar{\theta}^2 \nonumber 
\\ & & + (x^2+a_+^2)\sin^2\bar{\theta}\,d\varphi_+^2 +
(x^2+a_-^2)\cos^2\bar{\theta}\,d\varphi_-^2 + \frac{\rho^2x^2}{\Theta}\,dx^2\,
\ea
avec
\ba
\rho^2 &=& x^2 + a_+^2\cos^2\bar{\theta} + a_-^2\sin^2\bar{\theta}\,, \quad
\Theta = (x^2 + a_+^2)(x^2 + a_-^2) - \mu x^2,\\
 \bar{\theta}&=&\frac{\theta}{2}\, ,\quad \varphi_\pm=\frac{\varphi\pm\eta}{2}\,.
\ea

Autrement dit, dans le cas $\alpha^2=3$, la version magn\'etique de la solution (\ref{solnas}) peut-\^etre obtenue de deux mani\`eres diff\'erentes (voir fig 3.1): 

1) En appliquant la transformation $U_{m}$ sur la matrice $\chi_S$ associ\'ee \`a la m\'etrique de Schwarzschild plong\'ee dans cinq dimensions, nous obtenons une matrice $\chi_m$ associ\'ee \`a la solution (\ref{5m1}). Puis nous utilisons (\ref{gE5}) pour passer de la solution \`a cinq dimensions (\ref{5m1}) de E5 \`a la solution \`a quatre dimensions (\ref{tn3})-(\ref{tn3m}) de EMD3.

2) En r\'eduisant la m\'etrique de Myers et Perry statique (la solution de Tangherlini) par rapport au vecteur de Killing $\p_5$ ($x^5=r_0\eta$).
\begin{figure}
\centerline{\epsfxsize=470pt\epsfbox{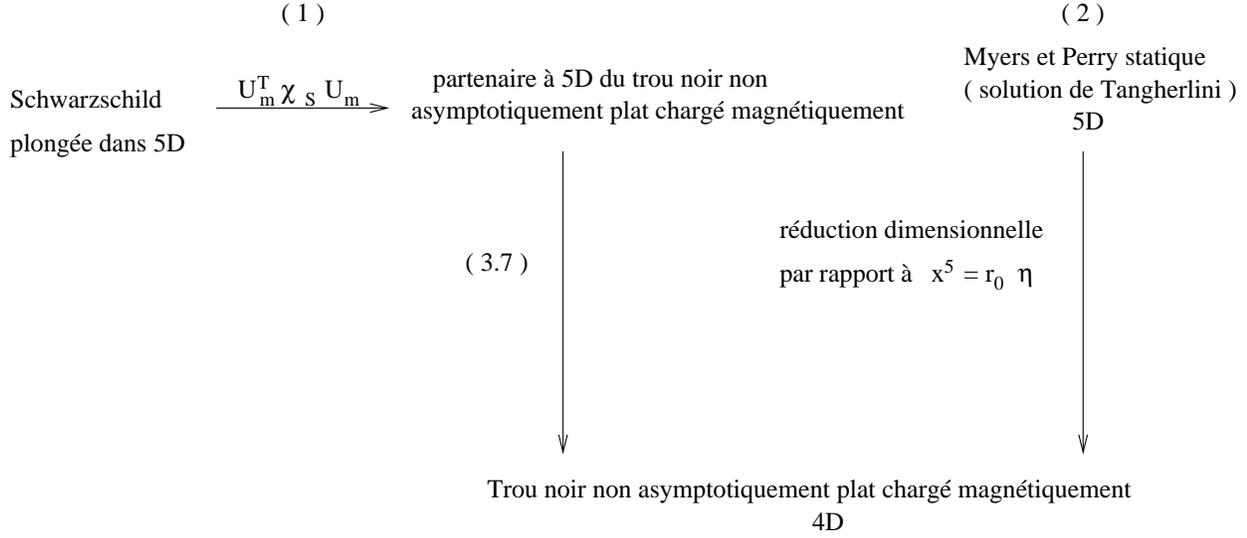}} 
\caption{Les deux fa\c{c}ons de construire la solution (\ref{tn3})-(\ref{tn3m}).}
\end{figure}

La transformation (\ref{transm}) fait passer d'une solution asymptotiquement plate \`a une solution non asymptotiquement plate, tout en g\'en\'erant une charge magn\'etique. Donc, nous pouvons pr\'evoir que, en appliquant $U_{m}$ sur la matrice $\chi_K$
\be
\chi_K =\frac1{f_0^2}\left(
\begin{array}{ccc}
-(r-b)^2-a_0^2\cos^2\theta & 0 &
a_0b\cos\theta \\ 0  & f_0^2 & 0 \\
a_0b\cos\theta & 0 & -\Sigma_0
\end{array}
\right)
\ee
repr\'esentant la solution de Kerr plong\'ee dans cinq dimensions (solution asymptotiquement plate en rotation qui se r\'eduit \`a Schwarzschild en prenant le param\`etre associ\'e \`a la rotation $a_0$ nul),
\ba
ds^2=-\frac{f_0^2}{\Sigma_0}(dt-\omega_0 d\varphi)^2+\Sigma_0\left(\frac{dr^2}{\Delta_0}+d\theta^2+\frac{\Delta_0\sin^2\theta}{f_0^2}d\varphi^2\right)+(dx^5)^2
\ea
avec
\ba
\Delta_0&=&r^2-br+a_0^2,\quad \Sigma_0=r^2+a_0^2\cos^2\theta,\\
f_0^2&=&\Delta_0-a_0^2\sin^2\theta,\quad \omega_0=-\frac{a_0br\sin^2\theta}{f_0^2},
\ea
nous obtiendrons une solution g\'en\'eralisant (\ref{tn3})-(\ref{tn3m}) par une rotation (une solution non asymptotiquement plate en rotation charg\'ee magn\'etiquement).
En appliquant la transformation (\ref{transm}) nous obtenons la nouvelle matrice $\chi_{Km}$
\be
\chi_{Km} =\frac1{f_0^2}\left(
\begin{array}{ccc}
-(r-b)^2-a_0^2\cos^2\theta &  -a_0b\sqrt{\frac{b}{r_0}}\cos\theta &
a_0\sqrt{br_0}\cos\theta \\  -a_0b\sqrt{\frac{b}{r_0}}\cos\theta  &
-\frac{b}{r_0}\Sigma_0 & \Sigma_0 \\ a_0\sqrt{br_0}\cos\theta
& \Sigma_0 & -r_0r
\end{array}
\right).
\ee
En comparant cette matrice avec (\ref{chi}) nous en d\'eduisons les quantit\'es $\tau$, $\lambda_{ab}$ et $\omega_a$. Puis, en utilisant la relation de dualit\'e (\ref{dual}), nous obtenons les $\mbox{\LARGE $a$}^a_i$ \`a partir des $\omega_a$. Nous pouvons alors reconstruire la m\'etrique associ\'ee \`a la matrice $\chi_{Km}$ et nous obtenons la m\'etrique \`a cinq dimensions suivante ($x^5=r_0\eta$): 
\ba \lb{UKerrU}
ds_5^2 & = & -\bigg(1-\frac{b}{r}\bigg)\psi^2
+\frac{a_0\sqrt{br_0}\cos\theta}{r} 2\psi\xi
+ \frac{r_0\Sigma_0}{r}\xi^2 \nonumber \\
&&\quad + \frac{r_0r}{f_0^2}\bigg(\frac{f_0^2}{\Delta_0}\,dr^2 +
f_0^2\,d\theta^2 + \Delta_0\sin^2\theta\,d\varphi^2
\bigg)\,. \lb{mrot5} 
\ea
o\`u
\be
\psi = dt+\frac{a_0\sqrt{br_0}r\sin^2\theta}{f_0^2}\,d\varphi\,,
\quad \xi = d\eta-\frac{\Delta_0\cos\theta}{f_0^2}\,d\varphi\,,
\ee
qui est le partenaire \`a cinq dimensions de la solution
\be \lb{solnasrot}
ds_4^2 = -
\frac{f_0^2}{\sqrt{r_0r\Sigma_0}}\bigg(dt +
\frac{\sqrt{br_0}a_0r\sin^2\theta}
{f_0^2}\,d\varphi\bigg)^2 
+\sqrt{r_0r\Sigma_0} \bigg(\frac{dr^2}{\Delta_0} + d\theta^2 +
\frac{\Delta_0\sin^2\theta}{f_0^2}\,d\varphi^2
\bigg),
\ee
\be
A =
-\frac{r^2+a_0^2}{\Sigma_0}\frac{r_0}2\cos\theta\bigg(d\varphi -
\frac{a_0\sqrt{b}}{\sqrt{r_0}(r^2+a_0^2)}\,dt\bigg)\,,\quad
\e^{2\phi/\sqrt3} = \sqrt{\frac{r_0r}{\Sigma_0}}\,,\lb{mrotp}
\ee
qui g\'en\'eralise effectivement (\ref{tn3})-(\ref{tn3m}) par l'ajout d'un nouveau param\`etre $a_0$ li\'e au moment angulaire. La charge magn\'etique
\be \lb{defP2}
P=\frac{1}{4\pi}\int F_{\theta\varphi}d\theta d\varphi=\frac{1}{2}\left. A_\varphi\right]_{\theta=\pi}^{\theta=0}=-\frac{r_0}{2}
\ee
est identique \`a celle du cas statique. La masse et le moment angulaire de cette solution sont (voir section 6 pour le calcul) 
\be
{\cal M}=\frac{3b}{8},\qquad J=\frac{a_0\sqrt{b r_0}}{2}.
\ee
La masse est la m\^eme que dans le cas statique.
Les diagrammes de Penrose de cette solution sont donn\'es dans la figure 3.2. Il sont identiques \`a ceux bien connus de la solution de Reissner-Nordstr\"om (le ``r\^ole'' de la charge \'electrique est ici tenu par le moment angulaire $a_0$).

\begin{figure}
\centerline{\epsfxsize=330pt\epsfbox{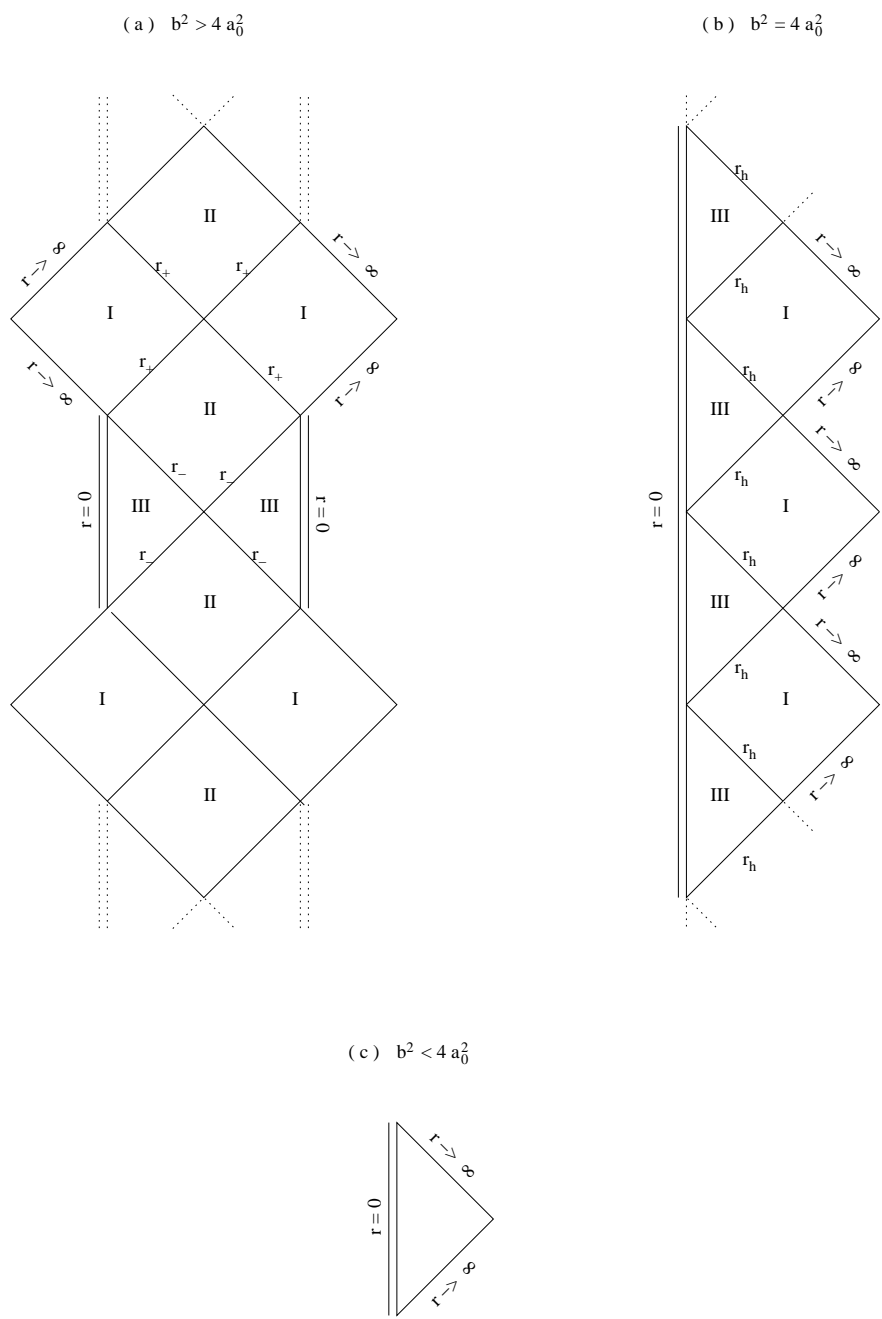}} 
\caption{Diagrammes de Penrose de la solution (\ref{solnasrot}). Dans les cas (a) et (b), les diagrammes sont des tours infinies constitu\'ees par un empilement infini du m\^eme motif.}
\end{figure}

Lorsque $b^2>4a_0^2$, la solution d\'ecrit un trou noir en rotation poss\'edant deux horizons situ\'es en $r_\pm$
\be
r_\pm=\frac{1}{2}\left(b\pm\sqrt{b^2-4a_0^2}\right).
\ee
Lorsque $b^2=4a_0^2$, la solution d\'ecrit un trou noir extr\^eme. Les deux horizons co\"incident en $r_h=r_+=r_-=b/2$. Enfin, lorsque $b^2<4a_0^2$, la solution d\'ecrit une singularit\'e nue. Quelque soit $b$, la singularit\'e de courbure, de genre temps, est situ\'ee en $r=0$. Nous voyons sur les diagrammes de Penrose (fig. 3.2) que les solutions stationnaires ont la m\^eme structure conforme \`a l'infini que dans le cas statique.  

D'autre part, comme la solution (\ref{5m1}), nous pouvons montrer que la solution (\ref{UKerrU}) est elle aussi li\'ee \`a la m\'etrique de Myers et Perry. En effet, en posant
\be
\bar{\theta}=\frac{\theta}{2},\quad \varphi_\pm=\frac{\varphi\pm\eta}{2},\quad x^2=4r_0r-\bar{a}^2,\quad \mu=4r_0b,\quad a_+=a_-=\bar{a}=-\frac{4r_0a_0}{\sqrt \mu}
\ee
la m\'etrique (\ref{UKerrU}) se r\'eduit \`a la m\'etrique de Myers et Perry avec les deux moments angulaires $a_\pm$ \'egaux. Donc (\ref{solnasrot})-(\ref{mrotp}), comme (\ref{tn3})-(\ref{tn3m}), peut-\^etre obtenue de deux fa\c{c}ons diff\'erentes. Soit par r\'eduction dimensionnelle de la m\'etrique de Myers et Perry avec deux moments angulaires \'egaux (par rapport \`a $\p_5=\p_\eta/r_0$), soit en utilisant la transformation (\ref{transchi}) sur la m\'etrique de Kerr plong\'ee dans cinq dimensions puis la formule (\ref{gE5}).

Nous avons donc vu que la version magn\'etique de (\ref{solnas})(dans le cas $\alpha^2=3$) s'obtient par r\'eduction dimensionnelle de la m\'etrique de Myers et Perry avec $a_\pm=0$ et que sa g\'en\'eralisation en rotation (\ref{solnasrot}) s'obtient par r\'eduction dimensionnelle de la m\'etrique de Myers et Perry avec $a_+=a_-$. Nous pouvons donc nous demander ce que nous obtiendrions en r\'eduisant dimensionnellement la m\'etrique de Myers et Perry la plus g\'en\'erale ($a_+\neq a_-$) par rapport \`a $\p_5=\p_\eta/r_0$. Comme nous pouvions le supposer, nous obtenons une solution encore plus g\'en\'erale que (\ref{solnasrot})-(\ref{mrotp}):
\ba \lb{dyonmag}
ds_4^2 & = & -
\frac{f^2}{\sqrt{\Pi A}}\bigg(dt - \bar{\omega}
\,d\varphi\bigg)^2 
+\sqrt{\Pi A} \bigg(\frac{dr^2}{\Delta} + d\theta^2 +
\frac{\Delta\sin^2\theta}{f^2}\,d\varphi^2
\bigg)\, \lb{drotg}, \\ {\cal A} & = &
-\frac{r_0}{2A}\left\{\bigg[(\Delta + br)\cos\theta
-\frac{b\beta\delta}{2r_0} -
\frac{\beta\delta}{2r_0}(r-b/2)\sin^2\theta\bigg]d\varphi\right. \nonumber \\
& & \left.\qquad - \frac{b}{2r_0}(\delta-\beta\cos\theta)\,dt\right\}\,, \lb{drota} \\
\e^{2\phi/\sqrt3} & = & \sqrt{\frac{\Pi}{A}}\,,\lb{drotp}
\ea
avec
\ba
\Delta & = & \left(r-\frac{b}{2}\right)^2-\frac{\bar{\beta}^2\bar{\delta}^2}{4r_0^2}\,, \quad f^2 = \Delta -
\frac{\beta^2\bar{\delta}^2}{4r_0^2}\sin^2\theta\,,\quad a_\pm=\beta\pm\delta\,,  \nonumber \\ A & = &
r^2 + \frac{b}{4r_0}(\delta - \beta\cos\theta)^2 -
\frac{\beta^2\delta^2}{4r_0^2}\cos^2\theta\,, \quad \Pi = r_0r
+\frac{\beta\delta}2\cos\theta\,, \lb{def1} \\
\bar{\omega} & = & \frac{\beta
b(r-\delta^2/2r_0)}{2f^2}\sin^2\theta \qquad (\bar{\beta}^2=br_0-\beta^2, \bar{\delta}^2=br_0-\delta^2)\,. \nonumber 
\ea
Cette solution d\'ecrit un trou noir dyonique (charg\'e \'electriquement et magn\'etiquement) en rotation qui poss\`ede deux horizons situ\'es en $r_\pm$
\be
r_\pm=\frac{b}{2}\pm\frac{\bar{\beta}\bar{\delta}}{2r_0}.
\ee
La charge magn\'etique (toujours identique au cas statique) et la charge \'electrique sont
\be
P=-\frac{r_0}{2},\qquad Q=\int \e^{-2\sqrt 3\phi}F^{tr}\sqrt{g}d\theta d\varphi =-\frac{b\delta}{2r_0}.
\ee
Ces deux r\'esultats peuvent \^etre compris en examinant le terme dominant de ${\cal A}_\varphi$ pour $P$ et le terme dominant de ${\cal A}_0$ pour $Q$. En effet, le terme dominant de ${\cal A}_\varphi$ est $-r_0\cos\theta/2$ qui conduit \`a $P=-r_0/2$ en utilisant (\ref{defP2}). Le terme dominant de ${\cal A}_0$ est proportionnel \`a $(\delta-\beta\cos\theta)$. Or, l'int\'egration du terme $\beta\cos\theta$ multipli\'e par le $\sin\theta$ provenant du $\sqrt g$ donne z\'ero. Il ne reste donc que la contribution du terme proportionnel \`a $\delta$ conduisant au r\'esultat ci-dessus.

Lorsque $\delta=0$, nous retrouvons (en posant $\beta=-2a_0\sqrt{r_0/b}$) la solution (\ref{solnasrot}). Lorsque $\beta=0$, la solution d\'ecrit un trou noir dyonique statique poss\'edant deux horizons si $br_0>\delta^2$, un trou noir statique extr\^eme ($r_+=r_-$) si $br_0=\delta^2$ et une singularit\'e nue si $br_0<\delta^2$. 

Finalement, en se souvenant que la transformation (\ref{transm}) fait passer d'une solution asymptotiquement plate \`a une solution non asymptotiquement plate tout en g\'en\'erant une charge magn\'etique, nous conjecturons que ce dyon en rotation peut aussi \^etre obtenu par application de $U_m$ sur la matrice associ\'ee \`a la m\'etrique de Kerr-Newman charg\'ee \'electriquement plong\'ee dans cinq dimensions.

\section{G\'en\'eration de solutions en rotation - Secteur \'electrique}%
Les versions \'electriques des solutions d\'eriv\'ees dans la section pr\'ec\'edente peuvent \^etre obtenues ais\'ement en utilisant la transformation de dualit\'e (\ref{transdual}). Cependant, comme nous allons le voir, les liens entre les versions \'electriques et leurs partenaires \`a cinq dimensions valent aussi la peine d'\^etre \'etudier.

Cette fois-ci, nous prenons comme point de d\'epart la version \'electrique (\ref{solnas})-(\ref{solnaselec}):
\ba \lb{tn4}
ds^2&=&-\frac{r^{-1/2}(r-b)}{\sqrt{r_0}}dt^2+\frac{\sqrt{r_0}}{r^{-1/2}(r-b)}[dr^2+r(r-b)d\Omega^2],\\
F&=&\frac{1}{2r_0} dr\wedge dt,\qquad \e^{2\sqrt 3\phi}=\left(\frac{r}{r_0}\right)^{\frac{3}{2}} \lb{tn4e}.
\ea
La m\'etrique de la version \'electrique \'etant la m\^eme que celle de la version magn\'etique, il en r\'esulte que, comme dans la section pr\'ec\'edente, le partenaire \`a cinq dimensions: 
\be\lb{5e1}
ds_5^2 = \frac{r_0}{r}\bigg(dx^5 +
\frac{r-b}{r_0}\,dt\bigg)^2-\frac{r-b}{r_0}\,dt^2 + \frac{r}{r-b}\left(dr^2 +
r(r-b)\right)d\Omega^2
\ee
poss\`ede la m\^eme m\'etrique r\'eduite $h_{ij}$ que la m\'etrique de Schwarzschild plong\'ee dans cinq dimensions (\ref{schw5}) (qui est contenue dans la solution de Rasheed \cite{rasheed}). Ces deux solutions sont donc reli\'ees entre elles par la transformation:
\be \lb{transe}
\chi_e=U_{e}^T\chi_SU_{e}
\ee
o\`u
\be\lb{chie}
\chi_e = \left( \begin{array}{ccc} -\frac{b(r-b)}{r_0r} &
\frac{r-b}{r}& 0
\\\frac{r-b}{r}  & \frac{r_0}{r} & 0 \\ 0 & 0 & -\frac{r}{r-b}  \end{array}
\right),\quad
U_{e} =\left(
\begin{array}{ccc}
\sqrt{\frac{b}{r_0}}&-\sqrt{\frac{r_0}{b}}&0\\
0&\sqrt{\frac{r_0}{b}}&0\\ 0&0&1
\end{array}
\right)\,
\ee
et $\chi_S$ est la matrice associ\'ee \`a la m\'etrique de Schwarzschild plong\'ee dans cinq dimensions (\ref{schw5}) dont l'expression est donn\'ee en (\ref{chim}).
D'autre part, (\ref{5e1}) peut se r\'e\'ecrire sous la forme
\be\lb{ts2}
ds_5^2 = -\frac{b}{r_0}\frac{r-b}{r}\bigg(dt -\frac{r_0}{b}
dx^5\bigg)^2 + \frac{r}{r-b}\,dr^2 + r^2\,d\Omega^2 +\frac{r_0}{b}(dx^5)^2
\ee
qui se d\'eduit de la m\'etrique de Schwarzschild plong\'ee dans cinq dimensions (\ref{schw5}) par la transformation
\be\lb{twist}
x^5\rightarrow \gamma x^5,\quad t\rightarrow \gamma^{-1}t-\gamma x^5,\quad \gamma=\sqrt{\frac{r_0}{b}}.
\ee




Effectuer une telle transformation n'est pas anodin. Premi\`erement, dans le contexte de la th\'eorie de Kaluza-Klein, la r\'eduction dimensionnelle se fait en supposant la coordonn\'ee $x^5$ p\'eriodique. Il s'ensuit que la nouvelle coordonn\'ee de temps dans la m\'etrique de Schwarzschild ``twist\'ee'' (\ref{ts2}) est elle aussi p\'eriodique. 

Deuxi\`emement, en prenant comme point de d\'epart une solution de Kaluza-Klein \`a sym\'etrie sph\'erique neutre (Schwarzschild plong\'ee dans cinq dimensions) dont la matrice $\chi$ est donn\'ee en (\ref{chim}), nous pouvons g\'en\'erer une solution de Kaluza-Klein \`a sym\'etrie sph\'erique charg\'ee \'electriquement en utilisant la matrice $V$
\be
V=\left( \begin{array}{ccc}
            \cosh\alpha&\sinh\alpha&0\\
	     \sinh\alpha&\cosh\alpha&0\\
             0&0&1\end{array} \right).
\ee
Nous obtenons la matrice $\bar{\chi}=V^{T}\chi_SV$, 
\be
\bar{\chi}=\left( \begin{array}{ccc}
            -f\cosh^2\alpha+\sinh^2\alpha&\frac{b}{r}\cosh\alpha\sinh\alpha &0\\
	     \frac{b}{r}\cosh\alpha\sinh\alpha&\cosh^2\alpha-f\sinh^2\alpha&0\\
             0&0&-f^{-1}\end{array} \right),
\ee
o\`u $f=1-b/r$, qui correspond \`a la m\'etrique suivante:
\ba
ds_5^2&=&(-f\cosh^2\alpha+\sinh^2\alpha)dt^2+2\frac{b}{r}\cosh\alpha\sinh\alpha dtdx^5\nonumber\\
&&\quad+(\cosh^2\alpha-f\sinh^2\alpha)(dx^5)^2+ f^{-1}\,dr^2 + r^2\,d\Omega^2.
\ea
Nous pouvons v\'erifier que cette solution contient bien une charge \'electrique en se rappelant que le coefficient du terme crois\'e $dt dx^5$ est proportionnel \`a la composante de temps $A_0$ du $4$-vecteur potentiel \'electromagn\'etique (\ref{gE5}).

En appliquant la transformation de twist (\ref{twist}) sur cette m\'etrique, nous obtenons 
\ba
ds_5^2&=&-f(\cosh\alpha\gamma^{-1}dt-\gamma(\cosh\alpha-\sinh\alpha)dx^5)^2+ f^{-1}\,dr^2 + r^2\,d\Omega^2\nonumber\\
&&\quad+(\sinh\alpha\gamma^{-1}dt+\gamma(\cosh\alpha-\sinh\alpha)dx^5)^2.
\ea
Or, en effectuant les transformations de jauge
\be
x^5\rightarrow x^5-\gamma^{-1}\sinh\alpha dt,\quad \gamma^{-1}(\cosh\alpha-\sinh\alpha)t\rightarrow \bar{\gamma}t, \quad\bar{\gamma}=\gamma(\cosh\alpha-\sinh\alpha)
\ee
la m\'etrique devient  
\be
ds_5^2 = -f\bigg(\bar{\gamma}^{-1}dt -\bar{\gamma}
dx^5\bigg)^2 + f^{-1}\,dr^2 + r^2\,d\Omega^2 +\bar{\gamma}^2(dx^5)^2.
\ee
Or cette m\'etrique est identique (en rempla\c{c}ant $\bar{\gamma}$ par $\gamma$) \`a la solution de Schwarzschild ``twist\'ee'' (\ref{ts2}) que nous avons obtenue en appliquant la transformation (\ref{twist}) sur une solution neutre de Kaluza-Klein.

En fait ce r\'esultat n'est pas \'etonnant. En effet, la transformation (\ref{transe}) g\'en\'ere une charge \'electrique tout en changeant le comportement asymptotique. Donc, en appliquant cette transformation sur une solution neutre asymptotiquement plate nous obtenons une solution charg\'ee \'electriquement non asymptotiquement plate. Mais si, comme nous venons de le faire, nous appliquons cette transformation sur une solution asymptotiquement plate charg\'ee \'electriquement, nous obtenons \`a nouveau une solution charg\'e \'electriquement non asymptotiquement plate. Donc, r\'eduire dimensionnellement une solution neutre asymptotiquement plate ``twist\'ee'' ou une solution asymptotiquement plate charg\'ee \'electriquement ``twist\'ee'' conduit au m\^eme r\'esultat.

En conclusion , comme pour la version magn\'etique, la version \'electrique de la solution non asymptotiquement plate (\ref{solnas}) peut \^etre obtenue (dans le cas $\alpha=\sqrt 3$) de deux fa\c{c}ons diff\'erentes (voir fig. 3.3).

1) En utilisant la transformation (\ref{transe}) sur la matrice associ\'ee \`a une solution de Kaluza-Klein neutre ou charg\'ee \'electriquement (qui sont contenues dans la solution de Rasheed \cite{rasheed}), nous obtenons une nouvelle matrice \`a partir de laquelle nous construisons la nouvelle solution \`a cinq dimensions. Puis, en utilisant (\ref{gE5}), nous obtenons la solution \`a quatre dimensions de EMD3. 

2) En r\'eduisant dimensionnellement une solution de Kaluza-Klein neutre ou charg\'ee \'electriquement par rapport au vecteur de Killing $\gamma(\p_5-\p_t)$. Ou bien ce qui est \'equivalent, en ``twistant'' la solution de Kaluza-Klein neutre ou charg\'ee \'electriquement puis en r\'eduisant par rapport \`a la nouvelle coordonn\'ee $x^5$.
\begin{figure}
\centerline{\epsfxsize=470pt\epsfbox{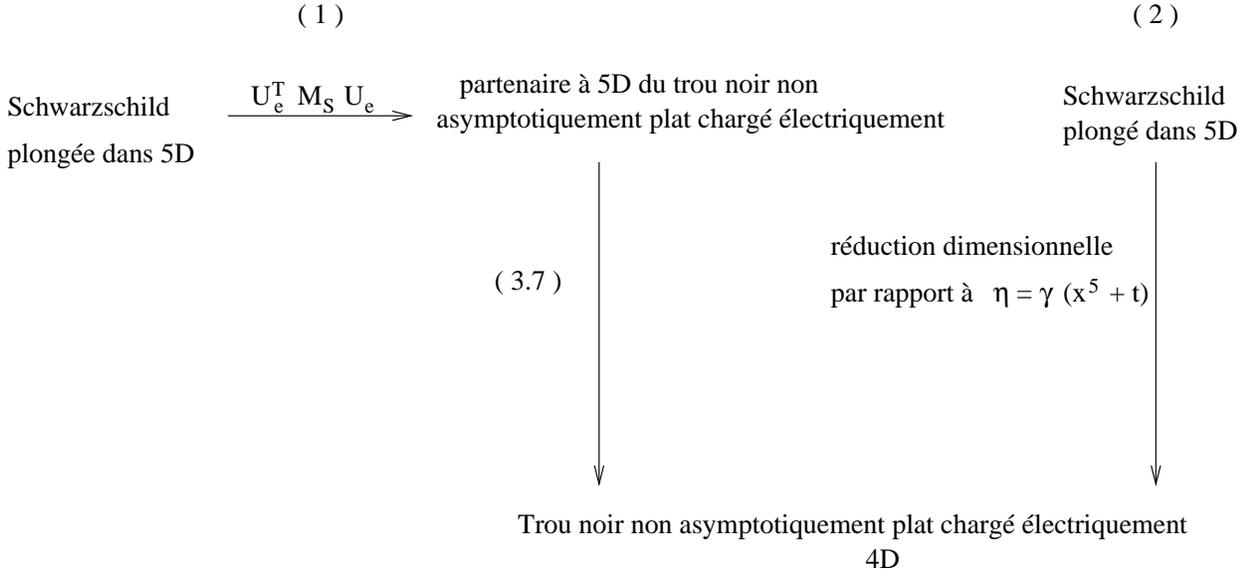}} 
\caption{Les deux fa\c{c}ons d'obtenir la solution (\ref{tn4})-(\ref{tn4e}).}
\end{figure}



Comme dans la section pr\'ec\'edente, en appliquant la transformation (\ref{transe}) sur la matrice associ\'ee \`a la m\'etrique de Kerr plong\'ee dans cinq dimensions (contenue dans la solution de Rasheed) puis en utilisant (\ref{gE5}) sur la nouvelle m\'etrique \`a cinq dimensions ainsi obtenue, nous obtenons la version \'electrique de la solution (\ref{solnasrot})-(\ref{mrotp}). La m\'etrique est toujours donn\'ee par (\ref{solnasrot}) et les champs de mati\`ere sont donn\'es par
\be
A  = \frac{1}{2r_0r}\bigg(f_0^2\,dt
+ \sqrt{br_0}a_0r\sin^2\theta\,d\varphi\bigg), \quad \e^{2\phi/\sqrt3} =  \sqrt{\frac{\Sigma_0}{r_0r}}. \lb{mrotpe}
\ee
Sans surprise, nous pourrions montrer que cette solution peut aussi \^etre obtenue en r\'eduisant dimensionnellement une solution de Kaluza-Klein en rotation (qui est contenue dans la solution de \cite{rasheed}) (neutre ou charg\'ee \'electriquement ) par rapport \`a $\gamma(\p_5-\p_t)$.

Calculons maintenant \`a l'aide de (\ref{transdual})(en choisissant le signe $-$), la version \'electrique du dyon en rotation (\ref{dyonmag}):
\ba \lb{dyonelec1}
ds_4^2 & = & -
\frac{f^2}{\sqrt{\Pi A}}\bigg(dt - \bar{\omega}
\,d\varphi\bigg)^2 
+\sqrt{\Pi A} \bigg(\frac{dr^2}{\Delta} + d\theta^2 +
\frac{\Delta\sin^2\theta}{f^2}\,d\varphi^2
\bigg)\,, \\ {\cal A} & = &\frac{1}{2\Pi}\bigg(\left(A-br-\frac{b\delta^2}{2r_0}\right)dt \nonumber\\
&&\qquad -\frac{b}{2r_0}\left(2\delta\Pi\cos\theta+\beta(r_0r+\delta^2/2)\sin^2\theta\right)d\varphi\bigg), \lb{dyonelec2} \\
\e^{2\phi/\sqrt3} & = & \sqrt{\frac{A}{\Pi}}\,.\lb{dyonelec3}
\ea
Nous obtenons une solution d\'ecrivant un trou noir dyonique en rotation mais qui, cette fois-ci, est une excitation sur le vide ($b=a=0$) charg\'e \'electriquement.

Nous pouvons \`a l'aide de (\ref{gE5}) obtenir le partenaire \`a cinq dimensions de cette m\'etrique. Puis, en appliquant la transformation de twist inverse
\be \lb{invtwist}
x^5\rightarrow \gamma^{-1} x^5,\quad t\rightarrow \gamma(t+ x^5),\quad \gamma=\sqrt{\frac{r_0}{b}}
\ee
nous obtenons une solution en rotation analogue \`a celle de Rasheed \cite{rasheed} mais avec en plus une charge NUT; cette charge NUT ayant une valeur oppos\'ee \`a la charge magn\'etique ($N=-P$). Essayons maintenant de comprendre pourquoi le dyon en rotation de ``type \'electrique'' est la r\'eduction dimensionnelle ``twist\'ee'' d'une solution analogue \`a la solution de Rasheed avec $P=-N$. Prenons comme point de d\'epart une solution de Kaluza-Klein statique charg\'ee magn\'etiquement:
\be
ds_5^2=(dx^5+2P\cos\theta d\varphi)^2-\frac{\Delta}{r^2}dt^2+\frac{r^2}{\Delta}dr^2+r^2d\Omega^2
\ee
(qui est le partenaire \`a cinq dimensions de la solution de Reissner-Nordstr\"om charg\'ee magn\'etiquement) o\`u $\Delta=r^2-2M r+P^2$.
En ``twistant'' cette solution nous obtenons comme pr\'evu une solution avec une charge \'electrique mais nous voyons aussi l'apparition d'une charge NUT proportionnelle \`a la charge magn\'etique $P$ ($N=\gamma P$)
\ba
ds_5^2&=&\gamma^2\left(1-\frac{\Delta}{r^2}\right)\left(dx^5+\frac{2P r^2\cos\theta}{\gamma(r^2-\Delta)}d\varphi+\frac{\Delta}{\gamma^2(r^2-\Delta)}dt\right)^2\nonumber\\
&&\quad -\frac{\Delta}{\gamma^2(r^2-\Delta)}(dt+2\gamma P \cos\theta d\varphi)^2+\frac{r^2}{\Delta}dr^2+r^2d\Omega^2.
\ea
Nous voyons donc que le ``twist'', en plus de g\'en\'erer une charge \'electrique et de changer le comportement asymptotique, g\'en\`ere aussi une charge NUT \`a partir de la charge magn\'etique.

D'autre part, nous avons vu que le trou noir dyonique en rotation de type \'electrique est la r\'eduction dimensionnelle ``twist\'ee'' d'une solution analogue \`a celle de Rasheed avec une charge NUT, $N$, telle que $N=-P$. De plus, le trou noir dyonique en rotation ne poss\`ede pas de charge NUT. Nous en d\'eduisons donc que la transformation de ``twist'' est telle que, appliqu\'ee sur une solution asymptotiquement plate avec $N=-P$, elle produit une solution non asymptotiquement plate avec une charge magn\'etique mais d\'epourvue de charge NUT.

Pour finir, nous conjecturons que si nous effectuions la r\'eduction dimensionnelle de la solution g\'en\'eralisant celle de Rasheed (avec $N\neq P$), nous obtiendrions une trou noir dyonique en rotation avec charge NUT ($N\neq P$) de type ``electrique''.
\section{Trous noirs avec charge NUT-Version \'electrique et version magn\'etique}%

Nous avons vu dans la section 2 que le dyon en rotation de type magn\'etique est la r\'eduction dimensionnelle par rapport \`a $\p_5=(\p_{\varphi_+}-\p_{\varphi_-})/(2r_0)$ de la m\'etrique de Myers et Perry (\ref{MP}). Dans la section 3 nous avons vu que le dyon en rotation de type \'electrique est la r\'eduction dimensionnelle ``twist\'ee'' d'une solution g\'en\'eralisant celle de Rasheed \cite{rasheed} par l'ajout d'une charge NUT; la charge NUT \'etant \'egale en valeur absolue \`a la charge magn\'etique mais avec un signe oppos\'e.

Nous avons conjectur\'e \`a la fin de la section pr\'ec\'edente que la r\'eduction dimensionnelle de la solution g\'en\'eralisant celle de Rasheed avec $N\neq P$ donnerait la version \'electrique d'un dyon en rotation avec charge NUT. Nous pouvons aussi nous demander quel serait le partenaire \`a cinq dimensions de la version magn\'etique de ce dyon. Dans cette section, nous allons r\'epondre \`a ces deux questions dans un cas particulier. Prenons l'exemple d'une m\'etrique \`a cinq dimensions avec charge NUT, c'est-\`a-dire la m\'etrique de Taub-NUT (\ref{taub-NUT}) plong\'ee dans cinq dimensions (voir fig. 3.4)
\be
ds_5^2 = -\frac{\Delta}{\Sigma}\bigg(dt' +
2l\cos\theta\,d\varphi\bigg)^2 \nonumber 
+ \frac{\Sigma}{\Delta}\bigg(dr^2 + \Delta\,d\Omega_2^2\bigg) +
(dx^{'5})^2\,,
\ee 
avec
\be
\Delta = r^2 - 2mr - l^2\,, \quad \Sigma = r^2 + l^2\,.
\ee
En ``twistant'' (\ref{twist}) cette m\'etrique puis en la r\'eduisant de cinq \`a quatre dimensions, nous obtenons la solution 
\ba
ds_4^2 & = & - \frac{\Delta}{\sqrt{\Pi\Sigma}}\bigg(dt + 2\gamma l\cos\theta
\,d\varphi\bigg)^2 +\sqrt{\Pi\Sigma} \bigg(\frac{dr^2}{\Delta} +
d\Omega_2^2 
\bigg)\,, \lb{nerotg} \\ A & = &
\frac{\Delta}{2\Pi}\bigg(dt + 2\gamma l\cos\theta\,d\varphi\bigg), \lb{nerota}
\\ \e^{2\phi/\sqrt3} & = & \sqrt{\frac{\Sigma}{\Pi}}\,,\lb{nerotp}
\ea
avec
\be
\Pi = 2\gamma^2(mr + l^2)\,.
\ee
qui est un trou noir non asymptotiquement plat avec une charge \'electrique et une charge magn\'etique ainsi qu'une charge NUT telle que $N=P=\gamma l$.

En utilisant la transformation de dualit\'e (\ref{transdual}), nous pouvons obtenir la version magn\'etique de cette solution. Puis, en utilisant (\ref{gE5}), nous obtenons son partenaire \`a cinq dimensions qui peut \^etre mis sous la forme
\ba
&& ds_5^2 = -\bigg(dt + \bar{a}\,d\eta\bigg)^2 +
\frac{\mu}{\rho^2}\bigg(dt + \bar{a}\,d\eta - \frac{\bar{a}}2(d\eta -
  \cos\theta\,d\varphi)\bigg)^2 \nonumber \\
& & + \frac{\rho^4}{\rho^4 - \mu\rho^2 + \mu\bar{a}^2}\,d\rho^2 +
  \frac{\rho^2}4\bigg(d\theta^2 + d\varphi^2 + d\eta^2 -
2\cos\theta\,d\varphi\,d\eta\bigg) \,, \lb{tmpe}
\ea
avec
\be
\rho^2 = 8\gamma^2(mr + l^2)\,, \; \eta = \frac1{2m\gamma^2}\,x^5\,,
\; \mu = 16\gamma^2(l^2 + m^2)\,, \; \bar{a} = 2\gamma l\,.  
\ee
\begin{figure}
\centerline{\epsfxsize=480pt\epsfbox{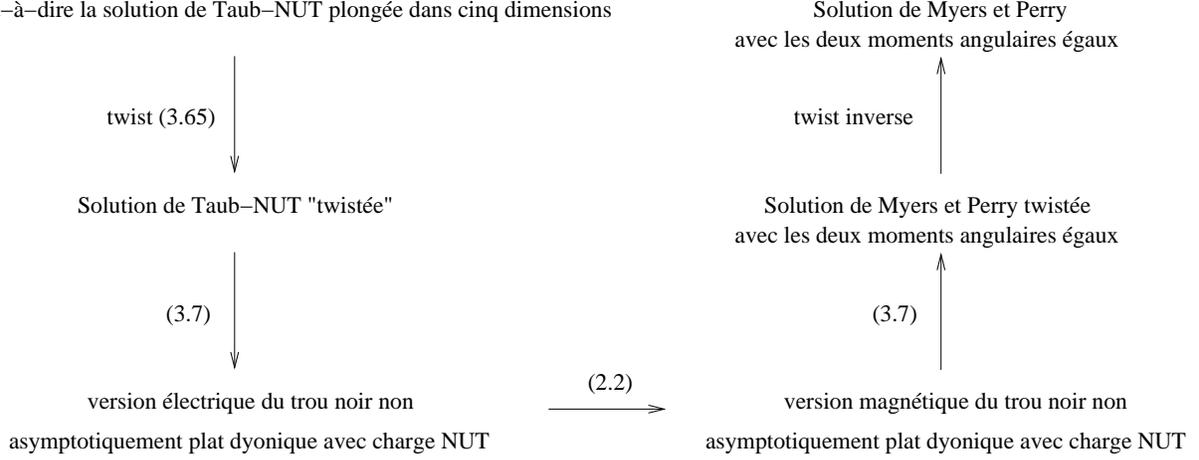}} 
\caption{En r\'eduisant dimensionnellement une solution de Kaluza-Klein avec charge NUT ``twist\'ee'', nous obtenons un trou noir non asymptotiquement plat dyonique de type \'electrique. Ensuite, nous calculons la version magn\'etique de ce dyon puis nous calculons son partenaire \`a cinq dimensions. Nous reconnaissons alors la m\'etrique de Myers et Perry avec $a_+=a_-$ ``twist\'ee'' (\ref{twist2}).}
\end{figure}
Nous reconnaissons alors la m\'etrique de Myers et Perry (\ref{MP}) avec les deux moments angulaires \'egaux ($a_+=a_-=\bar{a}$) qui aurait subi le ``twist''
\be \lb{twist2}
t\rightarrow t+\bar{a}\eta.
\ee
Nous en d\'eduisons que la version magn\'etique de la solution non asymptotiquement plate dyonique avec charge NUT est la r\'eduction dimensionnelle de la m\'etrique de Myers et Perry avec les deux moments angulaires \'egaux ``twist\'ee'' (\ref{twist2}) par rapport \`a $\p_\eta$.

En conclusion, nous avons prouv\'e la conjecture faite \`a la fin de la section pr\'ec\'edente  dans un cas particulier et nous avons montr\'e que la version magn\'etique du trou noir non asymptotiquement plat avec charge NUT est la r\'eduction dimensionnelle d'un cas particulier de la m\'etrique de Myers et Perry ``twist\'ee''. Nous faisons alors la deuxi\`eme conjecture suivante: la version magn\'etique du trou noir dyonique en rotation avec charge NUT peut-\^etre obtenue en r\'eduisant par rapport \`a $\p_5=\frac{\p_{\varphi_+}-\p_{\varphi_-}}{2r_0}$ la m\'etrique de Myers et Perry (\ref{MP}) g\'en\'erale (avec $a_+\neq a_-$) ``twist\'ee'' (\ref{twist2}).

\section{Solutions multi-centres}%
En prenant le carr\'e de la matrice $A$ de la solution (\ref{tn4})-(\ref{tn4e})
\be
A_e^2= \left(\begin{array}{ccc}
            \frac{b^2}{r_0^2}&-\frac{b}{r_0}&0\\
	     0&0&0\\
             0&0&\frac{b^2}{r_0^2}\end{array} \right),
\ee
nous voyons que $A_e^2=0$ lorsque $b=0$. Dans ce cas la matrice $\chi_e$ est simplement 
\be
\chi_e=\eta_e\e^{A_e\sigma}=\eta(I_3+\sigma A_e)=\left( \begin{array}{ccc}
            0&1&0\\
	     1&\frac{r_0}{r}&0\\
             0&0&-1\end{array} \right),\quad \sigma=\frac{r_0}{r}.
\ee
Ce cas correspond au fond charg\'e \'electriquement ($b=0$), sur lequel les versions \'electriques des trous noirs ($b>0$) se forment, dont la m\'etrique s'\'ecrit
\be
ds^2=-\frac{r}{r_0}dt^2+\frac{r_0}{r}\left[dr^2+r^2d\Omega^2\right].
\ee
Nous remarquons que la m\'etrique $h_{ij}$ est plate. Or, comme nous l'avons vu \`a la fin de la premi\`ere section, lorsque la m\'etrique r\'eduite $h_{ij}$ est plate, nous pouvons construire des solutions multi-centres. En effet, le potentiel $\sigma$ satisfait \`a l'\'equation de Lagrange $\Delta\sigma=0$ qui admet des solutions du type
\be
\sigma=\sum_i\frac{c_i}{|\vec{r}-\vec{r}_i|}.
\ee
Dans ce cas la matrice $\chi_e$ devient
\be
\chi_e=\eta_e\e^{A_e\sigma}=\eta_e(I_3+\sigma A_e)=\left( \begin{array}{ccc}
            0&1&0\\
	     1&\sigma&0\\
             0&0&-1\end{array} \right)
\ee
et la m\'etrique de la solution multi-centres est
\be
ds_5^2=\sigma(dx^5+\sigma^{-1}dt)^2-\sigma^{-1}dt^2+(dr^2+r^2d\Omega^2).
\ee
En utilisant (\ref{gE5}), nous en d\'eduisons la solution de EMD3 correspondante
\ba
ds^2=-\sigma^{-\frac{1}{2}}dt^2+\sigma^{\frac{1}{2}}\left[dr^2+r^2d\Omega^2\right]\\
\e^{2\phi/\sqrt 3}=\sigma^{-\frac{1}{2}}, \quad A_0=\frac{1}{2}\sigma^{-1},
\ea
qui d\'ecrit une superposition de fonds charg\'es.

\section{Energie quasilocale et thermodynamique}%

Nous allons dans cette section appliquer le formalisme quasilocal aux solutions que nous avons obtenues dans les sections pr\'ec\'edentes. En rassemblant les contributions gravitationnelles (\ref{equas})-(\ref{jquas}) et mat\'erielles (\ref{equasdil}) \`a l'\'energie quasilocale et au moment angulaire quasilocal, nous obtenons les formules suivantes:
\ba \lb{equas1}
E&=&\oint_{S^r_t}\left(N( \epsilon-\epsilon_0)+2N^i\pi_{ij}n^j+A_0(\bar{\Pi}^i-\bar{\Pi}_0^i)n_i\right)d^2x\\
J&=&-2\oint_{S^r_t}n_i\pi^{i}{}_\varphi d^2x-\int_{S^r_t}A_\varphi\bar{\Pi}^in_id^2x.\lb{jquas1}
\ea

D\'etaillons le calcul des quantit\'es quasilocales pour la solution (\ref{solnasrot})-(\ref{mrotpe}) qui d\'ecrit un trou noir en rotation charg\'e \'electriquement.
Tout d'abord r\'e\'ecrivons la m\'etrique sous la forme ADM
\be \lb{tnadm}
ds^2=-\frac{\Delta_0\sqrt{\Sigma_0}}{\sqrt{r_0r}(r^2+a_0^2)}dt^2+\sqrt{r_0r\Sigma_0}\left(\frac{dr^2}{\Delta_0}+d\theta^2+\frac{r^2+a_0^2}{\Sigma_0}\sin^2\theta\left(d\varphi-\frac{a_0\sqrt b}{\sqrt r_0(r^2+a_0^2)}dt\right)^2\right).
\ee
L'\est $M$ est alors repr\'esent\'e (voir fig. 1.3) par un empilement d'hypersurfaces $\Sigma_t$ de normale $u_\mu=-N\delta_\mu^0$. La m\'etrique induite par $g_{\mu\nu}$ sur ces hypersurfaces est: 
\be
h_{ij}dx^idx^j=g_{ij}+u_iu_j=\sqrt{r_0r\Sigma_0}\left(\frac{dr^2}{\Delta_0}+d\theta^2+\frac{r^2+a_0^2}{\Sigma_0}\sin^2\theta d\varphi^2\right).
\ee
Cet empilement est born\'e, pour une valeur de la coordonn\'ee radiale $r$ fix\'ee, par une hypersurface $\Sigma^r$ de normale $n_r=\sqrt{g_{rr}}$ de m\'etrique:
\be
-\frac{\Delta\sqrt{\Sigma}}{\sqrt{r_0r}(r^2+a_0^2)}dt^2+\sqrt{r_0r\Sigma}\left(d\theta^2+\frac{r^2+a_0^2}{\Sigma}\sin^2\theta\left(d\varphi-\frac{a_0\sqrt b}{\sqrt r_0(r^2+a_0^2)}dt\right)^2\right).
\ee 
L'\'energie quasilocale est donn\'ee par l'int\'egrale (\ref{equas1}) sur la surface $S^r_t=\Sigma_t\cap\Sigma^r$ dont la m\'etrique est donn\'ee par 
\be
\sigma_{ab}dx^adx^b=g_{ab}+u_au_b-n_an_b=\sqrt{r_0r\Sigma}\left(d\theta^2+\frac{r^2+a_0^2}{\Sigma}\sin^2\theta d\varphi^2\right)
\ee
o\`u $u_a=N\delta_a^0$ est la normale aux hypersurfaces $\Sigma_t$ et $n_a=\sqrt{g_{rr}}\delta_a^r$ est la normale aux hypersurfaces $\Sigma^r$.
La masse du trou noir est d\'efinie comme \'etant l'\'energie quasilocale \'evalu\'ee sur une surface \`a l'infini
\be \lb{mquas}
{\cal M}=\displaystyle{\lim_{r\rightarrow\infty}}E
\ee

En comparant (\ref{tnadm}) et (\ref{ADM}), nous voyons que 
\ba
N&=&\sqrt{\frac{\Delta}{r^2+a_0^2}}\left(\frac{\Sigma}{r_0r}\right)^{\frac{1}{4}}\simeq \left(\frac{r}{r_0}\right)^{\frac{1}{4}},\\
N^r&=& N^\theta=0,\quad N^\varphi=-\frac{a_0\sqrt b}{\sqrt r_0(r^2+a_0^2)}\simeq -\frac{a_0\sqrt b}{\sqrt{r_0}r^2}.
\ea
Les d\'eterminants des m\'etriques $h_{ij}$ et $\sigma_{ab}$ sont
\ba
h&=&\frac{(r_0r)^{3/2}(r^2+a_0^2)\sqrt{\Sigma}\sin^2\theta}{\Delta}\simeq r_0^{3/2}r^{5/2}\sin^2\theta\\
\sigma&=&r_0r(r^2+a_0^2)\sin^2\theta\simeq r_0r^3\sin^2\theta
\ea
o\`u nous avons aussi indiqu\'e leurs valeurs lorsque $r$ tend vers l'infini.

Calculons tout d'abord la masse de la solution. Pour \'evaluer le premier terme dans l'int\'egrale (\ref{equas1}), nous devons calculer $\varepsilon=\sqrt \sigma k/(8\pi)$. La courbure extrins\`eque $k$ de $S^r_t$ est 
\ba
k&=&-\sigma^{\mu\nu}D_\nu n_\mu=\left(\sigma^{\theta\theta}\,{}^{(3)}\Gamma^r_{\theta\theta}+\sigma^{\varphi\varphi}\,{}^{(3)}\Gamma^r_{\varphi\varphi})\right) n_r\\
&=&-\frac{1}{2}\frac{\sqrt{\Delta}(3r^2+a_0^2)}{r_0^{1/4}r^{5/4}(r^2+a_0^2)\Sigma^{1/4}}
\ea
o\`u les ${}^{(3)}\Gamma$ sont les symboles de Christoffel calcul\'es \`a partir de la m\'etrique $h_{ij}$. Lorsque $r\rightarrow\infty$, $k$ devient
\be \lb{kdl}
k\simeq -\frac{3}{2r_0^{1/4}r^{3/4}}\left(1-\frac{b}{2r}\right)
\ee
Comme $n_i=\sqrt g_{rr}\delta_i^r$ et $N^i=N^\varphi\delta_\varphi^i$, le deuxi\`eme terme dans l'int\'egrale (\ref{equas1}) est $N^\varphi\pi_{\varphi r}n^r$ avec:
\ba
K^{r\varphi}&=&-\frac{1}{2N}(h^{rr}\p_rN^\varphi+(h^{rr}\,{}^{(3)}\Gamma^\varphi_{r\varphi}+h^{\varphi\varphi}\,{}^{(3)}\Gamma^r_{r\varphi})N^\varphi)\\
&=&-\frac{\sqrt b a\sqrt \Delta r^{3/4}}{r_0^{3/4}(r^2+a^2)^{3/2}\Sigma^{3/4}}\\
\pi^{r\varphi}&=&\sqrt \sigma K^{r\varphi}=-\frac{\sqrt b a r^{3/2}\sin\theta}{(r^2+a^2)\sqrt\Sigma}\simeq -\frac{a\sqrt b}{r^{3/2}}\sin\theta.
\ea
Enfin la contribution \'electromagn\'etique est $A_0\Pi^r$ avec
\be \lb{pidl}
\Pi^r=-\frac{r_0(r^2+a^2)(r^2-a^2\cos^2\theta)\sin\theta}{8\pi\Sigma^2}\simeq-\frac{r_0}{8\pi}\left(1+\frac{a^2(1-3\cos^2\theta)}{r^2}\right)\sin\theta.
\ee
Finalement, les trois termes contribuant \`a la masse s'\'ecrivent (en ne gardant que le terme dominant lorsque $r\rightarrow\infty$):
\ba
N\epsilon=-\frac{3}{16\pi}r\sin\theta\\
2N^i\pi_{ij}n^j=2N^\varphi\pi_{\varphi r}n^r=\frac{b a^2r^2\sin^3\theta}{8\pi(r^2+a^2)\Sigma}\\
A_0\Pi^r=-\frac{(r^2+a^2)(r^2-a^2\cos^2\theta)\sin\theta}{16\pi r\Sigma}.
\ea
Nous voyons que le deuxi\`eme terme tend vers z\'ero et ne contribue donc pas \`a la masse. Au contraire, les deux autres termes sont lin\'eaires en $r$ et donc divergent lorsque $r$ tend vers l'infini. Donc nous obtenons un r\'esultat infini pour la masse. Comme dans le cas statique (voir chapitre 2 section 6), nous devons retrancher la contribution d'une solution de fond. La solution de fond doit \^etre telle que les $(N, N^i, \sqrt \sigma, A_t, \phi)$ et les $(N_0, N^i_0, \sqrt\sigma_0, A_t^0, \phi_0)$ co\"incident sur la surface $S^\infty_t$. Comme dans le cas statique, nous choisissons le fond charg\'e ($b=a=0$) sur lesquel les trous noirs se forment
\be \lb{tnadm0}
ds^2=-\sqrt{\frac{r}{r_0}}dt^2+\sqrt{\frac{r_0}{r}}\left(dr^2+r^2d\theta^2+r^2\sin^2\theta d\varphi^2\right).
\ee
La contribution de la solution de fond \'etant ajout\'ee pour annuler la contribution dominante \`a la masse, nous devons reprendre les calculs pr\'ec\'edents en tenant compte de l'ordre suivant dans les d\'eveloppements de $k$ (\ref{kdl}) et $\Pi^r$ (\ref{pidl}). Nous obtenons
\ba
N(\epsilon-\epsilon_0)=\frac{3b}{32\pi}\\
A_0(\Pi^r-\Pi^r_0)=-\frac{a^2(1-3\cos^2\theta)}{16\pi r}\sin\theta
\ea
ce qui donne pour la masse
\be
{\cal M}=\frac{3b}{8}.
\ee

Le calcul du moment angulaire est similaire \`a celui de la masse except\'e que nous n'avons pas \`a retrancher la contribution de la solution de fond puisque celle-ci est statique ($J_0=0$). Les quantit\'es intervenant dans l'int\'egrale (\ref{jquas1}) ont d\'ej\`a \'et\'e calcul\'ees lors du calcul de la masse. Les termes dominants des contributions gravitationnelles et mat\'erielles sont
\ba
2n_r\pi^r{}_\varphi\simeq -2a_0\sqrt{b r_0}\sin^3\theta\\
A_\varphi\Pi^r\simeq -a_0\sqrt{b r_0}\sin^3\theta.
\ea
Nous voyons que la contribution gravitationnelle est deux fois plus grande que la contribution mat\'erielle. En regroupant ces contributions, nous obtenons pour le moment angulaire
\be
J=\frac{\sqrt{br_0}\,a_0}{2}.
\ee
Nous pouvons faire le m\^eme calcul pour la solution dyonique en rotation (\ref{dyonelec1})-(\ref{dyonelec3}). La masse et le moment angulaire sont
\be
{\cal M}=\frac{3b}{8},\quad J=-\frac{b\beta}{4}.
\ee
Comme dans le cas statique (voir chapitre 1 section 6), nous pouvons v\'erifier que les versions magn\'etiques ont les m\^emes masses et moments angulaires que les versions \'electriques.


V\'erifions maintenant si la premi\`ere loi est satisfaite pour ces trous noirs non asymptotiquement plats. Prenons \`a nouveau l'exemple de la solution (\ref{solnasrot})-(\ref{mrotpe}).
Nous avons pour l'entropie, la temp\'erature, la vitesse angulaire de l'horizon et le potentiel \'electrosta-tique sur l'horizon:
\ba
S=\frac{A}{4}=\frac{1}{4}\int\sqrt{g_{\theta\theta}g_{\varphi\varphi}}d\theta d\varphi=\pi\sqrt{r_0r_+}\sqrt{r_+^2+a_0^2},\\
T=\frac{1}{2\pi}(n^r\p_r N)|_h=\frac{r_+-r_-}{4\pi\sqrt{r_0r_+}\sqrt{r_+^2+a_0^2}},\\
\Omega_h=-\frac{g_{t\varphi}}{g_{\varphi\varphi}}|_h=\frac{a_0\sqrt b}{\sqrt r_0(r_+^2+a_0^2)}\\
V_h=(A_0+\Omega A_\varphi)|_h=\frac{b}{2r_0}
\ea
ce qui donne (en variant $\cal M$, $S$, $J$ et $Q$ par rapport \`a $b$, $r_0$ et $a_0$)
\be
d{\cal M}=\frac{3 db}{8}, \quad TdS+\Omega_hdJ-V_h dQ=\frac{3 db}{8}-\frac{b}{8r_0}dr_0.
\ee
Donc la solution (\ref{solnasrot})-(\ref{mrotpe}) satisfait \`a la premi\`ere loi si $dr_0=0$, c'est-\`a-dire si l'on fixe la charge \'electrique. Il se passe donc la m\^eme chose que dans le cas statique. Le param\`etre $r_0$ \'etant un param\`etre du fond charg\'e et non un param\`etre du trou noir, il semble logique de ne pas varier ce param\`etre. Donc, nous concluons que la solution (\ref{solnasrot})-(\ref{mrotpe}) satisfait \`a la premi\`ere loi des trous noirs. Comme dans le cas statique, nous pouvons v\'erifier que la version magn\'etique satisfait aussi \`a la premi\`ere loi.

De m\^eme, nous pourrions v\'erifier que les versions \'electriques et magn\'etiques des trous noirs dyoniques en rotation satisfont aussi \`a la premi\`ere loi de la thermodynamique des trous noirs si la charge n'est pas vari\'ee.

\newpage
\thispagestyle{empty}
\null
\chapter{Trous noirs en th\'eorie d'Einstein-Maxwell dilato-axionique}

Dans le chapitre pr\'ec\'edent nous avons obtenu de nouvelles solutions des \'equations du mouvement (\ref{eqmvtemd1})-(\ref{eqmvtemd3}) de la th\'eorie d'Einstein-Maxwell dilatonique (EMD)  pour une valeur particuli\`ere de la constante de couplage du dilaton, $\alpha^2=3$. Pour se faire, nous avons utilis\'e le fait que EMD3 (EMD avec $\alpha^2=3$) est la r\'eduction dimensionnelle de la th\'eorie d'Einstein \`a cinq dimensions qui admet un mod\`ele $\sigma$. Nous avons ainsi obtenu de nouvelles solutions de EMD3 et mis en \'evidence les liens existants entre ces nouvelles solutions de EMD3 et les solutions de la th\'eorie d'Einstein \`a cinq dimensions.
Dans ce chapitre, nous allons nous int\'eresser au cas $\alpha=1$. Dans ce cas, EMD est la r\'eduction dimensionnelle de la th\'eorie des cordes h\'et\'erotiques. Mais nous avons mentionn\'e, dans la premi\`ere section du chapitre pr\'ec\'edent, que dans le cas g\'en\'eral ($\alpha\neq 0,\sqrt 3$), les transformations laissant le mod\`ele $\sigma$ de EMD invariant \'etaient difficiles \`a exploiter pour la g\'en\'eration de nouvelles solutions. Pour g\'en\'erer de nouvelles solutions de EMD pour $\alpha=1$, nous allons utiliser le fait que, dans ce cas, EMD est contenue dans une th\'eorie plus vaste: la th\'eorie d'Einstein-Maxwell dilato-axionique (EMDA). Cette th\'eorie admet un mod\`ele $\sigma$. 
Dans une premi\`ere partie nous allons introduire la th\'eorie EMDA puis dans une deuxi\`eme partie nous rappelerons les principales caract\'eristiques de son mod\`ele $\sigma$. Ensuite, nous utiliserons ce mod\`ele $\sigma$ pour construire une nouvelle solution en rotation non asymptotiquement plate et nous calculerons sa masse et son moment angulaire en utilisant le formalisme quasilocal. Apr\`es avoir montrer que EMDA peut \^etre obtenue par r\'eduction dimensionnelle de la th\'eorie d'Einstein \`a six dimensions, nous chercherons quelle est la solution \`a six dimensions correspondante \`a la solution en rotation. Pour finir, nous \'etudierons le mouvement g\'eod\'esique et le comportement d'un champ scalaire dans le champ de gravitation de la nouvelle solution en rotation. 

Dans ce chapitre, le syst\`eme d'unit\'e utilis\'e est tel que $G=1$.

\section{La th\'eorie d'Einstein-Maxwell dilato-axionique}%
La th\'eorie EMDA est d\'efinie par l'action
\begin{equation} \label{actionEMDA} S =
\frac{1}{16\pi}\int d^4x\sqrt{|g|}\left\{R -
2\partial_\mu\phi\partial^\mu\phi - \frac{1}{2}\e^{4\phi}
{\partial_\mu}\kappa\partial^\mu\kappa
-\e^{-2\phi}F_{\mu\nu}F^{\mu\nu}-\kappa F_{\mu\nu}{\tilde
F}^{\mu\nu}\right\}, \end{equation}
qui contient en plus du dilaton $\phi$ et du vecteur $A$, d\'ej\`a pr\'esents dans EMD, un champ pseudo-scalaire, l'axion $\kappa$.

\noindent Le tenseur $\tilde{F}^{\mu\nu}$ est le dual du tenseur \'electromagn\'etique d\'efini en (\ref{Fdual}).

L'action de EMDA (\ref{actionEMDA}) est invariante sous la transformation (S-dualit\'e) de $SL(2,R)$
\be \lb{sdu}
z=\kappa+i\,\e^{-2\phi}\rightarrow \tilde{z}=\frac{\alpha z+\beta}{\gamma z+\delta}, \quad \tilde{v}=\delta v+\gamma u,\quad \tilde{u}=\beta v+\alpha u\quad(\alpha\delta-\beta\gamma=1)
\ee
qui peut-\^etre utilis\'ee pour g\'en\'erer une charge \'electrique \`a partir d'une charge magn\'etique (et vice versa) ou encore pour passer d'une solution de type \'electrique \`a une solution de type magn\'etique (et vice versa).

En variant l'action (\ref{actionEMDA}) par rapport \`a $\phi$, $\kappa$, $A_\mu$ et $g_{\mu\nu}$, nous  obtenons les \'equations du mouvement de EMDA:
\ba \lb{eqmvtemda1}
\frac{1}{\sqrt{|g|}}\partial_\mu\left(\sqrt{|g|}\,\partial^\mu\phi\right)=\frac{1}{2}\e^{-2\phi}F^2+\frac{1}{2}\e^{4\phi}(\partial\kappa)^2\\
\nonumber\\\lb{eqmvtemda2}
\frac{1}{\sqrt{|g|}}\partial_\mu\left(\sqrt{|g|}\,\e^{4\phi}\partial^\mu\kappa\right)+F_{\mu\nu}\tilde{F}^{\mu\nu}=0\\
\nonumber\\\lb{eqmvtemda3}
\frac{1}{\sqrt{|g|}}\partial_\nu\left[\sqrt{|g|}\left(\e^{-2\phi}F^{\mu\nu}+\kappa\tilde{F}^{\mu\nu}\right)\right]=0\\
\nonumber\\\lb{eqmvtemda4}
R_{\mu\nu}=2\partial_\mu\phi\,\partial_\nu\phi+\frac{1}{2}\e^{4\phi}\partial_\mu\,\kappa\partial_\nu\kappa+\e^{-2\phi}\left(2F_{\mu\lambda}F^\lambda_\nu+\frac{1}{2}g_\mu\nu F^2\right).
\ea

Signalons que la limite $\phi\rightarrow 0$ et $\kappa\rightarrow 0$ ne redonne pas la th\'eorie d'Einstein-Maxwell puisque les \'equations du dilaton (\ref{eqmvtemda1}) et de l'axion (\ref{eqmvtemda2}) imposent respectivement les contraintes $F^2=0$ et $F_{\mu\nu}\tilde{F}^{\mu\nu}=0$.

\section{Le mod\`ele $\sigma$ de EMDA}%

Nous allons rappeler les principales \'etapes de l'obtention du mod\`ele $\sigma$ de EMDA en se basant sur l'article \cite{GK}. La notion de mod\`ele $\sigma$ a \'et\'e introduite dans la section 1 du chapitre pr\'ec\'edent.

\noindent Tout d'abord il est n\'ecessaire d'imposer que la m\'etrique poss\`ede un vecteur de Killing du genre temps afin de pouvoir mettre la m\'etrique sous la forme : 

\be
ds^2=-f(dt-\omega_i dx^i)(dt-\omega_j dx^j)+f^{-1}h_{ij}dx^i dx^j
\ee
o\`u la fonction $f$, la 1-forme $\omega=\omega_i dx^i$ et la m\'etrique r\'eduite $h_{ij}$ sont ind\'ependantes du temps. La m\'etrique $g_{\mu\nu}$ est alors stationnaire et la m\'etrique r\'eduite $h_{ij}$ est de genre espace.
 
\noindent Les composantes spatiales ($\mu=i$) de l'\'equation de Maxwell (\ref{eqmvtemda3}) et de l'identit\'e de Bianchi

\be \lb{idbemda}
\frac{1}{\sqrt{|g|}}\partial_\nu(\sqrt{|g|} \tilde{F}^{\nu\mu})=0
\ee

\noindent sont trivialement satisfaites en introduisant les quantit\'es $u$ et $v$ appel\'ees potentiel ``\'electrique'' et ``magn\'etique'' respectivement (par analogie avec l'\'electromagn\'etisme sans source):

\ba
F_{i0}=\frac{1}{\sqrt 2}\partial_i v \\
\e^{-2\phi}F^{ij}+\kappa \tilde{F}^{ij}=\frac{f}{\sqrt {2 h}}\epsilon^{ijk}\partial_k u.
\ea

En introduisant le $3$-vecteur $\tau^i$ et le potentiel $\chi$,
\be \lb{tauchi}
\tau^i=-\frac{f^2}{\sqrt h}\varepsilon^{ijk}\p_j\omega_k,\quad \tau_i=\p_i\chi+v\p_iu-u\p_iv
\ee
o\`u $\varepsilon^{ijk}$ est le symbole totalement antisym\'etrique \`a trois dimensions\footnote{Convention: $\varepsilon^{123}=1$}, la composante ($0i$) de l'\'equation d'Einstein est satisfaite.

\noindent Les \'equations (\ref{eqmvtemda1}), (\ref{eqmvtemda2}), les composantes ($00$) et ($ij$) de l'\'equation d'Einstein (\ref{eqmvtemda4}) ainsi que la composante $\mu=0$ des \'equations (\ref{eqmvtemda2}) et (\ref{idbemda}) constituent alors un syst\`eme de six \'equations pour les variables $f$, $\chi$, $u$, $v$, $\phi$ et $\kappa$. Or, ce syst\`eme de six \'equations d\'erivent de l'action

\be \lb{actionsemda}
S_\sigma=\frac{1}{2}\int \left({\cal R}_{ij} -{\cal G}_{AB} \partial_i \Phi^A\partial_j \Phi^B\right) h^{ij}\sqrt h \,d^3x
\ee

\noindent o\`u $\Phi^A=(f,\chi,v,u,\phi,\kappa)$
 est appel\'e espace cible, ${\cal G}_{AB}$ \'etant la m\'etrique de cet espace.
 Nous avons ainsi obtenu le mod\`ele $\sigma$ de EMDA.


En \'etudiant les isom\'etries (vecteurs de Killing) de la m\'etrique de l'espace cible ${\cal G}_{AB}$, les auteurs \cite{GK} ont montr\'e que ce mod\`ele $\sigma$ pouvait \^etre param\'etris\'e par la matrice $M\in Sp(4,R)/U(2)$ qui d\'epend des potentiels $\Phi^A$:
\begin{equation} \label{MPQ}
M=\left(\begin{array}{crc}
P^{-1}&P^{-1}Q\\
QP^{-1}&P+QP^{-1}Q\\
\end{array}\right),
\end{equation}
o\`u $P$ et $Q$ sont deux matrices r\'eelles
\begin{equation}
P=-{\rm e}^{-2\phi} \begin{pmatrix}  v^2-f{\rm e}^{2\phi} & v \\
v & 1 \end{pmatrix}, \quad Q= \begin{pmatrix} vw-\chi & w \\ w &
-\kappa \end{pmatrix},
\end{equation}
avec $w=u-\kappa v$.

L'action du mod\`ele $\sigma$
\be
S_\sigma=\frac{1}{2}\int\left({\cal R}+\frac{1}{4}Tr(\nabla M \nabla M^{-1})\right)\sqrt h d^3x \lb{eqmvtsigma1}
\ee
est invariante sous la transformation matricielle 
\be \lb{trans2}
M\rightarrow \bar{M}=U^T M U, \qquad U\in Sp(4,R),
\ee
que nous utiliserons dans la suite pour construire de nouvelles solutions de EMDA. Comme dans le cas du mod\`ele $\sigma$ de EMD, toutes solutions $M$ et $\bar{M}$ li\'ees par la transformation (\ref{trans2}) poss\`edent la m\^eme m\'etrique r\'eduite $h_{ij}$.


\section{Lien entre EMDA et la \RG \`a six dimensions}%

Comme nous l'avons vu dans la section pr\'ec\'edente, le mod\`ele $\sigma$ de EMDA est invariant sous le groupe $Sp(4,R)$ \cite{GK,GK2,gal}. D'autre part, la \RG \`a six dimensions sans source admet un mod\`ele $\sigma$ invariant sous le groupe $SL(4,R)$ \cite{Maison2} qui contient le groupe $Sp(4,R)$. Ceci am\`ene \`a penser qu'il pourrait exister un lien entre les deux th\'eories. Dans l'article \cite{CG}, les auteurs ont montr\'e que, \`a toute solution de EMDA correspond une solution de la Relativit\'e G\'en\'erale \`a six dimensions, g\'en\'eralisant ainsi le r\'esultat obtenu dans \cite{CGMS}.

Rappelons bri\`evement la proc\'edure suivie dans \cite{CG} afin de voir sous quelles conditions la \RG \`a six dimensions se r\'eduit \`a EMDA. Pour cela nous allons r\'eduire, \`a la Kaluza-Klein, la \RG sans source \`a six dimensions
\be \lb{act6}
S_6=\int d^6x \sqrt{g_6} R_6.
\ee
Quelque soit le nombre de dimensions, en imposant l'existence d'un vecteur de Killing du genre espace, la m\'etrique peut s'\'ecrire
\be \lb{gen}
ds_{n+1}^2=\e^{-2c\hat{\phi}_n}ds_n^2+\e^{2(n-2)c\hat{\phi}_n}(dx^{n+1}+(C_n)_\mu dx^\mu)^2
\ee
o\`u $c$ est une constante arbitraire que nous fixerons plus loin. En utilisant (\ref{gen}) nous pouvons exprimer la courbure scalaire \`a $n+1$ dimensions de la mani\`ere suivante
\be \lb{rn}
\sqrt{g_{n+1}}R_{n+1}=\sqrt{g_n}\left[R_n-(n-1)(n-2)c^2(\p\hat{\phi}_n)^2-\frac{1}{4}\e^{2(n-1)c\hat{\phi}_n}F^2(C_n)+2c\nabla^2\hat{\phi}_n\right].
\ee 

Pour r\'eduire l'action (\ref{act6}) de six \`a cinq dimensions, nous utilisons (\ref{gen}) et (\ref{rn}) avec $n=5$. En dualisant la $2$-forme $F$ et en fixant $c=1/\sqrt 6$, l'action (\ref{act6}) devient (en n\'egligeant la divergence)
\be \lb{act5}
S_5=\int d^5x\sqrt{g_5}\left[R_5-2(\p\hat{\phi}_5)^2-\frac{1}{12}\e^{-\alpha\hat{\phi}_5}H^2\right]
\ee
o\`u $H=d\hat{K}=\star F(C_5)$ et $\alpha=4\sqrt{2/3}$. Nous devons maintenant r\'eduire (\ref{act5}) de cinq \`a quatre dimensions. Pour cela nous utilisons (\ref{gen}) et (\ref{rn}) avec $n=4$ ainsi que
\be
\hat{K}=K_{\mu\nu}dx^\mu\wedge dx^\nu+E_\mu \,dx^5\wedge dx^\mu.
\ee
L'action (\ref{act5}) devient
\ba
S_4&=&\int d^4x \sqrt{g_4}\left[R_4-2(\p\phi)^2-(\p\psi)^2-\frac{1}{2}\e^{4\phi}(\p\kappa)^2-\frac{1}{4}\e^{2(\psi-\phi)}F^2(C_4)\nonumber\right.\\
&&\quad\left. -\frac{1}{4}\e^{-2(\psi+\phi)}F^2(E)-\frac{\kappa}{4}\left(F(C_4)\tilde{F}(E)+F(E)\tilde{F}(C_4)\right)\right]\lb{act4}
\ea
o\`u nous avons introduit $\kappa$ tel que $H=-\e^{4\phi}\star \kappa$ et 
\be
\phi=\sqrt{\frac{2}{3}}\phi_5-\sqrt{\frac{1}{3}}\phi_4,\quad \psi=\sqrt{2}\left(\sqrt{\frac{1}{3}}\phi_5+\sqrt{\frac{2}{3}}\phi_4\right).
\ee
Nous voulons que l'action (\ref{act4}), et les \'equations du mouvement qui en d\'ecoulent, se r\'eduisent \`a l'action de EMDA (\ref{actionEMDA}) et les \'equations du mouvement correspondantes (\ref{eqmvtemda1})-(\ref{eqmvtemda4}). Ceci peut-\^etre fait en imposant les conditions
\be
\psi=0 ,\quad (C_4)_\mu=E_\mu=\sqrt{2}A_\mu
\ee
compatibles avec toutes les \'equations du mouvement.

En r\'esum\'e l'action de la \RG \`a six dimensions (\ref{act6}) se r\'eduit \`a l'action de EMDA (\ref{actionEMDA}) en posant
\be \lb{ds621}
ds_6^2=ds_4^2+\e^{-2\phi}\theta^2+\e^{2\phi}(\zeta+\kappa\,\theta)^2
\ee
avec
\ba \lb{ds622}
\theta&=&dx^5+\sqrt 2 A_\mu dx^\mu , \quad \zeta=dx^6+\sqrt 2 B_\mu dx^\mu,\\
F_{\mu\nu}(B)&=&\e^{-2\phi}\tilde{F}_{\mu\nu}(A)-\kappa F_{\mu\nu}(A).\lb{ds623}
\ea
Nous voyons donc que, \`a toute solution ($ds_4^2$, $A$, $\phi$, $\kappa$) de EMDA correspond une solution $ds_6^2$ de la \RG sans source \`a six dimensions.

Pour finir signalons que nous pouvons mettre (\ref{ds621})-(\ref{ds623}) sous la forme plus compacte
\be
ds_6^2=G_{AB}dx^A dx^B=ds_4^2+\lambda_{ab}(dx^a+\sqrt 2 A^a_\mu dx^\mu)(dx^b+\sqrt 2 A^b_\mu dx^\mu)
\ee
avec 
\be
\lambda=\left(\begin{array}{cc}
\e^{-2\phi}+\kappa^2 \e^{2\phi}&\kappa \e^{2\phi}\\
\kappa\e^{2\phi}&\e^{2\phi}\\
\end{array}\right),\quad A^a_\mu=(A_\mu,B_\mu)
\ee
o\`u ($ds_4^2$, $A$, $\phi$, $\kappa$) est une solution de EMDA si
\be
det(\lambda)=1,\quad F^a_{\mu\nu}=-\varepsilon^{ab}\lambda_{bc}\tilde{F}^c_{\mu\nu}
\ee
avec
\be
\left(\begin{array}{c}F^1_{\mu\nu}\\F^2_{\mu\nu}\end{array}\right)=\left(\begin{array}{c}F_{\mu\nu}(A)\\F_{\mu\nu}(B)\end{array}\right).
\ee
Les indices $a, b, \ldots$ varient de $5$ \`a $6$ alors que les indices $\mu, \nu, \ldots$ prennent les valeurs $1, 2,3, 4$ et $\varepsilon^{ab}$ est le symbole totalement antisym\'etrique \footnote{o\`u nous avons adopt\'e la convention $\varepsilon^{12}=1$}.

Au d\'ebut de ce chapitre, nous avons vu que l'action de EMDA est invariante sous la transformation de $SL(2,R)$ (\ref{sdu}). Ici, nous voyons que cette sym\'etrie a pour origine l'invariance de la th\'eorie d'Einstein \`a six dimensions dans l'espace des deux vecteurs de Killing $\p_5$ et $\p_6$.

\section{Le formalisme quasilocal}%

La d\'erivation des formules pour l'\'energie quasilocale et le moment angulaire quasilocal dans le cas de EMDA est tr\`es similaire \`a ce qui a \'et\'e fait dans le cas de EMD (voir chapitre 2, section 5). La contribution gravitationnelle est toujours inchang\'ee et est donn\'ee par les formules (\ref{equas2}) et (\ref{jquas}). Concentrons nous sur la partie mati\`ere de l'action de EMDA:
\begin{equation} \label{actionEMDA2} S_m =-
\frac{1}{16\pi}\int d^4x\sqrt{|g|}\left\{2\partial_\mu\phi\partial^\mu\phi + \frac{1}{2}\e^{4\phi}{\partial_\mu}\kappa\partial^\mu\kappa
+\e^{-2\phi}F_{\mu\nu}F^{\mu\nu}+\kappa F_{\mu\nu}{\tilde
F}^{\mu\nu}\right\}, \end{equation}
Le moment conjugu\'e de $\kappa$ est
\be
p_\kappa=-\frac{\sqrt g}{16\pi}\e^{4\phi} \p^0\kappa.
\ee
Le moment conjugu\'e de $\phi$ est inchang\'e
\be
p_\phi=-\frac{\sqrt g}{4\pi}\p^0\phi
\ee
alors que le moment conjugu\'e de $A_i$ est modifi\'e par l'ajout d'une contribution provenant du terme de couplage entre l'axion et le champ \'electromagn\'etique:
\be \lb{piemda}
\Pi^i=\frac{\sqrt g}{4\pi}(\e^{-2\phi} F^{i0}+\kappa\tilde{F}^{i0})\equiv\frac{\sqrt g}{4\pi}E^i,
\ee
le champ \'electrique $E^i$ \'etant lui aussi modifi\'e.

Pour faire appara\^itre le Hamiltonien dans l'action (\ref{actionEMDA2}), nous utilisons les relations
\be \lb{relkappa}
\p_0\kappa=\frac{16\pi N^2}{\sqrt g}\e^{-4\phi}p_\kappa+N^i\p_i\kappa,\quad \p^i\kappa=h^{ij}\p_j\kappa+\frac{16\pi N^i}{\sqrt g}\e^{-4\phi}p_\kappa,
\ee
ainsi que les relations similaires pour $\phi$ (\ref{relphi})(en prenant $\alpha=1$). Les relations pour $F^{\mu\nu}$ (\ref{relF}) sont modifi\'ees par l'ajout d'une contribution de l'axion:
\ba\lb{relF2}
F_{0i}&=&N^j\bar{F}_{ji}+\frac{4\pi N}{\sqrt h}\e^{2 \phi}\Pi_i-\frac{\sqrt{g}}{2}\kappa\e^{2\phi}\varepsilon_{ikl} \bar{F}^{kl},\\
 F^{ij}&=&\bar{F}^{ij}+\frac{N^i}{\sqrt g}(4\pi\Pi^j-\frac{\kappa}{2}\varepsilon^{jkl}\bar{F}_{kl})\e^{2\phi}-\frac{N^j}{\sqrt g}(4\pi\Pi^i-\frac{\kappa}{2}\varepsilon^{ikl}\bar{F}_{kl})\e^{2\phi}.
\ea

Comme dans le cas de EMD (chapitre $2$, section $5$), nous obtenons 
\be
S_m =\int dt\left[\int_{\Sigma_t} d^3x \left(p_\phi\p_0\phi+\Pi^i\p_0A_i-N{\cal H}-N^i{\cal H}_i-A_0 {\cal H}_A\right)-\frac{1}{4\pi}\int_{S^r_t}\sqrt \sigma d^2x NA_0E^in_i\right]
\ee
o\`u ${\cal H}$, ${\cal H}_i$ et ${\cal H}_A$ sont les contraintes associ\'ees aux multitplicateurs de Lagrange $N$, $N^i$ et $A_0$
\ba
{\cal H}&=&\frac{2\pi}{\sqrt h}p_\phi^2+\frac{\sqrt h}{8\pi}h^{ij}\p_i\phi\p_j\phi+\frac{8\pi}{\sqrt h}\e^{-4\phi}p_\kappa^2+\frac{\sqrt h}{32\pi}\e^{4\phi}h^{ij}\p_i\kappa\p_j\kappa+\frac{2\pi}{\sqrt h}\e^{2\phi}\Pi_i\Pi^i\nonumber\\
&&\quad +\frac{\sqrt h}{16\pi}(\e^{-2\phi}+\kappa^2\e^{2\phi})\bar{F}_{ij}\bar{F}^{ij}-\frac{\sqrt{h}}{2}\kappa\e^{2\phi}\Pi^i\varepsilon_{ikl}\bar{F}^{kl}\\
{\cal H}_i&=&p_\phi\p_i\phi+p_\kappa\p_i\kappa+\bar{F}_{ij}\Pi^j\\
{\cal H}_A&=&-\p_i\Pi^i.
\ea
En particulier, nous voyons que l'axion, comme le dilaton, ne contribue qu'indirectement au Hamiltonien (dans le $\Pi^i$). Nous en d\'eduisons les formules suivantes pour les contributions de la mati\`ere \`a l'\'energie quasilocale et au moment angulaire quasilocal:
\be \lb{equasdilax}
E_m=\int_{S^r_t}A_0\bar{\Pi}^in_id^2x,\quad J_m=-\int_{S^r_t}A_\varphi\bar{\Pi}^in_id^2x
\ee
o\`u nous avons introduit $\bar{\Pi}^i=(\sqrt \sigma/\sqrt h) \Pi^i$.
\section{G\'en\'eration de solutions en rotation}%

Lorsque la constante de couplage du dilaton $\alpha$ est \'egale \`a un, nous avons vu que EMD \'etait contenue dans EMDA. Donc les solutions non asymptotiquement plates (\ref{solnas}) de EMD pour $\alpha=1$ sont aussi solution de EMDA. Dans l'article {\tt ``Linear Dilaton Black Holes''} (voir Appendice A), en utilisant l'invariance du mod\`ele $\sigma$ sous la transformation (\ref{trans2}), la version \'electrique de la solution (\ref{solnas}) a \'et\'e g\'en\'eralis\'ee. Dans cette section, nous allons faire le m\^eme travail pour la version magn\'etique
\ba \lb{solnas2}
ds^2=-\frac{r-b}{r_0}dt^2+\frac{r_0}{r-b}\left[dr^2+r(r-b)d\Omega^2\right],\\
F=\frac{r_0}{\sqrt 2} \sin\theta d\theta\wedge d\varphi,\quad \e^{2\phi}=\frac{r_0}{r}.
\ea
Nous remarquons que cette solution poss\`ede la m\^eme m\'etrique r\'eduite $h_{ij}$ que la solution de Schwarzschild
\ba \lb{schw4}
ds^2 &=& -\left(1-\frac{b}r\right)dt^2 +\left
(1-\frac{b}r\right)^{-1}\left(dr^2 + r(r-b) d\Omega^2\right), \\ e^{2\phi} &=& 1, \quad
\kappa = 0, \quad F = 0,\quad b=2M. \ea
Par cons\'equent, ces deux solutions sont li\'ees par la transformation de $Sp(4,R)$
\be
M_{\ell m}=U_m^TM_S U_m
\ee
avec
\be
M_S=\left(\begin{array}{cccc} \frac{r}{r-b}&0&0&0\\0&-1&0&0\\0&0&\frac{r-b}{r}&0\\0&0&0&-1 \end{array}\right),\quad
 M_{\ell m}=\left(\begin{array}{cccc} \frac{r_0}{r-b}&0&0&\frac{r}{r-b}\\0&-\frac{r_0}{r}&-1&0\\0&-1&-\frac{b}{r_0}&0\\\frac{r}{r-b}&0&0&\frac{br}{r_0(r-b)} \end{array}\right)
\ee
et
\be
U_m=\left(\begin{array}{cccc} \sqrt{\frac{r_0}{b}}&0&0&\sqrt{\frac{b}{r_0}}\\0& \sqrt{\frac{r_0}{b}}&\sqrt{\frac{b}{r_0}}&0\\0& \sqrt{\frac{r_0}{b}}&0&0\\ \sqrt{\frac{r_0}{b}}&0&0&0 \end{array}\right).
\ee
Cette transformation fait passer d'une solution asymptotiquement plate neutre \`a une solution non asymptotiquement plate charg\'ee magn\'etiquement. Donc, comme au chapitre pr\'ec\'edent, en appliquant cette transformation sur une m\'etrique asymptotiquement plate neutre en rotation, la solution de Kerr,
\ba ds^2 & = &
\frac{\Delta-a^2\sin^2\theta}{\Sigma}(dt-\omega d\varphi)^2 -
\Sigma\bigg(\frac{dr^2}{\Delta}+d\theta^2+\frac{\Delta\sin^2\theta}
{\Delta-a^2\sin^2\theta}\,d\varphi^2 \bigg), \\ e^{2\phi} &=&1,
\quad \kappa = 0, \quad F = 0, \ea
avec
\ba \Delta = r^2-2 M
r+a^2, \quad  \Sigma =  r^2+a^2\cos^2\theta, \quad \omega =
-2\,\frac{aMr\sin^2\theta} {\Delta -
a^2\sin^2\theta}.   \ea
nous obtenons une solution non asymptotiquement plate charg\'ee magn\'etiquement et en rotation.
La matrice (\ref{MPQ}) associ\'ee \`a la m\'etrique de Kerr \'etant
\begin{equation}
M_{K}=\left(
\begin{array}{cccc}
f^{-1}&0&-\chi f^{-1}&0\\ 0&-1&0&0\\ -\chi f^{-1}&0&f+\chi^2
f^{-1}&0\\ 0&0&0&-1\\
\end{array}
\right) , \label{mk}
\end{equation}
avec
\be
f=\frac{\Delta-a^2\sin^2\theta}{\Sigma}, \quad
\chi=-2\frac{aM\cos\theta}{\Sigma}\,.
\end{equation}
nous obtenons la nouvelle matrice 
\be \lb{MlKm}
M_{\ell K m}=U_m^TM_KU_m=\left(\begin{array}{cccc} \frac{r_0}{b}\frac{1-f}{f}&-\frac{r_0}{b}\frac{\chi}{f}&0&f^{-1}\\-\frac{r_0}{b}\frac{\chi}{f}&\frac{r_0}{b}\frac{\chi^2+f^2-f}{f}&-1&-\frac{\chi}{f}\\0&-1&-\frac{b}{r_0}&0\\f^{-1}&-\frac{\chi}{f}&0&\frac{b}{r_0f} \end{array}\right).
\ee
En comparant  (\ref{MPQ}) et (\ref{MlKm}), nous en d\'eduisons les $f$, $\phi$, $\kappa$, $u$, $v$ et $\chi$ de la nouvelle solution. Puis en utilisant (\ref{tauchi}), nous obtenons, \`a partir du potentiel $\chi$, les $\tau^i$ puis les $\omega_k$. Nous pouvons alors reconstituer la solution associ\'ee \`a la matrice $M_{\ell Km}$
\ba \lb{tna1}
ds^2  = 
\frac{\Delta-a^2\sin^2\theta}{r_0 r}(dt-\omega d\varphi)^2 -
r_0 r\bigg(\frac{dr^2}{\Delta}+d\theta^2+\frac{\Delta\sin^2\theta}
{\Delta-a^2\sin^2\theta}\,d\varphi^2 \bigg), \\ e^{2\phi} =\frac{r_0}{r},
\quad \kappa = -\frac{a\cos\theta}{r_0}, \quad A=-\frac{a\cos\theta}{\sqrt 2r}dt-\frac{r_0 \cos\theta}{\sqrt 2}d\varphi,\quad
\omega=\frac{a r_0 r\sin^2\theta}{\Delta-a^2\sin^2\theta}. \lb{tna12}
\ea
Nous avons vu dans la section 4.1 que EMDA \'etait invariante sous une transformation de dualit\'e (\ref{sdu}) permettant de passer de la version magn\'etique d'une solution \`a la version \'electrique de cette solution. Cependant, nous pouvons aussi utiliser la matrice $U_{me}$ qui fait passer des matrices $M_{m}$ associ\'ees aux versions magn\'etiques des solutions aux matrices $M_e$ associ\'ee aux versions \'electriques des solutions 
\be
M_{m}=U_{me}^T M_{e} U_{me}, \quad U_{me}=\left(\begin{array}{cccc}0&-1&-\frac{b}{r_0}&0\\\frac{r_0}{b}&-\frac{r_0}{b}&-1&1\\-\frac{r_0}{b}&-\frac{r_0}{b}&0&0\\0&1&0&0\end{array}\right).
\ee

Dans la section pr\'ec\'edente, nous avons adapt\'e le formalisme quasilocal \`a EMDA. En regroupant les contributions mat\'erielles et gravitationnelles, nous obtenons les m\^emes formules (\ref{equas1}) et (\ref{jquas1}) que dans le cas de EMD pour l'\'energie quasilocale et le moment angulaire quasilocal, mais avec le moment conjugu\'e $\Pi^i$ de $A_i$ modifi\'e par la contribution de l'axion (\ref{piemda}). Nous allons maintenant utiliser ces formules pour calculer la masse et le moment angulaire de la solution (\ref{tna1})-(\ref{tna12}). Le calcul des quantit\'es quasilocales a \'et\'e grandement d\'etaill\'e dans la section 6 du chapitre 2 et dans la section 5 du chapitre 3. De plus, le calcul de la masse et du moment angulaire de la solution en rotation de EMDA \'etant tr\`es similaire \`a celui conduit dans la section 5 du chapitre 3, nous n'entrerons pas dans les d\'etails. L'\'energie quasilocale est divergente et il est n\'ecessaire de retrancher la contribution d'une solution de fond pour obtenir un r\'esultat fini. Comme dans les chapitres pr\'ec\'edents, nous choisissons comme solution de fond le fond charg\'e sur lequel les trous noirs se forment, c'est-\`a-dire la solution (\ref{tna1})-(\ref{tna12}) avec $a=b=0$. Pour calculer la masse nous avons besoin de la courbure extrins\`eque de la surface $S^r_t$
\be
k=-\frac{1}{r}\sqrt{\frac{\Delta}{r_0r}}\simeq -\frac{1}{\sqrt{r_0 r}}\left(1-\frac{b}{2r}\right)
\ee
ce qui donne pour le premier terme de la contribution gravitationnelle
\be
N(\varepsilon-\varepsilon_0)=\frac{b}{16\pi}\sin\theta.
\ee
De m\^eme, en utilisant 
\be
\pi^r_\varphi=-\frac{ar_0\sin^3\theta}{32\pi}\sqrt{\frac{\Delta}{r_0r}}
\ee
la contribution du deuxi\`eme terme est
\be
n_r\pi^r_\varphi N^\varphi=-\frac{a^2\sin^3\theta}{32\pi r}
\ee
qui s'annule \`a l'infini et donc ne contribue pas \`a la masse.
Pour calculer la contribution mat\'erielle, nous avons besoin du moment conjugu\'e $\Pi^r$ de $A_r$ 
\be
\Pi^r=\frac{a}{2\sqrt 2 \pi}\cos\theta\sin\theta.
\ee 
La contribution mat\'erielle est alors ($\Pi^r_0=\Pi^r(a=b=0)=0$)
\be
A_0(\Pi^r-\Pi^r_0)=\frac{a^2\cos^\theta\sin\theta}{4\pi r}
\ee
qui s'annule \`a l'infini et donc ne contribue pas \`a la masse. En regroupant les trois contributions, nous obtenons pour la masse 
\be
{\cal M}=\frac{b}{4}.
\ee

Le calcul du moment angulaire fait intervenir les m\^emes quantit\'es que nous venons d'utiliser pour le calcul de la masse. De plus, nous n'avons pas \`a retrancher la contribution de la solution de fond puisque celle-ci est statique ($J_0=0$).
La contribution gravitationnelle et la contribution mat\'erielle au moment angulaire sont
\ba
2n_r\pi^r_\varphi=-\frac{a r_0}{16\pi}\sin^3\theta\\
A_\varphi\Pi^r=-\frac{ar_0}{4\pi}\cos^2\theta\sin\theta
\ea
et nous voyons que, comme pour la version \'electrique (voir l'article {\tt ``Linear Dilaton Black Holes''} dans l'Appendice A ), la contribution mat\'erielle est deux fois sup\'erieure \`a la contribution gravitationnelle
\be
J=J_g+J_m=\frac{ar_0}{6}+\frac{ar_0}{3}=\frac{ar_0}{2}.
\ee
Nous voyons aussi que la masse et le moment angulaire sont les m\^emes que ceux de la version \'electrique. 
Comme dans le cas des solutions de la th\'eorie d'Einstein-Maxwell dilatonique dans les chapitres 2 et 3, nous pouvons montrer que la version \'electrique et la version magn\'etique de la solution en rotation satisfont \`a la premi\`ere loi de la thermodynamique des trous noirs si la charge n'est pas vari\'ee. La charge n'\'etant pas un param\`etre du trou noir mais un param\`etre du fond, ce r\'esultat semble correct.

\section{Partenaire \`a six dimensions de la solution en rotation}%

D\'eterminons maintenant le partenaire \`a six dimensions de la solution (\ref{tna1})-(\ref{tna12}). Nous obtenons en utilisant (\ref{ds621})
\ba
ds_6^2&=&-2dt dx^6+2adtd\varphi+\frac{b}{r_0}dt^2+\frac{r_0}{r}(dx^6)^2+r_0rd\varphi^2-\frac{2a\cos\theta}{r}dx^6 dx^5\nonumber\\
&&\quad -2r\cos\theta d\varphi dx^5+\frac{r^2+a^2\cos^2\theta}{r_0r}(dx^5)^2+r_0rd\theta^2+\frac{r_0r}{\Delta}dr^2.
\ea
En posant 
\ba \lb{x5}
\rho&=&2\sqrt{r_0 r},\quad \bar{\theta}=\frac{\theta}{2},\quad \varphi_\pm=\frac{\varphi\pm \eta}{2}\\
d\tau&=&\sqrt{\frac{r_0}{b}}\left(dx^6-ad\varphi\right),\quad d\psi=\sqrt{\frac{b}{r_0}}\bigg(dt-\sqrt{\frac{r_0}{b}}d\tau\bigg),\quad d\eta=\frac{dx^5}{r_0} \lb{x6}
\ea
la m\'etrique se met sous la forme d'un produit direct 
\be \lb{pd}
ds_6^2 =  d\psi^2 + ds_5^2,
\end{equation}
o\`u $ds_5^2$ est la m\'etrique de Myers et Perry (\ref{MP}) avec les deux moments angulaires \'egaux ($a_+=a_-$):
\ba\label{mype} ds_5^2 & = & -d\tau^2 +
\frac{\mu}{\rho^2}\bigg(d\tau + (\bar{a}/2)(d\varphi -
\cos\theta\,d\eta)\bigg)^2 +
\frac{d\rho^2}{1-\mu\rho^{-2}+\mu\bar{a}^2\rho^{-4}} +
\rho^2\,d\Omega_3^2  \nonumber \\ & = & -d\tau^2 +
\frac{\mu}{\rho^2}\bigg(d\tau +
\bar{a}\sin^2\bar{\theta}d\varphi_+ +
\bar{a}\cos^2\bar{\theta}d\varphi_-\bigg)^2 \\ & & +
\frac{d\rho^2}{1-\mu\rho^{-2}+\mu\bar{a}^2\rho^{-4}} +
\rho^2\bigg(d\bar{\theta}^2 + \sin^2\bar{\theta}d\varphi_+^2 +
\cos^2\bar{\theta}d\varphi_-^2\bigg) \nonumber \ea
avec $\mu = 4br_0$ et $\bar{a} = 2(r_0/b)^{1/2}a$.
Donc la version magn\'etique de la solution (\ref{tna1}) peut-\^etre obtenue par une double r\'eduction dimensionnelle (\ref{ds621}) de (\ref{pd}) en utilisant les d\'efinitions (\ref{x5})-(\ref{x6}) pour $x^5$ et $x^6$.

Or, dans l'article ``Linear dilaton black holes'', le partenaire \`a six dimensions de la version \'electrique de la solution (\ref{tna1}) a \'et\'e calcul\'e et il a \'et\'e montr\'e que la m\'etrique \`a six dimensions peut-\^etre aussi mise sous la forme d'un produit direct (\ref{pd}) o\`u $ds_5^2$ est aussi la m\'etrique de Myers et Perry avec les deux moments angulaires \'egaux. La version \'electrique s'obtient elle aussi par double r\'eduction dimensionnelle de (\ref{pd}) mais par rapport \`a  des variables $x^5$ et $x^6$ diff\'erentes:
\ba 
\varphi_\pm&=&\frac{\varphi\pm x^6/r_0}{2}\\
d\psi&=&\sqrt{\frac{b}{r_0}}\bigg(dt+\frac{r_0}{b}dx^5\bigg),\quad d\tau=\sqrt{\frac{r_0}{b}}dx^5. 
\ea

Nous voyons que la diff\'erence dans la r\'eduction dimensionnelle conduisant \`a la version \'electrique ou \`a la version magn\'etique r\'eside principalement dans l'\'echange des variables $x^5$ et $x^6$. Le fait que l'\'echange des variables $x^5$ et $x^6$ n'affecte pas le r\'esultat de la r\'eduction dimensionnelle n'est pas \'etonnant. En effet, nous avons impos\'e d\`es le d\'epart que la th\'eorie d'Einstein \`a six dimensions poss\`ede les deux vecteurs de Killing $\p_5$ et $\p_6$. La th\'eorie EMDA obtenue par r\'eduction de la th\'eorie d'Einstein \`a six dimensions par rapport \`a ces deux vecteurs de Killing est alors invariante sous les transformations de $SL(2,R)$ agissant dans le plan $(\p_5,\p_6)$.


\section{Mouvement g\'eod\'esique et champ scalaire}%

Etudions le mouvement g\'eod\'esique des particules dans le champ de gravitation de la m\'etrique (\ref{tna1}) que nous \'ecrivons sous la forme ADM:
\be \lb{geo}
ds^2=-\frac{\Delta}{r_0r}dt^2+r_0 r\left[\frac{dr^2}{\Delta}+d\theta^2+\sin^2\theta\left(d\varphi-\frac{a}{r_0r}dt\right)^2\right].
\ee
La m\'etrique inverse est donn\'ee par:
\be \lb{ginv}
g^{\mu\nu}\p_\mu\p_\nu=-\frac{r_0r}{\Delta}\p_t\p_t-\frac{2a}{\Delta}\p_t\p_\varphi+\frac{\Delta-a^2\sin^2\theta}{r_0r\Delta\sin^2\theta}\p_\varphi\p_\varphi+\frac{\Delta}{r_0r}\p_r\p_r+\frac{1}{r_0r}\p_\theta\p_\theta.
\ee
La courbure scalaire
\be
R=\frac{\Delta+a^2\sin^2\theta}{2r_0r^3}
\ee
est finie dans tout l'espace except\'ee bien s\^ur en $r=0$ o\`u se trouve la singularit\'e.

Comme dans la section 3 du chapitre 2, nous allons utiliser l'\'equation de contrainte
\be \lb{cont}
g_{\mu\nu}\dot{x}^\mu \dot{x}^\nu=-\varepsilon.
\ee
pour \'etudier le mouvement des g\'eod\'esiques du genre temps ($\varepsilon=1$), lumi\`ere ($\varepsilon=0$) et espace ($\varepsilon=-1$) .

Le Lagrangien \'etant ind\'ependant de $t$ et $\varphi$, nous en d\'eduisons les deux constantes du mouvement $E$ et $L$:
\ba
\frac{\p {\cal L}}{\p \dot{t}}&=&-\frac{2\Delta}{r_0 r}\dot{t}+2r_0r\sin^2\theta \left(\dot{\varphi}-\frac{a}{r_0r}\dot{t}\right)\left(-\frac{a}{r_0r}\right)=-2E\\
\frac{\p {\cal L}}{\p \dot{\varphi}}&=&2r_0r\sin^2\theta \left(\dot{\varphi}-\frac{a}{r_0r}\dot{t}\right)=2L
\ea
ce qui donne, en inversant le syst\`eme,
\ba
\dot{t}&=&\frac{r_0r}{\Delta}E-\frac{a}{\Delta}L\\
\dot{\varphi}&=&\frac{a}{\Delta}E+\frac{\Delta-a^2\sin^2\theta}{r_0r\Delta\sin^2\theta}L.
\ea
En utilisant ces deux relations, la contrainte (\ref{cont}) devient
\be \lb{eqgeo}
\dot{r}^2=E^2-\frac{2a}{r_0r}EL+\frac{a^2\sin^2\theta-\Delta}{r_0^2r^2\sin^2\theta}L^2-\frac{\Delta}{r_0r}\varepsilon-\Delta\dot{\theta}^2.
\ee

Etudions le mouvement g\'eod\'esique dans le plan \'equatorial ($\theta=\pi/2$). L'\'equation (\ref{eqgeo}) devient:
\be
\dot{r}^2+V=E^2
\ee
avec le potentiel effectif
\be
V=\frac{2a}{r_0r}EL+\frac{r-b}{r_0^2r}L^2+\frac{r-b}{r_0}\varepsilon.
\ee
Lorsque $r$ tend vers l'infini le potentiel effectif est donn\'e par:
\be
V\rightarrow\left\{\begin{array}{ll}\frac{r}{r_0},&\quad\varepsilon=1\\\frac{L^2}{r_0^2},&\quad \varepsilon=0,\, E\neq\frac{L}{r_0}\\E^2+\frac{b-2a}{r_0^2r}L^2,&\quad \varepsilon=0,\, E=\frac{L}{r_0}\\-\frac{r}{r_0},&\quad\varepsilon=-1.\end{array}\right.
\ee 
Nous en d\'eduisons que les g\'eod\'esiques du genre espace peuvent se propager \`a l'infini au contraire des g\'eod\'esiques du genre temps qui sont r\'efl\'echies par le potentiel qui croit lin\'eairement avec $r$. Les g\'eod\'esiques du genre lumi\`ere pour lesquelles $E>\frac{L}{r_0}$ se propagent jusqu'\`a l'infini alors que celles pour lesquelles $E<\frac{L}{r_0}$ sont r\'efl\'echies par le potentiel. Lorsque $E=\frac{L}{r_0}$, le comportement des g\'eod\'esiques du genre lumi\`ere diff\`ere suivant que (\ref{geo}) repr\'esente une singularit\'e nue, un trou noir extr\^eme ou un trou noir. Elles sont confin\'ees lorsque $b<2a$ (singularit\'e nue), orbitent autour du trou noir lorsque $b=2a$ (trou noir extr\^eme) et se propagent jusqu'\`a l'infini lorsque $b>2a$ (trou noir).

Examinons, maintenant, le comportement d'un champ scalaire massif dans le champ de gravitation de la solution (\ref{geo}). Le champ scalaire $\psi=\psi(r,\theta)\e^{i(m\varphi-\omega t)}$ ob\'eit \`a l'\'equation de Klein-Gordon
\be
(\nabla_\mu\nabla^\mu-\mu^2)\psi=0
\ee
qui devient en utilisant (\ref{geo}) et (\ref{ginv})
\be
\p_r(\Delta\p_r\psi)+\frac{(r_0r\omega-am)^2}{\Delta}\psi-\mu^2r_0r\psi+\frac{1}{\sin\theta}\p_\theta(\sin\theta\p_\theta\psi)-\frac{m^2}{\sin^2\theta}\psi=0.
\ee
 Cette \'equation est s\'eparable en posant $\psi=R(r)\Theta(\theta)$
\ba \lb{eqR}
\p_r(\Delta\p_rR)+\frac{(r_0r\omega-am)^2}{\Delta}R-\mu^2r_0rR=\varkappa^2R\\
-\frac{1}{\sin\theta}\p_\theta(\sin\theta\p_\theta\Theta)+\frac{m^2}{\sin^2\theta}\Theta=\varkappa^2\Theta. \lb{eqT}
\ea
L'\'equation (\ref{eqT}) est l'\'equation diff\'erentielle des harmoniques sph\'eriques avec $\varkappa^2=l(l+1)$, $l=0,1\ldots$ et $l\geq |m|$.

Etudions maintenant l'\'equation radiale. 

i) \underline{$b=a=0$} (fond charg\'e):

\noindent L'\'equation (\ref{eqR}) se r\'eduit \`a
\be \lb{eqR00}
\p_r(r^2\p_rR)-(\nu^2-1/4+\mu^2r_0r)R=0
\ee
o\`u $\nu=\sqrt{(l+1/2)^2-r_0^2\omega^2}$.

\noindent En posant,
\be
r=\frac{x^2}{2\mu \sqrt{r_0}},\quad R=X/x
\ee
l'\'equation devient
\be
x^2\p^2_xX+x\p_xX-(\bar{\nu}^2+x^2)X=0
\ee
avec $\bar{\nu}=2\nu$. Cette \'equation est l'\'equation diff\'erentielle des fonctions de Bessel modifi\'ees
La solution de (\ref{eqR00}) est donc une combinaison lin\'eaire d'une fonction de Bessel de premi\`ere esp\`ece $I$ et d'une fonction de Bessel de seconde esp\`ece $K$:
\be \lb{sola0b0}
R=Ar^{-1/2}I_{2\nu}(2\mu\sqrt{r_0r})+Br^{-1/2}K_{2\nu}(2\mu\sqrt{r_0r}).
\ee
Le comportement asymptotique de cette solution est le suivant. A l'infini ($r\rightarrow \infty$) et pour $\mu\neq 0$, nous avons
\be \lb{mudiff0}
R\sim \bar{C}_1 \frac{\e^{2\mu\sqrt{r_0r}}}{r^{3/4}}+\bar{C}_2\frac{\e^{-2\mu\sqrt{r_0r}}}{r^{3/4}}.
\ee
Lorsque $r\rightarrow\infty$ et $\mu=0$, la solution d\'epend de la valeur de $\omega$: 
\ba \lb{mu01}
R\sim C_1r^{-1/2-\nu}+C_2r^{-1/2+\nu},&\quad \omega<(l+1/2)/r_0,\,\, \nu>0\\
R\sim C_1r^{-1/2}\e^{-iqx}+C_2r^{-1/2}\e^{iqx},&\quad \omega>(l+1/2)/r_0,\,\, \nu=iq\\
R\sim C_1r^{-1/2}+C_2r^{-1/2}\ln r,&\quad \omega=(l+1/2)/r_0,\,\, \nu=0\lb{mu02}
\ea
o\`u $x=\ln r$.
Par analogie avec le mouvement g\'eod\'esique nous pouvons interpr\'eter les solutions (\ref{mudiff0}) et (\ref{mu01})-(\ref{mu02}). Lorsque $\mu\neq 0$, comme pour les g\'eod\'esiques du genre temps, il y a une barri\`ere de potentiel et l'exponentielle d\'ecroissante repr\'esente l'onde exponentiellement supprim\'ee au fur et \`a mesure qu'elle p\'en\`etre la barri\`ere de potentiel. De plus, nous devons prendre $C_2=0$ pour avoir une solution physiquement acceptable, l'exponentielle croissante divergeant \`a l'infini. Lorsque $\mu=0$, comme pour les g\'eod\'esiques du genre lumi\`ere, il y a deux cas de figure. Lorsque $\nu>1/2$, c'est-\`a-dire $\omega^2<(l+1/4)(l+3/4)/r_0^2$ ($E<L/r_0$), l'onde est confin\'ee et nous devons prendre $C_2=0$ pour avoir une solution physiquement acceptable. Au contraire, lorsque $\nu<1/2$ c'est-\`a-dire $\omega^2>(l+1/4)(l+3/4)/r_0^2$ ($E>L/r_0$), l'onde est libre de se propager jusqu'\`a l'infini.

Il y a donc des modes provenant de la singularit\'e et d'autres provenant de l'infini. En interf\'erant, ces ondes peuvent alors donner naissance \`a un spectre d'\'energie $\omega$ discret. Dans ce cas la fonction d'onde doit \^etre de carr\'e sommable:
\be\lb{norme}
||\psi||^2=\int_\Sigma j^\mu d\Sigma_\mu=-i\int_\Sigma (\psi\p^\mu \psi^* -\psi^*\p^\mu\psi)d\Sigma_\mu.
\ee
Lorsque $a=b=0$ et $\nu\in R$ ($\omega\geq (l+1/2)/r_0$), cette condition devient
\ba
||\psi||^2 &=& -\int\omega g^{00}|\psi|^2\sqrt gd^3x=4\pi\int_0^\infty\omega r_0^2 R^2 dr\\
&=&4\pi\int_0^\infty\omega r_0^2 \left(Ar^{-1/2}I_{2\nu}(2\mu\sqrt{r_0r})+Br^{-1/2}K_{2\nu}(2\mu\sqrt{r_0r})\right)^2 dr.
\ea
Or, cette int\'egrale n'est finie que lorsque $A=B=0$ donc il n'y a pas de solution de carr\'e sommable. Lorsque $\nu\in iR$ ($\omega< (l+1/2)/r_0$), la solution n'est pas normalisable, mais est une onde sph\'erique. Donc, dans le cas $a=b=0$ qui correspond au fond sur lesquel les trous noirs se forment, le spectre est continu.

ii) \underline{$M<|a|$} (singularit\'e nue):

\noindent Dans ce cas la condition de normalisation (\ref{norme}) devient
\be
||\psi||^2 =-\int(\omega g^{00}-m g^{0\varphi})|\psi|^2\sqrt gd^3x=4\pi\int_0^\infty(r_0\omega r-a m)\frac{r_0r}{\Delta} R^2 dr.
\ee
Tout d'abord examinons le cas $a=0$ et $M<0$. L'\'equation radiale est:
\be
\p_r(r(r-b)\p_rR)+\left(\frac{(r_0 r \omega)^2}{r(r-b)}-\mu^2r_0r-\varkappa^2\right)R=0
\ee
que nous pouvons r\'esoudre exactement dans le cas $\mu=0$. Posons
\be
x=\frac{b}{r},\quad R=x^{\alpha}(1-x)^{\beta}X,\quad\alpha=\frac{1+\nu}{2},\quad\beta=ir_0\,\omega
\ee
o\`u $\nu=\sqrt{1+4(\varkappa^2-r_0^2\omega^2)}$. L'\'equation devient alors
\be
x(1-x)\p_x^2X+({\tt c}-(1+{\tt a}+{\tt b})x)\p_xX-{\tt a}{\tt b}X=0
\ee
avec 
\be
{\tt a}={\tt b}=\frac{1+\nu}{2}+ir_0\,\omega,\quad {\tt c}=1+\nu
\ee
qui est l'\'equation diff\'erentielle des fonctions hyperg\'eom\'etriques.
La solution est donn\'ee par \cite{stegun}
\be
R=x^\alpha(1-x)^\beta\left[C_1F({\tt a},{\tt b};{\tt c};x)+C_2x^{1-{\tt c}}F({\tt a}-{\tt c}+1,{\tt b}-{\tt c}+1;2-{\tt c};x)\right].
\ee
Lorsque $\nu\in R$, au voisinage de l'infini ($x\rightarrow 0$), la solution est 
\be
R\sim C_1x^{\frac{1+\nu}{2}}+C_2x^{\frac{1-\nu}{2}}.
\ee
Cette solution est de carr\'e sommable si $C_2=0$.
Au voisinage de la singularit\'e ($x\rightarrow -\infty$), la solution normalisable \`a l'infini (c'est-\`a-dire avec $C_2=0$) s'\'ecrit
\be
R=C_1\left(A\,x^{-{\tt a}}F[{\tt a},1- {\tt c}+{\tt a};1;x^{-1}]+B\,x^{-{\tt b}}F[{\tt b},1-{\tt c}+{\tt b};1;x^{-1}]\right).
\ee
Cette solution diverge logarithmiquement et n'est donc pas normalisable.
Donc, lorsque $\mu=0$, comme dans le cas $a=b=0$, il n'y a pas de solution de carr\'e sommable, et le spectre est continu.

Lorsque $\mu\neq 0$, la solution \`a l'infini est donn\'ee par (\ref{mudiff0}) alors que pr\`es de la singularit\'e l'\'equation radiale s'\'ecrit 
\be
r\p_r^2R+\p_rR=0
\ee
qui a pour solution
\be
R=C_3+C_4\ln R.
\ee
Or, $\ln R$ est une solution singuli\`ere de l'\'equation de Klein-Gordon (\ref{eqR}). Donc, pour avoir une solution r\'eguli\`ere, il faut poser $C_4=0$. Cette solution a une norme finie. A l'infini la solution est normalisable si $\bar{C}_1=0$. Donc il existe une solution normalisable si $\bar{C}_1=0$ et $C_4=0$. Le spectre est discret.

Examinons maintenant le cas $a> 0, M<a$. Dans le cas $\mu=0$, l'\'equation radiale peut-\^etre \`a nouveau r\'esolue. Posons
\be
r=M+i c\frac{\xi+1}{\xi-1},\quad R=\xi^\alpha(1-\xi)^\beta\Xi(\xi)
\ee
o\`u
\ba
c&=&\sqrt{a^2-M^2},\quad \alpha=\frac{\omega^{'}-im^{'}}{2},\quad\beta=\frac{1}{2}(1+\sqrt{1+4(\varkappa^2-r_0^2\omega^2)})\\
\omega^{'}&=&\frac{Mr_0\omega-am}{c},\quad m^{'}=r_0\omega.
\ea
\be
\xi(1-\xi)\p_\xi^2\Xi+({\tt c}-(1+{\tt a}+{\tt b})\xi)\p_\xi\Xi-{\tt a b}\Xi=0
\ee
avec
\ba
{\tt a}&=&\frac{1+2\omega^{'}+\sqrt{1+4(\varkappa^2-r_0^2\omega^2)}}{2},\  {\tt b}=\frac{1-2im^{'}+\sqrt{1+4(\varkappa^2-r_0^2\omega^2)}}{2}\\
{\tt c}&=&1+\omega^{'}-im^{'}.
\ea
La solution est 
\ba
R&=&\xi^\alpha(1-\xi)^\beta\bigg(C_1F({\tt a},{\tt b};{\tt a}+{\tt b}+{\tt c}-1;1-\xi)\nonumber\\
&&\ \quad\qquad\qquad+C_2(1-\xi)^{{\tt c}-{\tt a}-{\tt b}} F({\tt c}-{\tt b},{\tt c}-{\tt a};{\tt c}-{\tt a}-{\tt b}+1;1-\xi)\bigg).
\ea
Au voisinage de l'infini ($\xi\rightarrow 1$), la solution
\be
R\sim(1-\xi)^{2\beta}(C_1+C_2(1-\xi)^{{\tt c}-{\tt a}-{\tt b}})
\ee
est de carr\'e sommable si $C_2=0$ et $\sqrt{1+4(\varkappa^2-r_0^2\omega^2)}>0$. Au voisinage de la singularit\'e l'\'equation radiale devient:
\be \lb{vs}
a^2\p_r^2R+(m^2-\varkappa^2)R=0
\ee
qui a pour solution une s\'erie enti\`ere en $r$. Cependant la solution $R=r$ est une solution singuli\`ere de l'\'equation (\ref{eqR}) et pour avoir une solution r\'eguli\`ere il faut donc imposer la contrainte
\be\lb{cond}
\p_rR|_{r=0}=0.
\ee
Donc lorsque $\mu=0$, il existe un spectre semi-discret. Lorsque $\omega^2>(\varkappa^2+1/4)/r_0^2$, le spectre est continu alors que lorsque $\omega^2<(\varkappa^2+1/4)/r_0^2$, le spectre est discret. La r\'esolution num\'erique de l'\'equation (\ref{cond}) montre que le nombre de niveaux d'\'energie est \'egal \`a $m$ et que le signe de $\omega$ est oppos\'e \`a celui de $m$.

Lorsque $\mu\neq 0$, au voisinage de l'infini  la solution est donn\'ee par (\ref{mudiff0}) et nous devons prendre $\bar{C}_2=0$ pour avoir une solution finie \`a l'infini. Au voisinage de la singularit\'e, la solution est donn\'ee par (\ref{vs}) et nous devons, comme dans la cas $\mu=0$, imposer la condition de r\'egularit\'e (\ref{cond}). Donc, dans le cas $\mu \neq 0$, le spectre est discret.

iii) \underline{$M^2>a^2$} (trou noir):

Dans le cas d'un trou noir, il est bien connu qu'il ne peut exister de spectre discret. En effet, en introduisant la coordonnée ``tortue'',
\begin{equation}
dr_*=\frac{r_0r_+}{\Delta}dr,\quad r_*=\frac{r_0r_+}{r_+-r_-}\ln\left(\frac{r-r_+}{r-r_-}\right),
\end{equation}
l'\'equation radiale devient
\begin{equation}
\frac{d^2R}{dr^2_*}-VR=0,\quad V=\frac{\mu^2r\Delta}{r_0r_+^2}+\frac{\Delta l(l+1)}{r_0r_+^2}-\frac{r^2}{r_+^2}\left(\omega-\frac{am}{r_0r}\right)^2
\end{equation}
Or, au voisinage de l'horizon ($\Delta=0$), cette \'equation admet pour solution une onde sph\'erique
\begin{equation}
R=C_1\e^{ikr_*}+C_2\e^{-ikr_*},\quad k=\omega-m\Omega_h,
\end{equation}
qui n'est pas de carr\'e sommable. Le spectre est donc continu dans le cas d'un trou noir.

Cependant, dans le cas d'un trou noir en rotation, comme c'est le cas ici, un autre ph\'enom\`ene m\'erite d'\^etre \'etudi\'e: la super-radiance. Ce processus, d\'ecouvert par Penrose \cite{Pen69}, permet d'extraire de l'\'energie du trou noir. Il n\'ecessite la pr\'esence d'une zone o\`u $g_{tt}$ est positif, appel\'ee ergosph\`ere, qui est situ\'ee entre l'horizon $r_h$ (z\'ero du lapse $N$, voir \ref{ADM}) et $r_e$ (z\'ero de $g_{tt}$, appel\'ee limite statique).

L'extraction de l'\'energie du trou noir peut se faire par l'interm\'ediaire de particules \cite{Pen69,WaldEMDA} ou d'ondes \cite{WaldEMDA,MU,Z72,S73}.
Dans le cas d'une onde incidente $\psi=\psi(r,\theta)\e^{i(m\varphi-\omega t)}$ envoy\'ee sur le trou noir, nous pouvons montrer que, lorsque l'\'energie $\omega$ de l'onde incidente est telle que 
\be
\omega<m\Omega_h,
\ee 
(o\`u $\Omega_h$ est la vitesse angulaire de l'horizon) une partie de l'onde est absorb\'ee par le trou noir (onde transmise) alors que la partie r\'efl\'echie par l'horizon du trou noir a une \'energie sup\'erieure \`a l'onde incidente. Ceci peut \^etre fait, soit en calculant les coefficients de r\'eflexion et de transmission \cite{dewitt}, soit \`a partir du th\'eor\`eme de Hawking sur l'aire d'un trou noir \cite{beken73b,HRe}. 

Dans le cas d'une solution trou noir asymptotiquement plate, les modes super-radiants peuvent se propager jusqu'\`a l'infini en emportant une partie de l'\'energie du trou noir. Au contraire dans le cas de la solution non asymptotiquement plate (\ref{geo}), les modes super-radiants ne peuvent s'\'echapper \`a l'infini. En effet, nous avons vu plus haut que les modes massifs $\mu\neq 0$ ainsi que les modes non massifs tels que 
\be
\omega<\frac{l+1/2}{r_0}
\ee
sont r\'efl\'echis par une barri\`ere de potentiel avant d'atteindre l'infini. Or, comme la vitesse angulaire de l'horizon est 
\be
\Omega_h=\left. \frac{d\varphi}{dt}\right|_{r=r_h}=\left. -\frac{g_{t\varphi}}{g_{\varphi\varphi}}\right|_{r=r_h}=\frac{a}{r_0r_+}=\frac{a}{r_0(M+\sqrt{M^2-a^2})}
\ee
et que $M>a$ et $|m|<l$ nous avons 
\be
\frac{l+1/2}{r_0}<m\Omega_h.
\ee
Donc tous les modes super-radiants sont r\'efl\'echis par la barri\`ere de potentiel. L'\'energie prise au trou noir ne peut donc \^etre emport\'ee \`a l'infini dans notre cas. L'onde effectue des allers et retours entre l'horizon et la barri\`ere de potentiel pompant \`a chaque fois un peu plus de l'\'energie du trou noir. En raisonnant par analogie avec le cas des trous noirs asymptotiquement plats, le processus pourrait se poursuivre jusqu'\`a \'epuisement du moment angulaire du trou noir. La solution trou noir en rotation serait donc instable et l'\'etat final du trou noir serait la solution (\ref{geo}) avec $a=0$, c'est-\`a-dire la solution statique (\ref{solnas2}).
Cependant, avant de tirer des conclusions et pour savoir si un tel processus peut avoir lieu pour nos solutions, il faudrait faire une \'etude plus rigoureuse de la quantification du champ scalaire dans un espace-temps non asymptotiquement Minkowskien et non asymptotiquement AdS, puis \'etudier la super-radiance \`a l'image de ce qui a \'et\'e fait pour des espaces asymptotiquement AdS \cite{AIS,winstanley}


\newpage
\thispagestyle{empty}
\null
\chapter{Trous noirs \`a plus de quatre dimensions et branes noires}

Dans les chapitres pr\'ec\'edents nous nous sommes int\'eress\'es \`a des solutions (d\'ecrivant des trous noirs) \`a quatre dimensions, alors que dans le chapitre suivant nous chercherons des solutions \`a trois dimensions. Dans ce chapitre, nous allons \'elargir notre recherche \`a un nombre quelconque $D$ de dimensions d'espace-temps ($D>2$). Ce chapitre regroupe une partie d'un travail en cours sur la construction et l'\'etude de solutions trou noir et branes noires non asymptotiquement plates.

Nous allons tout d'abord introduire la th\'eorie, puis nous construirons de nouvelles solutions statiques non asymptotiquement plates de cette th\'eorie. Apr\`es avoir adapt\'e le formalisme quasilocal \`a la th\'eorie, nous l'utiliserons pour calculer la masse de ces solutions. Les deux derni\`eres parties de ce chapitre seront consacr\'ees \`a montrer comment nous pouvons g\'en\'eraliser les solutions statiques \`a des solutions en rotation, pour des valeurs particuli\`eres de la constante de couplage du dilaton.

Dans ce chapitre, nous utilisons un syst\`eme d'unit\'e tel que $G=1/16\pi$.

\section{Trous noirs et branes noires}%

La th\'eorie de Kaluza-Klein \cite{kaluza,klein} fut la premi\`ere \`a postuler l'existence de dimensions suppl\'ementaires dans une tentative d'unification de la gravitation et de l'\'electromagn\'e-tisme. En prenant comme point de d\'epart la gravitation sans source \`a cinq dimensions et en compactifiant la th\'eorie le long de la cinqui\`eme dimension, nous obtenons la \RG \`a quatre dimensions coupl\'ee au champ \'electromagn\'etique (plus un champ scalaire non d\'esir\'e coupl\'e au champ \'electromagn\'etique, nomm\'e plus tard le dilaton). Les solutions de cette th\'eorie d\'ecrivant des trous noirs ont \'et\'e largement \'etudi\'ees \cite{GW6,rasheed5,MP5,tang5,RE5}.

L'id\'ee des dimensions suppl\'ementaires a retrouv\'e un second souffle avec l'apparition de la th\'eorie des cordes, qui est le candidat moderne \`a l'unification des quatre interactions fondamentales et \`a la formulation d'une th\'eorie quantique de la gravitation. Il existe cinq th\'eories de supercordes ($D=10$) diff\'erentes reli\'ees entre elles (ainsi qu'\`a la supergravit\'e \`a $11$ dimensions) par de nombreuses dualit\'es \cite{W6,Sen6,FILQ,HT6}, ce qui laisse penser qu'elles ne sont que diff\'erentes facettes d'une seule et m\^eme th\'eorie: la th\'eorie $M$ ($D=11$).
Les trous noirs sont suppos\'es jouer un r\^ole aussi important en gravitation quantique qu'en Relativit\'e G\'en\'erale. Il est donc important d'\'etudier les solutions trou noir des diff\'erentes th\'eories des cordes. Un exemple de trous noirs \`a plus de quatre dimensions est fourni par la solution de Tangherlini \cite{tang5}
\be
ds^2=-\left(1-\frac{2M}{r^{D-3}}\right)dt^2+\left(1-\frac{2M}{r^{D-3}}\right)^{-1}dr^2+r^2 d\Omega_{D-2}^2.
\ee
Cette solution est la g\'en\'eralisation de la solution de Schwarzschild pour la \RG sans source \`a $D$ dimensions. 

\noindent Cependant l'existence de dimensions suppl\'ementaires fait appara\^itre un autre type d'objet noir: les $p$-branes noires qui sont des objets \'etendus \`a $p$ dimensions d'espace (la $1$-brane \'etant une corde) entour\'es par un horizon. Par exemple, la solution
\be
ds^2=-\left(1-\frac{2M}{r}\right)dt^2+\left(1-\frac{2M}{r}\right)^{-1}dr^2+r^2 d\Omega_2^2+dx^idx_i,
\ee
qui est le produit de la solution de Schwarzschild \`a quatre dimensions et d'un objet \'etendu \`a $D-4$ dimensions, repr\'esente une ($D-4$)-brane entour\'ee d'un horizon. D'autres solutions trou noir et $p$-branes noires ont \'et\'e construites dans les r\'ef\'erences \cite{GM6,HS6}.

\section{Pr\'esentation de la th\'eorie}%

Dans la suite, nous allons chercher des solutions d\'ecrivant des $p$-branes noires de la th\'eorie d\'efinie par l'action:
\begin{equation}\label{actionemdD}
S = \int d^D x \sqrt{-g} \left( R - \frac12 \partial_\mu \phi
\partial^\mu \phi - \frac1{2\, q!} \, {\rm e}^{a\phi} \, F_{[q]}^2
\right)
\end{equation}
qui d\'ecrit la gravitation \`a $D$ dimensions coupl\'ee \`a un champ scalaire $\phi$ (le dilaton) et \`a une $q$-forme $F=dA$. Les solutions obtenues dans cette th\'eorie peuvent avoir de nombreuses applications. En effet, cette th\'eorie, suivant les valeurs de $a$ et $q$ peut repr\'esenter, par exemple, une partie du secteur bosonique de la supergravit\'e ($D=11$, $q=4$ et $a=0$) ou bien l'action effective \`a basse \'energie des diff\'erentes th\'eories des cordes \cite{argurio}.
D'autre part, dans le cas $D=4$ et $q=2$, (\ref{actionemdD}) se r\'eduit (moyennant le remplacement $\phi\rightarrow -2\phi$) \`a l'action de EMD (\ref{actionEMD}).

Contrairement \`a l'action de EMD (2.1), l'action (5.3) n'est pas invariante sous la tranformation de dualit\'e (nomm\'ee S-dualit\'e) (sauf dans le cas o\`u $d=\tilde{d}$, auquel cas les formes $F$ et $\hat{F}$ ont le m\^eme rang):
\be
(g_{\mu\nu},\phi,F_{\mu\nu})\rightarrow (\hat{g}_{\mu\nu}=g_{\mu\nu},\hat{\phi}=-\phi,\hat{F}=\e^{a\phi}\star F)
\ee
o\`u la $(D-q)$-forme $\star F$ est le dual  de la $q$-forme $F$:
\be \lb{sdual}
(\star F)^{\mu_1\ldots \mu_{D-q}}=\frac{1}{q!\sqrt g}\varepsilon^{\mu_1\ldots\mu_{D-q}\ldots\mu_D}F_{\mu_{D-q+1}\ldots\mu_D}
\ee
avec $\varepsilon$ le tenseur totalement antisym\'etrique\footnote{La convention adopt\'ee pour le symbole antisym\'etrique est $\varepsilon^{1\cdots (D-1)t}=1$}. Cette transformation fait passer d'une th\'eorie o\`u $F$ est une $q$-forme \`a la th\'eorie duale o\`u $\hat{F}$ est une $(D-q)$-forme.

Remarquons que, de mani\`ere similaire au cas de la th\'eorie d'Einstein-Maxwell dilatonique, nous pouvons librement effectuer une translation de $\phi$ accompagn\'ee d'un changement d'\'echelle de $F$
\be
\phi\rightarrow\phi+\phi_0,\quad F\rightarrow \e^{-a\phi_0/2}F
\ee
sans que l'action (\ref{actionemdD}) en soit modifi\'ee.

En variant l'action par rapport \`a $g_{\mu\nu}$, $A_\mu$ et $\phi$, nous obtenons les \'equations du mouvement associ\'ees:
\begin{eqnarray}
R_{\mu\nu} - \frac12 \partial_\mu \phi \partial_\nu \phi -
\frac{{\rm e}^{a\phi}}{2(q-1)!} \left[
F_{\mu\alpha_2\cdots\alpha_q} F_\nu{}^{\alpha_2\cdots\alpha_q}-
\frac{q-1}{q(D-2)} F_{[q]}^2 \, g_{\mu\nu} \right] &=& 0,
\label{Ein} \\
\partial_\mu \left( \sqrt{-g} \, {\rm e}^{a\phi} \,
F^{\mu\nu_2\cdots\nu_q} \right) &=& 0, \label{form} \\
\frac1{\sqrt{-g}}\, \partial_\mu \left( \sqrt{-g} \partial^\mu
\phi \right) - \frac{a}{2\, q!} {\rm e}^{a\phi} F_{[q]}^2 &=& 0.
\label{dil}
\end{eqnarray}

\section{R\'esolution des \'equations du mouvement}%

Nous allons chercher des solutions de la th\'eorie (\ref{actionemdD}) en r\'esolvant les \'equations du mouvement correspondantes (\ref{Ein})-(\ref{dil}). Nous donnerons seulement les grandes \'etapes de cette r\'esolution. Plus de d\'etails seront donn\'es dans un article en pr\'eparation (voir aussi \cite{CGL}).

Nous souhaitons obtenir des solutions statiques d\'ecrivant des $p$-branes donc des objets \'etendus dans $p$ dimensions avec un volume d'univers \`a $d=p+1$ dimensions. Nous allons donc r\'esoudre les \'equations (\ref{Ein})-(\ref{dil}) en utilisant l'ansatz m\'etrique suivant:
\begin{equation}\label{ansbra}
ds^2 = - {\rm e}^{2B} dt^2 + {\rm e}^{2D} (dx_1^2 + \cdots +
dx_p^2) + {\rm e}^{2C} \, d\Sigma_{q,\sigma}^2 + {\rm e}^{2A}
d\rho^2,
\end{equation}
o\`u les fonctions $A$, $B$, $C$ et $D$ ne d\'ependant que de $\rho$ et 
\begin{equation}
d\Sigma_{q,\sigma}^2 = \bar g_{ab} dy^a dy^b = \left\{
 \begin{array}{ll}
 d \psi^2 + \sinh^2\psi \, d\Omega_{\td}^2, \qquad & \sigma=-1,\\
 d \psi^2 + \psi^2 \, d\Omega_{\td}^2, \qquad & \sigma=0,\\
 d \psi^2 + \sin^2\psi \, d\Omega_{\td}^2, \qquad & \sigma=+1.
 \end{array} \right.
\label{gmetric}\end{equation}
o\`u $\td=q-1$.
Le cas $\sigma=1$ est le cas \`a sym\'etrie sph\'erique o\`u $\Sigma$ est une $q$-sph\`ere. Les solutions avec $\sigma=0$ ($\Sigma$ est un hyperplan de dimension $q$) et $\sigma=-1$ ($\Sigma$ est un espace hyperbolique de dimension $q$) sont appel\'ees solutions topologiques. La dimension de l'espace-temps est $D=p+q+2=d+\td+2$.

L'\'equation du mouvement pour la $q$-forme $F$ (\ref{form}) est trivialement r\'esolue en posant:
\begin{equation}\label{solF}
F_{[q]} = b  \,\, \mbox{vol}(\Sigma_{\tilde{d} +1,\sigma}),
\end{equation}
ce qui correspond \`a une solution du type ``magn\'etique'' o\`u le param\`etre $b$ est reli\'e \`a la charge ``magn\'etique'' et $\mbox{vol}(\Sigma_{q,\sigma})$ est le volume de $\Sigma_{q,\sigma}$:
\be
\mbox{vol}(\Sigma_{q,\sigma})=\frac{\sqrt{\bar{g}}}{q!}\,\varepsilon_{\mu_1\ldots\mu_q}\,dx^{\mu_1}\wedge\ldots\wedge dx^{\mu_q}.
\ee
La version \'electrique peut bien s\^ur \^etre obtenue \`a l'aide de la transformation de dualit\'e (\ref{sdual}).

En utilisant l'ansatz (\ref{ansbra}), les composantes du tenseur de Ricci sont:
\ba
R_{tt} &=& {\rm e}^{2B-2A} \left[   B'' + B' (B'-A'+q\,C'+pD' )  \right], \label{Rtt}\\
R_{xx} &=& \e^{2D-2A}\left[- D'' - D'(B'-A'+q\,C'+p
D')\right],\\
R_{\rho\rho} &=& - B'' - B'(B'-A') - q\,(C''+C'^2-A' C')
\nonumber \\
& & \qquad\qquad - p\,(D''+D'^2-A' D'), \lb{Rrr} \\
R_{ab} &=&  \left\{- {\rm e}^{2C-2A} \left[  C'' +
C'(B'-A'+q\,C'+pD' )  \right] + \sigma \td \right\}
\, \bar g_{ab}. \label{Rab}
\ea

Nous voyons alors que pour r\'esoudre l'\'equation d'Einstein (\ref{Ein}) et l'\'equation du dilaton (\ref{dil}), il est commode de se placer dans la jauge 
\be
A-B-qC-pD=0.
\ee
Les \'equations (\ref{Ein}) et (\ref{dil}) donnent quatre \'equations pour les fonctions $B$, $C$, $D$ et $\phi$, 
\ba
B'' &=& \frac{\td b^2}{2(D-2)}\e^G, \label{EqB}\\
C'' &=& -\frac{db^2}{2(D-2)}\e^G+\sigma \td {\rm e}^{2(A-C)},
\lb{EqC}\\
D'' &=& \frac{\td b^2}{2(D-2)}\e^G, \label{EqD} \\
\phi'' &=& \frac{a b^2}2\e^G,\label{EqPhi}
\ea
plus une contrainte,
\be
-(B'+qC'+pD')^2 + B'^2 + qC'^2 + pD'^2 +
\frac12  \phi'^2 = \frac{b^2}2\e^G - \sigma \td(\td+1)\e^{2(A-C)}\lb{cons}
\ee
avec 
\be\lb{G}
G=a\phi+2B+2(d-1)D.
\ee 
Ce syst\`eme d'\'equations admet pour solution \cite{CGL}
\ba
B&=&\frac{\td}{\Delta(D-2)}\left(G-g_1\rho-g_0\right),\lb{Sol1}
\\ D&=&\frac{\td}{\Delta(D-2)}\left(G-g_1\rho-g_0\right)+d_1\rho+d_0,\lb{Sol2}
\\ C&=&\frac{1}{2\td}\;H-\frac{d}{\Delta
(D-2)}\; G+c_1\rho+c_0,\lb{Sol3}
\\ A&=& \frac{(1+\td)}{2\td}\,H-\frac{d}{\Delta(D-2)}\;G+c_1\rho+c_0,\lb{Sol5}\\
\phi&=&\frac{a}{\Delta}\;G+f_1\rho+f_0,\ea
o\`u
\ba
\lb{SolG} 
G&=&\ln\left(\frac{\alpha^2}{\Delta b^2}\right)
-\ln\left[\sinh^2\left(\frac{\alpha}{2}
(\rho-\rho_0)\right)\right],\\
\lb{SolH} 
H &=& \left\{ \begin{array}{ll}
 2\ln\beta/2\td-\ln\left( \sinh^2[\beta (\rho-\rho_1)/2
   ] \right), \qquad & \sigma=1, \\
 \pm \beta (\rho-\rho_1), & \sigma=0, \\
 2\ln\beta/2\td-\ln\left( \cosh^2[\beta (\rho-\rho_1)/2
   ] \right), & \sigma=-1, \end{array}\right.
\ea
et
\ba
\Delta&=&a^2+\frac{2d\td}{D-2}\\
c_{0,1}&=&\frac{a}{\Delta}\left(\frac{d}{D-2}\phi_{0,1}-\frac{(d-1)a}{\td}d_{0,1}\right)\\
f_{0,1}&=&\frac{2\td}{a}c_{0,1}\lb{Solc}\\
g_{0,1}&=&a\phi_{0,1}+2(d-1)d_{0,1}.
\ea
La contrainte (\ref{cons}) s'\'ecrit alors:
\ba\lb{cons1} 
&&\frac{(\td+1)\beta^2}{4\td}-\frac{\alpha^2}{2\Delta}-
\frac{d\td}{\Delta(D-2)}\phi_1^2  + \frac{2a(d-1)}{\Delta}\phi_1d_1
\nonumber \\ && \qquad
-\frac{d-1}{\Delta(D-2)\td} \bigg(a^2(D-2)(D-3) + 2\td^2\bigg)d_1^2  = 0.
\ea
La solution d\'epend de neuf param\`etres $b,\,d_0,\,d_1,\,\phi_0,\,\phi_1,\,\rho_0,\,\rho_1,\,\alpha$ et $\beta$ ce qui laisse huit param\`etres ind\'ependants en tenant compte de la contrainte (\ref{cons1}).

Nous pouvons cependant fixer certains de ces param\`etres. Tout d'abord, nous pouvons effectuer une translation de $\rho$ sans changer la signification physique de la m\'etrique. Ceci nous permet de fixer $\rho_1$,
\be\lb{rho1}
\rho_1=0.
\ee
Les fonctions $G$ (\ref{SolG}) et $H$ (\ref{SolH}) ne d\'ependent pas du signe de $\rho$. En effet, nous pouvons passer de $\rho$ positif \`a $\rho$ n\'egatif en changeant $\alpha$ et $\beta$ en $-\alpha$ et $-\beta$. Nous choisissons $\alpha$ et $\beta$ positif. Lorsque $\rho$ tend vers l'infini, le comportement asymptotique de $G$ et $H$ est 
\be
G\simeq -\alpha\rho,\quad H\simeq -\beta\rho
\ee
et $B$ tend vers
\be
B\simeq \frac{\td}{\Delta(D-2)}(-\alpha\rho-g_1\rho).
\ee
Or, nous sommes int\'eress\'es par une solution poss\'edant un horizon, c'est-\`a-dire un z\'ero de $g_{tt}$. Nous voyons que $g_{tt}=\e^{2B}$ s'annule en $\rho\rightarrow\infty$ lorsque
\be
\alpha+g_1>0.
\ee
De plus, cet horizon doit \^etre r\'egulier (les invariants de courbure doivent \^etre finis) si nous voulons que la solution d\'ecrive un trou noir et non une singularit\'e nue. Une condition suffisante pour que l'horizon soit r\'egulier est que $D$ et $\phi$ soient finis sur l'horizon. Lorsque $\rho$ tend vers l'infini, nous avons
\ba
D\simeq -\frac{\td}{\Delta(D-2)}(\alpha+g_1)\tau+d_1\tau\\
\phi\simeq -\frac{a}{\Delta}\alpha\tau+f_1\tau.
\ea
Nous voyons que $D$ et $\phi$ sont finis sur l'horizon si les coefficients de $\tau$ dans $D$ et $\phi$ sont nuls, ce qui impose les deux relations suivantes
\ba \lb{alp}
\alpha=\frac{\Delta}{a}f_1\\
d_1=\frac{\td}{a(D-2)}\phi_1.\lb{d1}
\ea
En translatant $\rho$ et en imposant que l'horizon soit r\'egulier, nous avons donc fix\'e trois nouveaux param\`etres ce qui laisse cinq param\`etres libres. Mais nous pouvons encore changer l'\'echelle de $\rho$, $t$ et $x$ sans changer la signification physique de la m\'etrique ce qui nous permet de fixer deux autres param\`etres
\be
d_0=0,\qquad d_1=1.\lb{d0}
\ee
En combinant toutes les conditions pr\'ec\'edentes (\ref{rho1}), (\ref{alp})-(\ref{d1}), (\ref{d0}) et la contrainte (\ref{cons}), nous avons
\be
\alpha=\beta=2,\quad f_{1}=\frac{2a}{\Delta},\quad \phi_{1}=\frac{a}{\td},\quad g_{1}=\frac{\Delta(D-2)}{\td}-2,\quad c_{1}=\frac{a^2}{\Delta\td}
\ee
et il ne reste que trois param\`etres libres $\rho_0$, $c_0$ et $b$.
Les solutions du syst\`eme (\ref{EqB})-(\ref{cons}), suivant les valeurs de $\sigma$, qui admettent un horizon r\'egulier sont

$\centerdot$ $\sigma=1$:
\ba
ds^2&=&\left(\frac{\e^\rho}{2\sinh(\rho-\rho_0)}\right)^{\frac{4\td}{\Delta(D-2)}}\left(-\e^{-2\rho}dt^2+dx_1^2+\ldots+dx_p^2\right)\nonumber\\
&&\quad+\mu^2\left(\frac{\e^\rho}{2\sinh\rho}\right)^{\frac{2}{\td}}\left(\frac{2\sinh(\rho-\rho_0)}{\e^\rho}\right)^{\frac{4d}{\Delta(D-2)}}\left(d\Sigma^2+\frac{d\rho^2}{\td^2\sinh^2\rho}\right),
\ea

$\centerdot$ $\sigma=-1$:
\ba
ds^2&=&\left(\frac{\e^\rho}{2\sinh(\rho-\rho_0)}\right)^{\frac{4\td}{\Delta(D-2)}}\left(-\e^{-2\rho}dt^2+dx_1^2+\ldots+dx_p^2\right)\nonumber\\
&&\quad+\mu^2\left(\frac{\e^\rho}{2\cosh\rho}\right)^{\frac{2}{\td}}\left(\frac{2\sinh(\rho-\rho_0)}{\e^\rho}\right)^{\frac{4d}{\Delta(D-2)}}\left(d\Sigma^2+\frac{d\rho^2}{\td^2\cosh^2\rho}\right),
\ea

\vspace{.5cm}
$\centerdot$ $\sigma=0$ et $H=-\beta\rho$:
\ba
ds^2&=&\left(\frac{\e^\rho}{2\sinh(\rho-\rho_0)}\right)^{\frac{4\td}{\Delta(D-2)}}\left(-\e^{-2\rho}dt^2+dx_1^2+\ldots+dx_p^2\right) \nonumber\\
&&\qquad +\mu^2\left(\frac{2\sinh(\rho-\rho_0)}{\e^\rho}\right)^{\frac{4d}{\Delta(D-2)}}\left(d\Sigma^2+\e^{-2\rho}d\rho^2\right),
\ea

\vspace{.5cm}
$\centerdot$ $\sigma=0$ et $H=\beta\rho$:
\ba
ds^2&=&\left(\frac{\e^\rho}{2\sinh(\rho-\rho_0)}\right)^{\frac{4\td}{\Delta(D-2)}}\left(-\e^{-2\rho}dt^2+dx_1^2+\ldots+dx_p^2\right) \nonumber\\
&&\qquad +\mu^2\left(2\sinh(\rho-\rho_0)\right)^{\frac{4d}{\Delta(D-2)}}\e^{2\frac{\Delta+a^2}{\Delta\td}\rho}\left(d\Sigma^2+\e^{-2\rho}d\rho^2\right),
\ea
o\`u nous avons introduit $\mu$ d\'efini par
\be
\ln\mu=c_0+\frac{a^2}{\Delta\td}\ln2-\frac{1}{\td}\ln\td-\frac{d}{\Delta(D-2)}\ln\left(\frac{4}{\Delta b^2}\right).
\ee
L'expression du dilaton est la m\^eme quelque soit $\sigma$:
\be
\e^{a\phi}=\frac{4\td^2}{\Delta b^2}\mu^{2\td}\left(\frac{\e^\rho}{2\sinh(\rho-\rho_0)}\right).
\ee

\section{La solution non asymptotiquement plate \`a sym\'etrie sph\'erique ($\sigma=1$)}%

Dans cette section, nous allons nous int\'eresser plus particuli\`erement \`a la solution avec $\sigma=1$, c'est-\`a-dire lorsque $\Sigma_{q,\sigma}$ est une sph\`ere de dimensions $q$. 
La solution d\'epend de trois param\`etres $b$, $\mu$ et $\rho_0$. Le syst\`eme de coordonn\'ees utilis\'e jusqu'ici \'etant peu intuitif (l'infini \'etant ``situ\'e'' en $\rho=0$ et l'horizon en $\rho\rightarrow\infty$), nous allons utiliser un autre syst\`eme de coordonn\'ees:
\be \lb{syscoo}
\e^{2\rho}=\frac{r}{r-\mu}
\ee
o\`u $r=\mu$ ($r>\mu$) est l'horizon et $r=\infty$ correspond \`a l'infini. Dans ce nouveau syst\`eme de coordonn\'ees et en posant,
\be
\e^{2\rho_0}=\frac{r_0}{\mu-r_0},
\ee
avec $0<r_0<\mu$, $g_{tt}$ s'\'ecrit:
\be
g_{tt}=-\left(\frac{r_0(\mu-r_0)r^2}{[(\mu-2r_0)r+r_0\mu]^2}\right)^{\frac{2\td}{\Delta(D-2)}}\frac{r-\mu}{r}.
\ee
En particulier, nous remarquons que dans le cas g\'en\'eral ($\mu\neq 2r_0$), $g_{tt}$ tend vers une valeur finie lorsque $r$ tend vers l'infini. La solution correspondante est donc asymptotiquement plate et nous pouvons v\'erifier que cette solution est identique \`a celle donn\'ee par Horowitz et Strominger \cite{HS6,duff}. Par contre, dans le cas particulier $\mu=2r_0$, $g_{tt}$ diverge lorsque $r$ tend vers l'infini. La solution correspondante est non asymptotiquement plate et a \'et\'e donn\'ee par \cite{CHM6} dans le cas $p=0$ ($d=1$).
Dans la suite, nous allons nous int\'eresser uniquement \`a la solution \`a sym\'etrie sph\'erique non asymptotiquement plate ($\sigma=1$ et $\mu=2r_0$).

Dans le nouveau syst\`eme de coordonn\'ees (\ref{syscoo}) et pour $r_0=\mu/2$, la solution non asymptotiquement plate \`a sym\'etrie sph\'erique s'\'ecrit:
\ba
ds^2&=&\left(\frac{r}{\mu}\right)^{\frac{4\td}{\Delta(D-2)}}\left(-\frac{r-\mu}{r}dt^2+dx_1^2+\ldots +dx_p^2\right)\nonumber\\
&&\quad+\mu^2\left(\frac{r}{\mu}\right)^{\frac{2a^2}{\Delta\td}}\left(d\Sigma^2+\frac{dr^2}{\td^2r(r-\mu)}\right)\lb{solnasD1}\\
\e^{a\phi}&=&\frac{4\td^2}{\Delta b^2}\mu^{2\td}\left(\frac{r}{\mu}\right)^{\frac{2a^2}{\Delta}},\quad F_{\mu_1\ldots\mu_q}=b\sqrt{\bar{g}}\lb{solnasD2}
\ea
et d\'epend de deux param\`etres $\mu$ et $b$.


\section{Le formalisme quasilocal}%

Nous pouvons maintenant appliquer le formalisme quasilocal aux solutions trouv\'ees pr\'ec\'edemment afin de calculer leur masse et leur moment angulaire. La d\'erivation des formules pour l'\'energie quasilocale et le moment angulaire quasilocal est tr\`es similaire \`a celle expos\'ee dans les chapitres pr\'ec\'edents, avec cependant quelques diff\'erences.

Le point de d\'epart est l'action (\ref{actionemdD}) \`a laquelle nous ajoutons le terme de surface habituel
\begin{equation}\label{actionemdDquas}
S = \int_M d^D x \sqrt{-g} \left( R - \frac12 \partial_\mu \phi
\partial^\mu \phi - \frac1{2\, q!} \, {\rm e}^{a\phi} \, F_{[q]}^2
\right)+2\int_{\p_M} K\sqrt{h}\,d^{D-1}x.
\end{equation}
o\`u $K$ est la courbure extrins\`eque  de la fronti\`ere $\p_M$ de l'espace-temps $M$ qui est cette fois-ci une hypersurface \`a $D-1$ dimensions.

Les moments conjugu\'es de $\phi$, $h_{ij}$ et $A_{i_1\ldots i_{q-1}}$ sont
\ba
p_\phi&=&-\sqrt{g}\p^0\phi,\quad p^{ij}=\sqrt h(h^{ij}K-K^{ij}),\\
\Pi^{i_1\ldots i_{q-1}}&=&-\sqrt g \e^{a\phi}F^{0i_1\ldots i_{q-1}}\equiv -\sqrt g E^{i_1\ldots i_{q-1}}.
\ea
Nous obtenons pour le Hamiltonien de la th\'eorie
\ba
H&=&\int_{\Sigma_t}\left(N{\cal H}+N^i{\cal H}_i+A_{0i_2\ldots i_{q-1}}{\cal H}_A^{i_2\ldots i_{q-1}}\right)d^{D-1}x +2\int_{S^r_t}\sqrt{\sigma}\left(Nk+\frac{n_\mu p^{\mu\nu}N_\nu}{\sqrt h}\right)d^{D-2}x\nonumber\\
&&\quad +(q-1)\int_{S^r_t}N\sqrt{\sigma}A_{0i_2\ldots i_{q-1}}E^{ji_2\ldots i_{q-1}}n_jd^{D-2}x
\ea
avec les contraintes
\be
{\cal H}=-R_{D-1}+K_{\mu\nu}K^{\mu\nu}-K^2+\frac{p_\phi^2}{2\sqrt h}+\frac{\sqrt h}{2}(\p\phi)^2+\frac{\e^{-a\phi}}{2(q-1)!\sqrt h}\Pi^2+\frac{\sqrt{h}}{2q!}\e^{a\phi}F^2\\
\ee
\ba
{\cal H}_j&=&-D_\mu\left(\frac{p^\mu{}_j}{\sqrt h}\right)+p_\phi\p_j\phi+\frac{1}{(q-1)!}\Pi^{i_1\ldots i_{q-1}}F_{ji_1\ldots i_{q-1}}\\
{\cal H}_A^{i_2\ldots i_{q-1}}&=&-(q-1)\p_j\Pi^{ji_2\ldots i_{q-1}}.
\ea
L'hypersurface $S^r_t$ est constitu\'ee du produit de l'espace $\Sigma_{k,\sigma}$ et de la $p$-brane.

Nous en d\'eduisons donc les formules suivantes pour l'\'energie et le moment angulaire:

\ba 
E&=&2\int_{S^r_t}\sqrt{\sigma}\left(N(k-k_0)+\frac{n_\mu p^{\mu\nu}N_\nu}{\sqrt h}\right)d^{D-2}x \nonumber\\
 &&\quad +(q-1)\int_{S^r_t}A_{0i_2\ldots i_{q-1}}(\bar{\Pi}^{ji_2\ldots i_{q-1}}-\bar{\Pi}^{ji_2\ldots i_{q-1}}_0)n_jd^{D-2}x\lb{eD}
\ea
\be
J_i=-2\int_{S^r_t}\frac{n_\mu p^{\mu}{}_i}{\sqrt h}\sqrt{\sigma}d^{D-2}x
 -(q-1)\int_{S^r_t}A_{ii_2\ldots i_{q-1}}\bar{\Pi}^{ji_2\ldots i_{q-1}}n_jd^{D-2}x.
\ee
o\`u $\bar{\Pi}^{ji_2\ldots i_{q-1}}=(\sqrt \sigma/\sqrt h)\Pi^{ji_2\ldots i_{q-1}}$.

Appliquons ces formules \`a la solution (\ref{solnasD1})-(\ref{solnasD2}). Comme pr\'ec\'edemment, pour obtenir un r\'esultat fini, il faut retrancher la contribution d'une solution de fond. Le candidat naturel est le fond charg\'e ($\mu=0$). Cependant nous voyons que l'on ne peut pas prendre $\mu=0$ dans cette m\'etrique. Introduisons un syt\`eme de coordonn\'ees plus adapt\'e. Posons
\be
r=\frac{\mu}{c}\bar{r}^{\td},\quad \mu=b^{\frac{2d}{\Delta(D-2)}}c^{\frac{a^2}{\Delta\td}},
\ee
accompagn\'e d'un changement d'\'echelle de $t$ et de $x$
\be
t\rightarrow\left(\frac{b}{c}\right)^{-\frac{2\td}{\Delta(D-2)}}t,\quad x\rightarrow\left(\frac{b}{c}\right)^{-\frac{2\td}{\Delta(D-2)}}x.
\ee
La solution (\ref{solnasD1})-(\ref{solnasD2}) devient:
\ba
ds^2&=&b^{-\frac{4\tilde{d}}{\Delta(D-2)}}r^{\frac{4\tilde{d}^2}{\Delta(D-2)}}\left[-\left(1-\frac{c}{r^{\tilde{d}}}\right)dt^2+dx_1^2+\cdots +dx_p^2\right]\nonumber\\
&&\quad +b^{\frac{4d}{\Delta(D-2)}}r^{-\frac{4d\tilde{d}}{\Delta(D-2)}}\left[\frac{dr^2}{1-\frac{c}{r^{\tilde{d}}}}+r^2d\Sigma_{k,1}\right]\\
\e^{a\phi}&=&\frac{4\td^2}{\Delta}\left(\frac{r^{\td}}{b}\right)^\frac{2a^2}{\Delta} ,\quad F_{\mu_1\ldots\mu_k}=b\sqrt{\bar{g}}\varepsilon_{\mu_1\ldots\mu_k}
\ea
o\`u nous avons renomm\'e $\bar{r}$ en $r$. La solution de fond est maintenant $ds^2(c=0)$.
C'est une solution statique donc son moment angulaire est nul et $N^i=0$. De plus, elle est du type ``magn\'etique'' c'est-\`a-dire $A_{0i_2\ldots i_{q-1}}=0$. Le calcul de sa masse en utilisant la formule (\ref{eD}) donne
\be
\frac{{\cal M}}{\mbox{vol}(\mbox{p-brane})}=2\left(\frac{(1+\tilde{d})a^2}{\Delta}+\frac{2\tilde{d}^2(d-1)}{\Delta(D-2)}\right)c \;\mbox{vol}(\Sigma_{k,\sigma}).
\ee
Cette masse est toujours positive puisque $d\geq 1$, $\Delta>0$ et $D>2$.

\section{G\'en\'eration de branes noires en rotation}%

Dans le chapitre 3 (section 1), nous avons remarqu\'e que l'action de EMD, dans le cas $\alpha=\sqrt 3$, est la r\'eduction de la \RG sans source \`a cinq dimensions. De m\^eme, pour une certaine valeur de la constante de couplage du dilaton, l'action \`a $D$ dimensions (\ref{actionemdD}) peut \^etre obtenue par r\'eduction dimensionnelle de la th\'eorie d'Einstein \`a $D+1$ dimensions:
\be \lb{actionRGDp1}
S=\int \sqrt{g_{D+1}}R_{D+1}d^{D+1}x.
\ee
Pour r\'eduire l'action de la \RG \`a $D+1$ dimensions, nous supposons que la m\'etrique $g^{D+1}_{\mu\nu}$ poss\`ede un vecteur de Killing $\p_{D+1}$ de genre espace. La m\'etrique peut alors \^etre mise sous la forme:
\be \lb{gDp1}
ds_{D+1}^2=\e^{-2c\phi}ds_D^2+\e^{2(D-2)c\phi}(dx^{D+1}+A_\mu dx^\mu)^2
\ee
o\`u $\phi$, $A_\mu$ et $ds_D^2$ sont ind\'ependants de la coordonn\'ee $x^{D+1}$ et $c$ est une constante que nous fixerons plus loin. En utilisant (\ref{gDp1}), la courbure scalaire $R_{D+1}$ peut \^etre d\'ecompos\'ee en fonction de la courbure scalaire \`a $D$ dimensions $R_{D}$, d'un champ scalaire $\phi$ et d'une $2$-forme $F=dA$:
\be
\sqrt{g_{D+1}}R_{D+1}=\sqrt{g_D}\left(R_{D}-(D-1)(D-2)c^2(\p\phi)^2-\frac{1}{4}\e^{2(D-1)c\phi}F^2+2c\nabla^2\phi\right)
\ee
L'action (\ref{actionRGDp1}) devient alors
\be
S=\int \sqrt{g_D}d^Dx\left(R_{D}-(D-1)(D-2)c^2(\p\phi)^2-\frac{1}{4}\e^{2(D-1)c\phi}F^2\right)
\ee
o\`u nous reconnaissons l'action (\ref{actionemdD}) dans le cas $q=2$ et $a=a_0=\sqrt{2\frac{D-1}{D-2}}$: 
\be \lb{action2f}
S=\int \sqrt{g_D}d^Dx\left(R_{D}-\frac{1}{2}(\p\phi)^2-\frac{1}{4}\e^{a_0\phi}F_{[2]}^2\right).
\ee
avec la normalisation $c=\frac{1}{\sqrt{2(D-1)(D-2)}}$.

L'action (\ref{actionemdD}) peut donc \^etre obtenue par r\'eduction dimensionnelle de la gravitation sans source \`a $D+1$ dimensions dans le cas o\`u $F$ est une $2$-forme et pour une valeur particuli\`ere $a_0$ de la constante de couplage $a$.

Dans ce cas particulier, la solution (\ref{solnasD1})-(\ref{solnasD2}) s'\'ecrit ($\td=1$, $\Delta=4$ et $d=D-3$)
\ba
ds^2&=&\left(\frac{r}{\mu}\right)^{\frac{1}{D-2}}\left(-\frac{r-\mu}{r}dt^2+dx_1^2+\ldots +dx_{D-4}^2\right)\nonumber\\
&&\quad+\mu^2\left(\frac{r}{\mu}\right)^{\frac{D-1}{D-2}}\left(d\Omega^2+\frac{dr^2}{r(r-\mu)}\right)\lb{solnasD5}\\
\e^{a_0\phi}&=&\frac{\mu^{2}}{b^2}\left(\frac{r}{\mu}\right)^{\frac{D-1}{D-2}},\quad F=b\sin\theta d\theta\wedge d\varphi .\lb{solnasD6}
\ea
Elle d\'ecrit une $(D-4)$-brane noire charg\'ee magn\'etiquement.

En utilisant (\ref{gDp1}) et en posant 
\ba
R^2=\frac{\nu}{\mu}r,\qquad \tau =\left(\frac{b}{\mu}\right)^{\frac{1}{D-1}}t,\qquad y=\left(\frac{b}{\mu}\right)^{\frac{1}{D-1}}x\\
dx^{D+1}=b\,\eta,\qquad \nu=4\left(\mu^{D-2}b\right)^{\frac{2}{D-1}}
\ea
nous obtenons facilement la m\'etrique \`a $D+1$ dimensions qui lui est associ\'ee:
\be
ds^2=-\left(1-\frac{\nu}{R^2}\right)d\tau^2+\left(1-\frac{\nu}{R^2}\right)^{-1}dR^2+R^2d\Omega_3^2+dy_1^2+\ldots+dy_p^2.
\ee
De mani\`ere non surprenante, il se passe la m\^eme chose que dans le cas de la th\'eorie d'Einstein-Maxwell dilatonique avec $\alpha=\sqrt 3$. Le partenaire \`a $D+1$ dimensions est le produit de la m\'etrique de Myers et Perry \`a cinq dimensions statique (la solution de Tangherlini) \cite{MP5,tang5} par un espace euclidien de dimension $p$. En poursuivant l'analogie avec le cas $D=4$ et $\alpha=\sqrt 3$, nous pouvons obtenir des branes noires dyoniques en rotation en r\'eduisant (par rapport au vecteur de Killing $\p_5=(\p_{\phi_+}-\p_{\phi_-})/(2r_0)$) la m\'etrique \`a $D+1$ dimensions obtenue en multipliant la m\'etrique de Myers et Perry g\'en\'erale (\ref{MP}) par un espace euclidien \`a $p=D-4$ dimensions. En suivant la m\^eme proc\'edure que dans le chapitre 3 (section 1) pour r\'eduire la m\'etrique de Myers et Perry, nous obtenons la solution suivante:
\ba \lb{dyonmagD}
ds_D^2 & = & \left(\frac{A}{\Pi}\right)^\frac{1}{D-2}\bigg[-
\frac{f^2}{A}\bigg(dt - \bar{\omega}
\,d\varphi\bigg)^2 
+\Pi \bigg(\frac{dr^2}{\Delta} + d\theta^2 +
\frac{\Delta\sin^2\theta}{f^2}\,d\varphi^2
\bigg)\,\,\,\,\,\, \lb{drotgD}\nonumber \\ &&\qquad\qquad\qquad\qquad\qquad +(dy_1^2+\cdots dy_p^2)\bigg]\\{\cal A} & = &
-\frac{r_0}{2A}\left\{\bigg[(\Delta + br)\cos\theta
-\frac{b\beta\delta}{2r_0} -
\frac{\beta\delta}{2r_0}(r-b/2)\sin^2\theta\bigg]d\varphi\right. \nonumber \\
& & \left.\qquad - \frac{b}{2r_0}(\delta-\beta\cos\theta)\,dt\right\}\,,\\
\e^{\sqrt{2}\sqrt{\frac{D-2}{D-1}}\phi}& =& \frac{A}{\Pi}\,,\lb{drotpD}
\ea
o\`u les fonctions $f$, $\Pi$, $A$, $\Delta$ et $\bar{\omega}$ sont d\'efinies par (\ref{def1}).
\section{Trous noirs multi-dimensionnels}%

Nous avons obtenu, dans la section pr\'ec\'edente, des solutions d\'ecrivant des branes noires dyoniques en rotation non asymptotiquement plates. Pour obtenir des solutions d\'ecrivant des trous noirs multi-dimensionnels, nous allons prendre comme point de d\'epart l'action (\ref{action2f}) et nous allons dualiser la $2$-forme $F$:
\be \lb{hodge}
\e^{a_0\phi}F=-\star H
\ee
ce qui donne 
\be\lb{FH}
F^2=-\frac{2}{(D-2)!}\e^{-2a_0\phi}H^2.
\ee

Les \'equations d'Einstein (\ref{Ein}) et l'\'equation du dilaton (\ref{dil}) deviennent ($q_0=D-2$)
\begin{eqnarray}
R_{\mu\nu} - \frac12 \partial_\mu \phi \partial_\nu \phi -
\frac{{\rm e}^{-a_0\phi}}{2(q_0-1)!} \left[
H_{\mu\alpha_2\cdots\alpha_{q_0}} H_\nu{}^{\alpha_2\cdots\alpha_{q_0}}-
\frac{q_0-1}{q_0(D-2)} H^2 \, g_{\mu\nu} \right] &=& 0,
\label{Ein2}  \\
\frac1{\sqrt{-g}}\, \partial_\mu \left( \sqrt{-g} \partial^\mu
\phi \right) + \frac{a_0}{2\, q_0!} {\rm e}^{-a_0\phi} H^2 &=& 0.
\label{dil2}
\end{eqnarray}
alors que l'\'equation de la $q$-forme $F$ est trivialement satisfaite. L'\'equation pour la $q_0$-forme $H$ provient de l'identit\'e de Bianchi
\be
\partial_\mu \left( \sqrt{-g} \, {\rm e}^{-a_0\phi} \,
H^{\mu\nu_2\cdots\nu_{q_0}} \right) = 0. \label{form2}
\ee
Ces \'equations d\'erivent de l'action
\be
S_D=\int\sqrt{g_D}\left(R_D-\frac{1}{2}(\p\phi)^2-\frac{1}{2(D-2)!}\e^{-a_0\phi}H^2\right).
\ee
Remarquons que cette action diff\`ere de celle obtenue en rempla\c{c}ant simplement $F$ par son expression (\ref{FH}) \cite{DT,DHT} (le signe devant le terme proportionnel \`a $H$ est diff\'erent) 
\be \lb{actionp}
S_D=\int\sqrt{g_D}\left(R_D-\frac{1}{2}(\p\phi)^2+\frac{1}{2(D-2)!}\e^{-a_0\phi}H^2\right).
\ee

Dans ce cas particulier la solution (\ref{solnasD1})-(\ref{solnasD2}) devient ($d=1$, $\td=D-3$, $\Delta=4$)
\be
ds^2=-\left(\frac{r}{\mu}\right)^{\frac{D-3}{D-2}}\frac{r-\mu}{r}dt^2+\mu^2\left(\frac{r}{\mu}\right)^{\frac{D-1}{(D-2)(D-3)}}\left(d\Omega_{D-2}^2+\frac{dr^2}{(D-3)^2r(r-\mu)}\right)\lb{solnasD7}\\
\ee
\be
\e^{-a\phi}=(D-3)^2\frac{\mu^{2(D-3)}}{b^2}\left(\frac{r}{\mu}\right)^{\frac{D-1}{D-2}},\qquad H_{\alpha_1\ldots\alpha_{(D-2)}}=b\sqrt{g_\Omega}.\lb{solnasD8}
\ee
et d\'ecrit un trou noir statique charg\'e magn\'etiquement.

Quel est le partenaire \`a $D+1$ dimensions de cette solution? Tout d'abord utilisons (\ref{hodge}) pour calculer la $2$-forme $F$. Nous obtenons 
\be
F^{r0}=-\frac{1}{(D-2)!}\frac{\e^{-a\phi}}{\sqrt{g_D}}\varepsilon^{r0\alpha_1\ldots\alpha_{D-2}}H_{\alpha_1\ldots\alpha_{D-2}}=-\frac{(-1)^D}{\sqrt{g_D}}\e^{-a\phi}b\sqrt{g_\Omega}.
\ee
ce qui donne pour $F$ 
\be
F=(-1)^D\frac{D-3}{b}\mu^{D-4}dr\wedge dt
\ee
qui d\'erive du potentiel
\be \lb{AD}
A=(-1)^D\frac{D-3}{b}\mu^{D-4}(r-\mu) \,dt.
\ee
En utilisant (\ref{solnasD7})-(\ref{solnasD8}) et (\ref{AD}) dans (\ref{gDp1}), nous obtenons la m\'etrique \`a $D+1$ dimensions suivante
\be
ds^2_{D+1}=d\eta^2-\left(1-\frac{\nu}{R^{D-3}}\right)(d\tau-d\eta)^2+\left(1-\frac{\nu}{R^{D-3}}\right)^{-1}dR^2{\bf +}R^2d\Omega^2
\ee
o\`u nous avons introduit
\be
R=\left(\frac{D-3}{b}\right)^{\frac{1}{D-1}}\mu^{2\frac{D-2}{D-1}}\left(\frac{r}{\mu}\right)^{\frac{1}{D-3}},\quad \eta=(-1)^D\left(\frac{D-3}{b}\right)^{\frac{2-D}{D-1}}\mu^{-\frac{(D-3)(D-2)}{D-1}}dx^{D+1},\;\;\ \ \ \ \ \
\ee
\be
\tau=\left(\frac{D-3}{b}\right)^{\frac{1}{D-1}}\mu^{\frac{D-3}{D-1}} t,\quad \nu^{-1}=\left(\frac{b}{D-3}\right)^{\frac{D-3}{D-1}}\mu^{-2\frac{(D-3)(D-2)}{D-1}}.
\ee
Nous reconnaissons la solution de Tangherlini (Myers et Perry statique) \`a $D$ dimensions \cite{MP5,tang5} plong\'ee dans $D+1$ dimensions et ``twist\'ee'' c'est-\`a-dire o\`u nous avons pos\'e
\be
t=\tau-\eta.
\ee

Donc, en r\'eduisant la m\'etrique de Myers et Perry g\'en\'erale (avec les moments angulaires non nuls) plong\'ee dans $D+1$ dimensions et ``twist\'ee'', puis en dualisant la $2$-forme $F$, nous devrions obtenir une solution g\'en\'eralisant la solution (\ref{solnasD7})-(\ref{solnasD8}).

\newpage
\thispagestyle{empty}
\null
\chapter{Trous noirs en Gravitation Topologiquement Massive}
Dans ce chapitre, nous allons construire et \'etudier une famille de solutions d\'ecrivant des trous noirs dans une th\'eorie \`a trois dimensions, qui est une extension de la Relativit\'e G\'en\'erale. 

Il peut sembler \'etrange de s'int\'eresser \`a une th\'eorie \`a trois dimensions alors que la recherche d'une th\'eorie englobant toutes les interactions fondamentales conduit plut\^ot \`a supposer l'existence d'un certain nombre de dimensions suppl\'ementaires. Pourtant, en g\'en\'eral, il est habituel de commencer par l'\'etude de mod\`eles simples. Puis, \`a la lumi\`ere de ce qui a \'et\'e appris pr\'ec\'edemment, la th\'eorie compl\`ete est examin\'ee. 

Cependant, la Relativit\'e G\'en\'erale \`a trois dimensions (RG3) est une th\'eorie triviale. En effet, \`a trois dimensions, les composantes du tenseur de Riemann $R^\mu{}_{\nu\rho\sigma}$ sont des combinaisons lin\'eaires des composantes du tenseur d'Einstein $G_{\mu\nu}\equiv R_{\mu\nu}-g_{\mu\nu}R$. Or, pour toute solution de la Relativit\'e G\'en\'erale, les composantes du tenseur d'Einstein sont nulles, $G_{\mu\nu}=0$, ce qui implique que toute solution de RG3 est plate ($R^\mu{}_{\nu\rho\sigma}=0$).
De plus, RG3 ne contient aucun degr\'e de libert\'e dynamique (pas d'onde gravitationnelle). Donc pour que le mod\`ele \`a $2+1$ dimensions ait un int\'er\^et, il faut consid\'erer une th\'eorie plus g\'en\'erale que RG3. Pour cela, nous ajoutons \`a l'action de RG3 un terme de Chern-Simons \cite{jackiw}, terme que l'on trouve aussi dans l'action de la supergravit\'e, par exemple.


\section{La Gravitation Topologiquement Massive}%

En 1982, Deser, Jackiw et Templeton \cite{DJT} ont introduit la Gravitation Topologiquement Massive (GTM) qui g\'en\'eralise RG3 en lui ajoutant un terme dit de Chern-Simons. L'action de cette th\'eorie est
\be \lb{actiontmg}
S=\frac{1}{16\pi G}\int d^3x\left(\sqrt{|g|}R+\frac{1}{2\mu}\varepsilon^{\lambda\mu\nu}\Gamma^\beta_{\lambda\sigma}\left[\p_\mu\Gamma^\sigma_{\beta\nu}+\frac{2}{3}\Gamma^\sigma_{\mu\tau}\Gamma^\tau_{\nu\beta}\right]\right)
\ee
o\`u $R$ est le scalaire de Ricci \`a trois dimensions, $g$ le d\'eterminant de la m\'etrique $g_{\mu\nu}$, $\varepsilon$ le symbole totalement antisym\'etrique \`a trois dimensions, $\Gamma^\lambda_{\mu\nu}$ les symboles de Christoffel, $G$ la constante de Newton et $\mu$ la constante de couplage du terme de Chern-Simons. Le terme de Chern-Simons est aussi appel\'e terme topologique car il ne d\'epend pas explicitement de la m\'etrique. D'autre part, l'ajout de ce terme fait appara\^itre un degr\'e de libert\'e dynamique: une particule de spin $2$ et de masse $\mu$. Dans la limite $\mu\rightarrow\infty$, cette particule devient infiniment massive et se d\'ecouple, les \'equations de GTM se r\'eduisant \`a celles la th\'eorie d'Einstein \`a trois dimensions.

\noindent En variant l'action par rapport \`a $g_{\mu\nu}$, nous obtenons l'\'equation du mouvement suivante
\be \lb{eqmvttmg}
G^\mu{}_\nu+\frac{1}{\mu}C^\mu{}_\nu=0
\ee
o\`u
\be
C^\mu{}_\nu=\frac{1}{\sqrt g}\varepsilon^{\mu\alpha\beta}\nabla_\alpha(R_{\nu\beta}-\frac{1}{4}g_{\nu\beta}R)
\ee
est appel\'e le tenseur de Cotton.
Nous voyons ais\'ement que sa trace $C^\mu{}_\mu$ est nulle, ce qui conduit en prenant la trace de (\ref{eqmvttmg}) \`a $R=0$. Donc, pour toute solution de GTM, la courbure scalaire est nulle.

Les auteurs de \cite{DJT} ont montr\'e que, pour un espace asymptotiquement plat, l'\'energie est positive \`a condition de prendre le ``mauvais'' signe pour le terme d'Einstein dans l'action (\ref{actiontmg}), c'est-\`a-dire prendre $G$ n\'egatif. De m\^eme, en \'etudiant la th\'eorie lin\'earis\'ee autour de la m\'etrique de Minkowski, ils ont montr\'e que cette th\'eorie d\'ecrivait la propagation d'une particule de masse $\mu$ et de spin $2$ avec un Hamiltonien d\'efini positivement \`a condition de prendre \`a nouveau $G$ n\'egatif.

De nombreuses solutions de GTM ont \'et\'e d\'eriv\'ees \cite{exact,vuo,part,nutku,GCJ,desa}. Signalons en particulier la solution de Vuorio \cite{vuo}, telle que $g_{tt}$ est constant, 
\be\lb{vuorio}
ds^2 = -\bigg[d{\tilde t} - (2\cosh\sigma + \tilde{\omega})d\tilde{\varphi}\bigg]^2 +
d\sigma^2 + \sinh^2\sigma\,d\tilde{\varphi}^2.
\ee
Nous avons \'ecrit cette solution dans un syst\`eme de coordonn\'ees tel que $\mu=3$ et nous avons r\'eintroduit la constante d'int\'egration, $\tilde{\omega}$, que Vuorio avait prise \'egale \`a $-2$.
Cette solution pr\'esente l'int\'er\^et d'admettre le groupe d'isom\'etrie $SL(2,R)\times U(1)$. De plus, c'est la seule solution (en l'absence de constante cosmologique) \`a admettre un horizon des \'ev\`enements (voir \cite{nutku,gurses} pour les solutions avec constante cosmologique).  Cependant, cette solution ne d\'ecrit pas un ``bon'' trou noir comme nous allons le voir dans la section suivante.


\section{Une famille de trous noirs}%

Pour mettre en \'evidence les horizons de la solution de Vuorio \cite{vuo}, posons
\be
\rho=\rho_0 \cosh\sigma,
\ee
avec $\rho_0>0$. La m\'etrique devient
\be
ds^2=\frac{\rho^2-\rho_0^2}{\rho_0^2 \tilde{r}^2}d\tilde{t}^2+\frac{d\rho^2}{\rho^2-\rho_0^2}-\tilde{r}^2\left(d\tilde{\varphi}+\frac{2\rho+\rho_0\tilde{\omega}}{\rho_0\tilde{r}^2}d\tilde{t}\right)^2
\ee
o\`u
\be
\tilde{r}^2=\frac{3\rho^2}{\rho_0^2}+\frac{4\tilde{\omega}\rho}{\rho_0}+\tilde{\omega}^2+1.
\ee
Les horizons sont situ\'es en (z\'ero de $N$ (\ref{ADM}))
\be
\rho=\pm\rho_0.
\ee
Nous voyons que $g_{\tilde{\varphi}\tilde{\varphi}}$ s'annule lorsque
\be
\rho=\rho_{c\pm}=\frac{\rho_0}{3}(-2\tilde{\omega}\pm\sqrt{\tilde{\omega}^2-3}).
\ee
Donc $g_{\tilde{\varphi}\tilde{\varphi}}$ est toujours n\'egatif lorsque $\tilde{\omega}^2<3$ et est n\'egatif pour $\rho<\rho_{c-}$ et $\rho>\rho_{c+}$ lorsque $\tilde{\omega}^2>3$. Or, lorsque $g_{\tilde{\varphi}\tilde{\varphi}}$ est n\'egatif, le vecteur de Killing $\p_{\tilde{\varphi}}$ est du genre temps. La coordonn\'ee angulaire $\tilde{\varphi}$ \'etant p\'eriodique, les orbites de $\p_{\tilde{\varphi}}$ sont alors des courbes ferm\'ees du genre temps. Donc la solution de Vuorio d\'ecrit un trou noir avec des courbes ferm\'ees de genre temps situ\'ees, en particulier, \`a l'ext\'erieur de l'horizon des \'ev\`enements $\rho=\rho_0$. Les violations de causalit\'e provoqu\'ees par ces courbes ferm\'ees de genre temps sont donc ``visibles'' puisque situ\'ees \`a l'ext\'erieur de l'horizon. La solution de Vuorio n'est donc pas une solution trou noir acceptable.  

Cependant, en effectuant une continuation analytique,
\be \lb{continuation}
\tilde{t} = i\sqrt{3}t, \quad \tilde{\varphi} =i(\rho_0/\sqrt{3})\varphi,
\ee
de la m\'etrique de Vuorio, nous pouvons \'eliminer une partie des singularit\'es causales. Nous obtenons alors une solution \`a deux param\`etres ($\omega$ et $\rho_0$) d\'ecrivant des trous noirs acceptables
\be\lb{bh2}
ds^2 = -\frac{\rho^2-\rho_0^2}{r^2}\,dt^2 +
\frac{d\rho^2}{\rho^2-\rho_0^2}  + r^2\bigg(d\varphi -
\frac{2\rho + 3\omega}{r^2}\,dt\bigg)^2
\ee
avec
\be\lb{r2}
r^2 = \rho^2 + 4\omega\rho + 3\omega^2 + \rho_0^2/3
\ee
o\`u nous avons pos\'e $\omega=\rho_0\,\tilde{\omega}/3$.
Cette solution poss\`ede toujours deux horizons situ\'es en 
\be
\rho_\pm=\pm\rho_0.
\ee
La m\'etrique n'est pas asymptotiquement plate mais est d\'epourvue de singularit\'e de courbure. En effet, nous avons signal\'e plus haut que pour toute solution de GTM, la courbure scalaire est nulle. Les autres invariants de courbure $R_{\mu\nu}R^{\mu\nu}$ et $R_{\mu\nu\lambda\sigma}R^{\mu\nu\lambda\sigma}$
\be
R_{\mu\nu}R^{\mu\nu}=6,\qquad R_{\mu\nu\lambda\sigma}R^{\mu\nu\lambda\sigma}=24
\ee
sont constants.

Pour interp\'eter physiquement la solution (\ref{bh2}), il faut distinguer plusieurs cas en fonction des valeurs des deux param\`etres $\omega$ et $\rho_0$. Les diagrammes de Penrose correspondants sont donn\'es dans la figure 6.1.

\noindent Dans le cas g\'en\'eral $\omega\neq\pm2\rho_0/3$, (\ref{bh2}) d\'ecrit un trou noir avec deux horizons. Son diagramme de Penrose est similaire \`a celui de la m\'etrique de Kerr (voir fig. 6.1). Cependant, la continuation analytique (\ref{continuation}) n'a pas totalement supprim\'e les courbes ferm\'ees de genre temps. En effet, $g_{\varphi\varphi}$ est n\'egatif lorsque $\rho_{c-}<\rho<\rho_{c+}$ avec
\be
\rho_{c\pm}=-2\omega\pm\sqrt{\omega^2-\frac{\rho_0^2}{3}}.
\ee
Donc, lorsque $\omega^2<\rho_0^2/3$, (\ref{bh2}) est d\'epourvue de courbes ferm\'ees de genre temps alors que pour $\omega^2\geq\rho_0^2/3$, (\ref{bh2}) contient des courbes ferm\'ees de genre temps.
\begin{figure}
\centerline{\epsfxsize=400pt\epsfbox{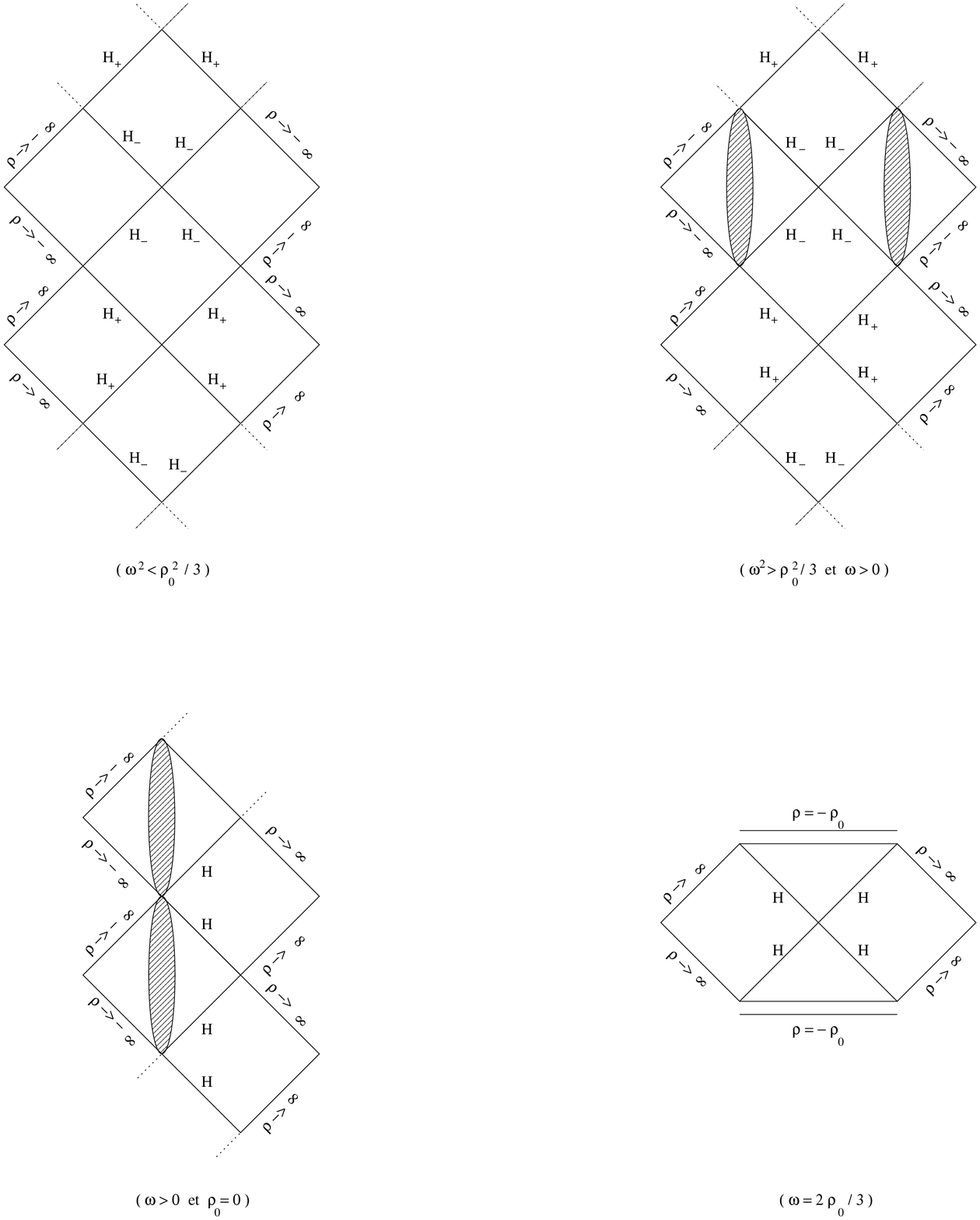}} 
\caption{Repr\'esentation des diagrammes de Penrose de (\ref{bh2}), except\'e pour le cas $\omega=\rho_0=0$ pour lequel on ne peut pas construire un tel diagramme. Les zones hachur\'ees indiquent la pr\'esence de courbes ferm\'ees de genre temps. Enfin, signalons que dans le cas $\omega=\frac{2\rho_0}{3}$, la double ligne n'est pas une singularit\'e de courbure (voir texte).}
\end{figure}
De plus, lorsque $\omega^2\geq\rho_0^2/3$ et $\omega<0$, il est facile de voir que $\rho_{c+}>\rho_{c-}>\rho_+$ et donc que les courbes ferm\'ees de genre temps sont situ\'ees \`a l'ext\'erieur de l'horizon. La pr\'esence de courbes ferm\'ees de genre temps ``visibles'' par un observateur nous conduit \`a rejeter cette solution. Au contraire, lorsque $\omega^2\geq\rho_0^2/3$ et $\omega>0$, on a $\rho_{c-}<\rho_{c+}<\rho_-$ donc les courbes ferm\'ees de genre temps sont cach\'ees derri\`ere l'horizon  int\'erieur ($\rho_-$) et, par cons\'equent, invisibles pour un observateur. Dans ce cas (\ref{bh2}) est donc une solution trou noir acceptable.

\noindent Lorsque $\rho_0=0$, les deux horizons $\rho_\pm$ co\"incident en $\rho=0$. (\ref{bh2}) d\'ecrit un trou noir extr\^eme et le diagramme de Penrose est similaire \`a celui de la m\'etrique de Kerr extr\^eme. Encore une fois, lorsque $\omega>0$, les courbes ferm\'ees de genre temps sont cach\'ees derri\`ere l'horizon et, lorsque $\omega<0$, les courbes ferm\'ees de genre temps sont situ\'ees \`a l'ext\'erieur de l'horizon. Les seuls trous noirs extr\^emes acceptables sont donc ceux pour lesquels $\omega>0$.

\noindent Examinons maintenant le cas particulier $\omega=\pm\frac{2\rho_0}{3}$. Lorsque $\omega$ est n\'egatif les courbes ferm\'ees de genre temps sont situ\'ees \`a l'ext\'erieur de l'horizon et donc la solution correspondante doit \^etre rejet\'ee. Dans le cas $\omega=\frac{2\rho_0}{3}$, (\ref{bh2}) devient 
\be\lb{sch}
ds^2 = -\frac{\rho-\rho_0}{\rho +5\rho_0/3}\,dt^2 +
\frac{d\rho^2}{\rho^2-\rho_0^2} + (\rho
+5\rho_0/3)(\rho+\rho_0)\bigg(d\varphi - \frac{2dt}{\rho
+5\rho_0/3}\bigg)^2.
\ee
Cette solution poss\`ede un horizon en $\rho=\rho_0$ et les courbes ferm\'ees de genre temps sont situ\'ees dans la zone $-5\rho_0/3<\rho<-\rho_0$. En essayant de construire le diagramme de Penrose de cette solution, nous voyons que les g\'eod\'esiques sont prolongeables \`a travers l'horizon mais nous rencontrons un probl\`eme en $\rho=-\rho_0$. En effet, au voisinage de $\rho=-\rho_0$, (\ref{sch}) devient
\be
ds^2\simeq 3dt^2-\frac{d\rho^2}{2\rho_0(\rho+\rho_0)}+\frac{2\rho_0}{3}(\rho+\rho_0)d\hat{\varphi}^2
\ee
o\`u nous avons pos\'e $\hat{\varphi}=\varphi-3 t$. Nous voyons que la continuation analytique est probl\'ematique dans le secteur ($\rho,\hat{\phi}$). En effet, la variable $\hat{\phi}$ est une variable p\'eriodique, donc la continuation analytique au-del\`a de l'hypersurface $\rho=-\rho_0$ va entra\^iner l'identification de points de l'espace-temps les uns avec les autres. Nous ne savons pas comment continuer la m\'etrique ni comment construire le diagramme de Penrose dans une telle situation. Par cons\'equent, nous consid\'erons $\rho=-\rho_0$ comme une singularit\'e de la m\'etrique.

\noindent Enfin, lorsque $\omega=0$ et $\rho_0=0$, la m\'etrique s'\'ecrit
\be
ds^2=-dt^2+\frac{d\rho^2}{\rho^2}+\rho^2\left(d\varphi -\frac{2}{\rho}dt\right)^2.
\ee
Pour construire le diagramme de Penrose, les coordonn\'ees angulaires sont n\'eglig\'ees. Dans le cas d'un trou noir en rotation, on se place dans le rep\`ere tournant avec le trou noir afin d'\'eliminer le terme crois\'e $dt d\varphi$ (en posant $d\tilde{\varphi}=(d\varphi -\omega dt)$). La coordonn\'ee angulaire $\tilde{\varphi}$ est alors n\'eglig\'ee et l'on se concentre sur la partie ($t,\rho$). Or, ici, nous voyons que nous ne pouvons pas le faire car, lorsque $\rho\rightarrow 0$, la vitesse angulaire $\omega$ diverge et nous ne pouvons pas introduire une nouvelle coordonn\'ee $\tilde{\varphi}$. Donc, dans le cas $\omega=\rho_0=0$, nous ne pouvons pas construire le diagramme de Penrose.
   
En effectuant une continuation analytique de la solution de Vuorio, nous avons obtenu une solution \`a deux param\`etres qui, pour certaines valeurs de ces param\`etres, d\'ecrit des trous noirs acceptables o\`u les singularit\'es causales ont \'et\'e rejet\'ees derri\`ere l'horizon. Notons que la m\'etrique de Kerr poss\`ede, elle aussi, des courbes ferm\'ees de genre temps cach\'ees derri\`ere l'horizon, pr\`es de l'anneau singulier. Cette situation n'est donc pas unique.

Cependant, l'espace-temps d\'ecrit par (\ref{bh2}) est quelque peu sp\'ecial. En effet, $g_{tt}$ est positif, ce qui implique que la norme du vecteur de Killing $\xi=\p_t$ est positive ($|\xi|^2=g_{tt}$) et donc que $\p_t$ est du genre espace. L'espace-temps n'est donc pas statique et il ne peut exister d'observateur au repos. L'absence d'observateur statique n'est pas inhabituelle. En effet, reprenons l'exemple de la m\'etrique de Kerr
\be ds^2  =- 
\frac{\Delta-a^2\sin^2\theta}{\Sigma}(dt-\omega d\varphi)^2 +
\Sigma\bigg(\frac{dr^2}{\Delta}+d\theta^2+\frac{\Delta\sin^2\theta}
{\Delta-a^2\sin^2\theta}\,d\varphi^2 \bigg) \ee avec \be \Delta  =  r^2-2 M
r+a^2, \quad  \Sigma =  r^2+(a\cos\theta)^2, \quad\omega =
-2\,\frac{aMr\sin^2\theta} {\Delta -
a^2\sin^2\theta}.\ee
Il existe une zone appel\'ee ergosph\`ere (voir fig. 6.2)
\be
r_h<r<r_e\quad \mbox{avec} \quad r_h=M+\sqrt{M^2-a^2},\quad r_e=M+\sqrt{M^2-a^2\cos^2\theta}
\ee
comprise entre l'horizon $r_h$ (p\^ole de $g_{rr}$) et $r_e$ (z\'ero de $g_{tt}$) o\`u $g_{tt}$ est positif. 
\begin{figure}
\centerline{\epsfxsize=200pt\epsfbox{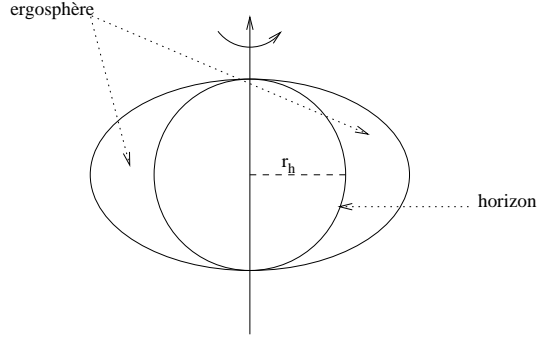}} 
\caption{Repr\'esentation de l'ergosph\`ere de la m\'etrique de Kerr. L'axe de rotation du trou noir est dans le plan de la figure.}
\end{figure}
Dans cette zone, il ne peut exister d'observateur statique puisque le vecteur de Killing $\p_t$ est du genre espace. Mais il peut exister des observateurs stationnaires, se d\'epla\c{c}ant avec la vitesse angulaire $\Omega$, lorsque le vecteur de Killing $\chi=\p_t+\Omega\p_\varphi$ est de genre temps, c'est-\`a-dire lorsque $\chi.\chi=g_{tt}+2\Omega g_{t\varphi}+\Omega^2 g_{\varphi\varphi}$ est n\'egatif.

La diff\'erence avec la m\'etrique de Kerr est qu'ici $g_{tt}$ est positif de l'horizon \`a l'infini et donc l'ergosph\`ere s'\'etend de l'horizon \`a l'infini. Il n'existe aucun endroit permettant l'existence d'observateur statique. Mais, comme dans le cas de la m\'etrique de Kerr, il peut exister des observateurs stationnaires, se d\'epla\c{c}ant avec la vitesse angulaire $\Omega$ le long des orbites du vecteur de Killing $\chi=\p_t+\Omega\p_\varphi$ lorsque celui-ci est de genre temps, c'est-\`a-dire lorsque
\be
\frac{2\rho + 3\omega - \sqrt{\rho^2-\rho_0^2}}{r^2} < \Omega < 
\frac{2\rho + 3\omega + \sqrt{\rho^2-\rho_0^2}}{r^2}.
\ee

\section{Masse, moment angulaire et thermodynamique}%

Dans la section 1.3, nous avons vu que la variation de l'action d'Einstein-Hilbert
\be \lb{var}
\delta S=\int (\mbox{\'eqs du mouvement})\delta g^{\mu\nu}\sqrt{|g|}\,d^4x+ \mbox{terme de surface}(\delta g^{\mu\nu},\delta \dot{g}^{\mu\nu})
\ee
ne conduisait aux \'equations d'Einstein que si l'on pouvait imposer \`a la fois des conditions aux limites du type de Dirichlet ($\delta g^{\mu\nu}=0$) et du type de Neumann ($\delta \dot{g}^{\mu\nu}=0$), ce qui est impossible en g\'en\'eral. Nous avons aussi remarqu\'e qu'en ajoutant un terme de surface \`a l'action d'Einstein-Hilbert, la contribution propotionnelle \`a $\delta \dot{g}^{\mu\nu}$ disparaissait et que le principe variationnel \'etait alors bien d\'efini. En variant l'action
\be
S=\frac{1}{16\pi G}\int_M R\sqrt{|g|} \,d^4x+\int_{\partial M} K \sqrt{h}\,d^3x
\ee
et en imposant des conditions aux limites de type Dirichlet, nous obtenions bien les \'equations d'Einstein.

Dans le cas de GTM le probl\`eme est le m\^eme. En effet, en variant l'action (\ref{actiontmg}) par rapport \`a $g_{\mu\nu}$, nous obtenons (\ref{var}).
La diff\'erence avec la th\'eorie d'Einstein est que le terme de Chern-Simons contribue lui aussi au $\mbox{terme de surface}$ et que nous ne savons pas, pour l'instant, quel terme de surface ajouter \`a l'action (\ref{actiontmg}) pour \'eliminer cette contribution. Nous ne pouvons donc pas utiliser le formalisme quasilocal pour calculer la masse et le moment angulaire de la solution (\ref{bh2}).

Pour calculer la masse et le moment angulaire de la solution (\ref{bh2}), nous allons utiliser la m\'ethode introduite dans \cite{GCTMG}. Nous allons bri\`evement introduire cette m\'ethode en prenant l'exemple de la \RG sans source \`a trois dimensions
\be \lb{act3}
S=\frac{1}{16\pi G}\int d^3x \sqrt{g} R_3.
\ee
La premi\`ere \'etape consiste \`a reformuler la th\'eorie, en supposant l'existence de deux vecteurs de Killing, en un probl\`eme unidimensionnel en suivant \cite{G93,G94}. Nous utilisons pour la m\'etrique l'ansatz suivant
\be \lb{gtmg}
ds^2=\lambda_{ab}(\rho)dx^adx^b+\zeta^{-2}(\rho)R^{-2}(\rho)d\rho^2
\ee
o\`u $a,b=t,\varphi$.
La pr\'esence du $\zeta$ assure l'invariance de cet ansatz sous une reparam\'etrisation de $\rho$. L'invariance de la partie $\lambda_{ab}(\rho)dx^adx^b$ sous les transformations du groupe $SL(2,R)$, localement isomorphe au groupe $SO(2,1)$, sugg\`ere la param\'etrisation suivante pour la matrice $\lambda_{ab}$
\be\lb{la}
\lambda = \left( \begin{array}{cc}
   T+X & Y \\
    Y & T-X \end{array} \right).
\ee
L'invariance de $\lambda_{ab}$ se traduit par l'invariance sous les rotations du groupe de Lorentz du vecteur ${\bf X}=(T,X,Y)$ d'un espace cible de Minkowski muni de la  m\'etrique $\eta_{AB}=-++$. La coordonn\'ee $\rho$ est choisie telle sorte que $R^2=-\mbox{det}(\lambda)=\eta_{AB}X^AX^B=-T^2+X^2+Y^2$.

En utilisant la param\'etrisation (\ref{gtmg}), l'action (\ref{act3}) devient
\be \lb{act32}
S=\frac{1}{32\pi G}\int d^2x\int d\rho\, \zeta {\bf \dot{X}}^2 .
\ee
Nous avons donc reformul\'e la th\'eorie d'Einstein \`a trois dimensions sous la forme d'un probl\`eme \`a une dimension o\`u toute solution (\ref{gtmg}) est repr\'esent\'ee par un point, rep\'er\'e par le vecteur ${\bf X}$, dans l'espace cible.

En examinant la variation de (\ref{act32}) par rapport \`a une rotation infinit\'esimale $\delta{\bf X}={\bf \Omega}\wedge{\bf X}$
\be
\delta S=\int d^2x \int d\rho \frac{d{\bf J}}{d\rho}. {\bf \Omega}
\ee 
nous en d\'eduisons la quantit\'e conserv\'ee ${\bf J}$
\be
{\bf J}=\frac{\zeta}{16\pi G} {\bf X}\wedge {\bf \dot{X}}
\ee
appel\'ee super moment angulaire\footnote{La d\'efinition du produit vectoriel \'etant $({\bf U}\wedge {\bf V})^A=\eta^{AB}\epsilon_{BCD}U^CV^D$ avec $\epsilon_{TXY}=1$.}. Bien que le vecteur ${\bf J}$ ait trois composantes, il n'y a que deux quantit\'es conserv\'ees, la troisi\`eme \'etant associ\'ee \`a la libert\'e que nous avons d'effectuer une rotation infinit\'esimale $\delta\varphi=-\delta\Omega \,t$ dans (\ref{gtmg}) \cite{GCTMG}. Nous pouvons alors montrer \cite{GCTMG} que ces quantit\'es conserv\'ees sont reli\'ees \`a l'\'energie et au moment angulaire
\ba \lb{e}
E=-2\pi J^Y\\
J=2\pi(J^T-J^X) \lb{j}
\ea
de la solution (\ref{gtmg}). Dans l'article \cite{GCTMG}, il a \'et\'e montr\'e que les d\'efinitions (\ref{e}) et (\ref{j}) sont compatibles avec les quantit\'es quasilocales obtenues en ajoutant \`a (\ref{act3}) des termes de surface appropri\'es et pour des conditions aux limites m\'elangeant des conditions aux limites de type Dirichlet et de type Neumann. Cette d\'emonstration a \'et\'e effectu\'ee pour les th\'eories d'Einstein-champ scalaire et d'Einstein-Maxwell.

Cette m\'ethode peut aussi s'appliquer en principe \`a GTM. 

Le super moment angulaire conserv\'e de GTM s'\'ecrit (dans le cas $\zeta=1$) \cite{KAT,GCJ}
\be\lb{superam}
{\bf J} = \frac1{16\pi G}\bigg({\bf X}\wedge\dot{\bf X} + 
\frac{1}{2\mu}\bigg[\dot{\bf X}\wedge({\bf X}\wedge\dot{\bf X}) -
2{\bf X}\wedge({\bf X}\wedge\ddot{\bf X})\bigg]\bigg).
\ee
Nous admettrons ici que la masse et le moment angulaire des solutions de GTM sont encore donn\'es par les formules (\ref{e})-(\ref{j}) (avec {\bf J} donn\'e par (\ref{superam})). Cette conjecture peut \^etre v\'erifi\'ee dans le cas de la m\'etrique de BTZ \cite{BTZ7}, qui est solution des \'equations de GTM avec constante cosmologique \cite{kaloper}. Dans l'article \cite{GHHM}, les auteurs ont calcul\'e la masse et le moment angulaire de la solution de BTZ dans une th\'eorie contenant GTM. Or, nous pouvons v\'erifier que les valeurs obtenues dans l'article \cite{GHHM} et celles obtenues en appliquant les formules (\ref{e}) et (\ref{j}) sont identiques.

Le vecteur ${\bf X}$ de la solution (\ref{bh2}) est 
\be
{\bf X}=\left(\begin{array}{c}\rho^2/2+2\omega\rho+3(\omega^2+1)/2+\rho_0^2/6\\-\rho^2/2-2\omega\rho-3(\omega^2-1)/2-\rho_0^2/6\\-2\rho-3\omega\end{array}\right)
\ee
ce qui donne pour la masse et le moment angulaire
\be
{\cal M}=\frac{\omega}{8G},\qquad {\cal J}=\frac{\omega^2-5\rho_0^2/9}{8G}.
\ee
Nous remarquons que la masse ${\cal M}$ n'est pas reli\'ee au rayon de l'horizon comme c'est le cas habituellement. D'autre part, nous voyons que ${\cal M}$ et ${\cal J}$ s'annulent pour la solution repr\'esentant le vide des trous noirs ($\rho_0=\omega=0$).


V\'erifions maintenant si les trous noirs satisfont \`a la premi\`ere loi de la thermodynamique des trous noirs. Nous pouvons calculer la temp\'erature de ces trous noirs en utilisant la formule habituelle
\be
T=\frac{1}{2\pi}n^\rho\p_\rho N|_{\rho=\rho_h}=\left. \frac{\zeta R \dot{R}}{2\pi\sqrt{V}}\right|_{\rho=\rho_h}=\frac{\sqrt 3}{2\pi}\frac{\rho_0}{2\rho_0+3\omega}.
\ee
Par contre, les lois de la thermodynamique n'ont jamais \'et\'e d\'eriv\'ees dans le cas de GTM donc nous ne savons pas quelle est, dans ce cas, la forme de la premi\`ere loi, ni d'ailleurs si l'entropie est donn\'ee par la formule habituelle (un quart de l'aire de l'horizon du trou noir).

Nous allons supposer que la premi\`ere loi est de la forme 
\be \lb{fl}
d{\cal M}=T dS+z\Omega_h d{\cal J}
\ee
o\`u $z$ est une constante que nous d\'eterminerons plus loin et $\Omega_h$ est la vitesse angulaire de l'horizon
\be
\Omega_h=\frac{3}{2\rho_0+3\omega}\, .
\ee
En utilisant cette premi\`ere loi nous pouvons d\'eterminer l'entropie $S$ par l'int\'egration de la relation
\be
\left.\frac{\p S}{\p {\cal M}}\right|_{\cal J}=T^{-1}.
\ee
L'entropie est donc donn\'ee par
\be
S=\frac{\pi}{12\sqrt 3 G}(5\rho_0+6\omega)+f({\cal J})
\ee
o\`u $f({\cal J})$ est une fonction de ${\cal J}$ que nous prendrons \'egale \`a z\'ero.
Nous voyons alors que la solution (\ref{bh2}) satisfait \`a la premi\`ere loi (\ref{fl}) si $z=1/2$.

Nous avons fait remarquer dans la premi\`ere section que la constante de gravitation $G$ doit \^etre n\'egative pour que GTM soit d\'epourvue de fant\^omes et que l'\'energie soit d\'efinie positive \cite{DJT}. La masse et l'entropie des trous noirs (\ref{bh2}) seraient donc n\'egatives (ce qui pose, entre autre, le probl\`eme de l'existence d'un \'etat de plus basse \'energie). Signalons qu'il en est de m\^eme pour la masse des trous noirs de BTZ consid\'er\'es comme solutions de GTM avec constante cosmologique. Cependant, la condition $G$ n\'egative a \'et\'e obtenue en lin\'earisant la th\'eorie GTM autour de la solution de Minkowski. Or, nous avons vu que, de m\^eme que le vide de BTZ n'est pas l'espace-temps de Anti-deSitter, le vide des trous noirs (\ref{bh2}) n'est pas l'espace-temps de Minkowski. Donc, avant de conclure que la masse et l'entropie des trous noirs (\ref{bh2}) sont n\'egatives, il faudrait refaire une \'etude similaire \`a celle conduite dans \cite{DJT}, en prenant comme fond non pas Minkowski mais le vide ($\rho_0=\omega=0$) des trous noirs (\ref{bh2}).


\newpage
\thispagestyle{empty}
\null
\fancyhead[RO]{Conclusion}
\fancyhead[LE]{Conclusion}
\pagestyle{plain}
\chapter*{Conclusion}%
\addcontentsline{toc}{chapter}{Conclusion}

Tout au long de cette th\`ese, nous avons construit et \'etudi\'e de nouvelles solutions trou noir non asymptotiquement plates, dans le cadre de th\'eories d\'ecrivant la gravitation coupl\'ee \`a des champs de mati\`eres et pour des dimensions allant de $3$ \`a $D$ dimensions d'espace-temps ($D>2$).

Dans cette th\`ese, nous nous sommes int\'eress\'es \`a des solutions avec un comportement asymptotique atypique. En effet, ces solutions ne sont ni asymptotiquement Minkowskiennes ni asymptotiquement AdS.
En \'etudiant le mouvement g\'eod\'esique des particules dans le champ de gravitation de ces solutions, nous avons vu que l'une des caract\'eristiques de ces espace-temps non asymptotiquement plats est la pr\'esence d'une barri\`ere de potentiel r\'efl\'echissant les g\'eod\'esiques du genre temps, les emp\^echant d'aller jusqu'\`a l'infini. Un autre aspect inhabituel de ces trous noirs est qu'il sont des excitations sur un fond charg\'e. En fait, il existe une infinit\'e de fonds charg\'es et sur chaque fond peut exister une famille de trous noirs caract\'eris\'es par une masse et un moment angulaire. Donc la charge \'electrique (ou magn\'etique) est un param\`etre associ\'e au fond sur lesquel les trous noirs se forment, et non aux trous noirs eux-m\^emes. Nous avons calcul\'e les masses et les moments angulaires de ces solutions non asymptotiquement plates en utilisant l'approche moderne au calcul de l'\'energie en Relativit\'e G\'en\'erale, le formalisme quasilocal. Pour obtenir un r\'esultat fini, nous avons retranch\'e la contribution du fond charg\'e sur lequel les trous noirs existent. Nous avons ensuite v\'erifi\'e que ces solutions satisfont \`a la premi\`ere loi de la thermodynamique des trous noirs si la charge n'est pas vari\'ee. Nous avons conclu qu'il \'etait logique de ne pas varier la charge puisque celle-ci n'est pas un param\`etre du trou noir. Cependant, nous avons indiqu\'e qu'il serait int\'eressant d'\'etudier de mani\`ere plus approfondie la thermodynamique de ces solutions, ce qui nous permettrait de valider ou non ce r\'esultat. 


Dans le chapitre 3, nous avons construit de nouvelles solutions de la th\'eorie d'Einstein-Maxwell dilatonique, de type \'electrique ou magn\'etique, pour une valeur particuli\`ere de la constante de couplage du dilaton: $\alpha^2=3$. En utilisant le lien existant entre cette th\'eorie et la th\'eorie d'Einstein \`a cinq dimensions (la th\'eorie d'Einstein-Maxwell dilatonique est la r\'eduction dimensionnelle de la th\'eorie d'Einstein sans source \`a cinq dimensions par rapport \`a un vecteur de Killing du genre espace), nous avons montr\'e que les versions \'electriques et magn\'etiques sont des r\'eductions dimensionnelles inhabituelles de solutions de la gravitation \`a cinq dimensions. Les versions magn\'etiques sont des r\'eductions dimensionnelles de la solution de Myers et Perry, alors que les versions \'electriques sont des r\'eductions dimensionnelles de solutions de Rasheed.

Dans le chapitre 4, nous avons construit une nouvelle solution en rotation de la th\'eorie d'Einstein-Maxwell dilato-axionique. Cette th\'eorie peut \^etre obtenue par r\'eduction d'un secteur de la th\'eorie d'Einstein sans source \`a six dimensions. Nous avons montr\'e que la version \'electrique et la version magn\'etique de la solution en rotation  sont toutes les deux des r\'eductions dimensionnelles de la solution de Myers et Perry \`a cinq dimensions (avec les deux moments angulaires \'egaux) trivialement plong\'ee dans six dimensions. Nous avons ensuite \'etudi\'e les modes d'un champ scalaire dans le champ de gravitation de la solution en rotation. Nous avons vu que, suivant les cas, le spectre d'\'energie est continu, semi-discret ou discret. Lorsque la solution d\'ecrit un trou noir en rotation, il est bien connu que le spectre est continu. Cependant, dans ce cas, le trou noir poss\`ede une ergosph\`ere. Il peut alors y avoir super-radiance, c'est-\`a-dire, extraction d'une partie de l'\'energie du trou noir \`a l'aide d'une onde envoy\'ee sur l'horizon. Or, dans le cas de notre solution non asymptotiquement plate, la barri\`ere de potentiel emp\^eche l'onde d'atteindre l'infini, et la renvoie sur l'horizon. Les allers et retours effectu\'es par l'onde entre l'horizon et la barri\`ere de potentiel peuvent alors conduire \`a une instabilit\'e des solutions en rotation.
Cependant, ceci est un raisonnement par analogie avec le cas des trous noirs asymptotiquement plats ou asymptotiquement AdS. Pour bien comprendre ce qui se passe pour les solutions construites dans cette th\`ese, il faudrait faire une \'etude plus rigoureuse de la quantification du champ scalaire et de la super-radiance, \`a l'image de ce qui a \'et\'e fait dans le cas des espaces asymptotiquement plats ou asymptotiquement AdS.

Dans le chapitre 5, nous avons construit des solutions trou noir et branes noires statiques d'une th\'eorie \`a $D$ dimensions d'espace-temps qui g\'en\'eralise la th\'eorie d'Einstein-Maxwell dilatonique. Pour une valeur particuli\`ere de la constante de couplage du dilaton, cette th\'eorie est la r\'eduction dimensionnelle de la th\'eorie d'Einstein sans source \`a $D+1$ dimensions par rapport \`a un vecteur de Killing du genre espace. En utilisant une approche similaire \`a celle utilis\'ee dans le chapitre 3, nous avons montr\'e que nous pouvions construire des dyons en rotation pour cette valeur de la constante de couplage. Le travail expos\'e dans ce chapitre est une partie d'un travail en cours. Dans la suite, plusieurs pistes sont \`a explorer. Premi\`erement, il serait int\'eressant d'examiner si les solutions non asymptotiquement plates pr\'eservent au moins une partie des supersym\'etries. D'autre part, des solutions dyons statiques asymptotiquement plates ont \'et\'e obtenues pour une valeur de la constante de couplage du dilaton diff\'erente de celle que nous avons consid\'er\'ee ici. Il serait int\'eressant de construire les solutions non asymptotiquement plates correspondantes.

Dans le chapitre 6, nous avons travaill\'e dans le cadre d'une th\'eorie \`a $2+1$ dimensions g\'en\'eralisant la th\'eorie d'Einstein sans source par l'ajout d'un terme de Chern-Simons. Nous avons construit une famille de trous noirs de cette th\'eorie en prenant comme point de d\'epart la solution de Vuorio. La solution de Vuorio poss\`ede un horizon mais contient des courbes ferm\'ees de genre temps \`a l'ext\'erieur de l'horizon. Cette violation de causalit\'e visible fait que la solution de Vuorio ne peut-\^etre consid\'er\'ee comme une solution trou noir acceptable. Nous avons montr\'e que, en effectuant une continuation analytique de cette solution, nous pouvions r\'esoudre cette violation de causalit\'e en rejetant les courbes ferm\'ees de genre temps \`a l'int\'erieur de l'horizon. Nous avons ensuite construit les diagrammes de Penrose et calcul\'e la masse et le moment angulaire de la nouvelle solution ainsi obtenue. 

\noindent La solution de Vuorio a \'et\'e g\'en\'eralis\'ee au cas de la Gravitation Topologiquement Massive avec constante cosmologique. Or, ces solutions, comme la solution de Vuorio, poss\`edent des courbes ferm\'ees de genre temps \`a l'ext\'erieur de l'horizon. Ce ne sont donc pas de bons trous noirs. Il devrait \^etre possible d'effectuer, comme pour la solution de Vuorio, une continuation analytique pour obtenir de bons trous noirs. D'autre part, il serait int\'eressant de g\'en\'eraliser la famille de solutions que nous avons construite au cas de la Gravito-\'electrodynamique Topologiquement Massive qui d\'ecrit la Gravitation Topologiquement Massive coupl\'ee \`a l'Electrodynamique Topologiquement Massive (c'est-\`a-dire l'\'electrodynamique g\'en\'eralis\'ee par l'ajout d'un terme de Chern-Simons).

\setcounter{chapter}{0}
\setcounter{equation}{0}
\setcounter{figure}{0}
\newpage
\thispagestyle{empty}
\null
\chapter*{Conventions et D\'efinitions}%
\addcontentsline{toc}{chapter}{Conventions et d\'efinitions}

Dans cet appendice sont regroup\'ees les conventions et les d\'efinitions de terme qui sont utilis\'ees tout au long de la th\`ese. Les d\'efinitions d'un trou noir, d'un horizon et d'une singularit\'e sont donn\'ees \`a la fin de la partie 1.1.

\vspace{2cm}
\underline{Conventions}:

\vspace{1cm}
Le syst\`eme d'unit\'e utilis\'e est sp\'ecifi\'e au d\'ebut de chaque chapitre.

\vspace{0.5cm}
Pour la signature de la m\'etrique, nous prenons la convention avec une majorit\'e de signes $+$, c'est-\`a-dire $-++\cdots$.

\vspace{0.5cm}
Pour le tenseur de Riemann, nous adoptons la convention de Landau et Lifchitz
\be
R^\mu{}_{\nu\lambda\sigma}=\p_\lambda\Gamma^\mu_{\nu\sigma}-\p_\sigma\Gamma^\mu_{\nu\lambda}+\Gamma^\mu_{\tau\lambda}\Gamma^\tau_{\nu\sigma}-\Gamma^\mu_{\tau\sigma}\Gamma^\tau_{\nu\lambda}.
\ee

\vspace{0.5cm}
Le tenseur de Ricci est d\'efini en contractant le tenseur de Riemann par rapport au premier et au troisi\`eme indice
\be
R_{\mu\nu}=R^\lambda{}_{\mu\lambda\nu}=\p_\lambda\Gamma^\lambda_{\mu\nu}-\p_\nu\Gamma^\lambda_{\mu\lambda}+\Gamma^\lambda_{\tau\lambda}\Gamma^\tau_{\mu\nu}-\Gamma^\lambda_{\tau\nu}\Gamma^\tau_{\mu\lambda}
\ee
et la courbure scalaire par
\be
R=R^\mu_\mu.
\ee

\newpage
\underline{D\'efinitions}:

\vspace{1cm}
$\bullet$
\underline{Asymptotiquement plat}: Un espace-temps est dit asymptotiquement plat si la m\'etrique de cet espace-temps, dans le syst\`eme de coordonn\'ees o\`u $r$ est le rayon de la sph\`ere,
\be
ds^2=-B(r)(dt+\omega_i dx^i)^2+A(r)dr^2+r^2d\Omega^2,
\ee
tend \`a l'infini ($r \rightarrow\infty$) vers la m\'etrique de Minkowski plus une contribution en $1/r$,
\be
g_{\mu\nu}\simeq\eta_{\mu\nu}+O\left(\frac{1}{r}\right).
\ee
Nous pouvons montrer que ceci implique que les composantes du tenseur de Riemann et du tenseur de Ricci ainsi que la courbure scalaire doivent alors d\'ecro\^itre au moins aussi vite que $r^{-3}$. De m\^eme, les invariants de courbure $R^{\mu\nu}R_{\mu\nu}$ et $R^{\mu\nu\lambda\sigma}R_{\mu\nu\lambda\sigma}$ doivent d\'ecro\^itre au moins aussi vite que $r^{-6}$.

\vspace{0.5cm}
$\bullet$
\underline{Charge NUT}: En r\'eduisant la th\'eorie d'Einstein \`a la Kaluza-Klein (de quatre \`a trois dimensions), c'est-\`a-dire en utilisant l'ansatz
\be
ds^2=-f(dt-\omega_idx^i)^2+f^{-1}h_{ij}dx^idx^j,
\ee
la composante $(0i)$ de l'\'equation d'Einstein s'\'ecrit
\be
\epsilon^{ijk}\tau_{k,j}=0,\quad \mbox{avec}\ \tau^i=-\frac{f^2}{\sqrt h}\epsilon^{ijk}\partial_j\omega_k.
\ee
On en d\'eduit les deux r\'esultats suivants
\be
\tau_i=\partial_i\chi,\quad \partial_i\left(\epsilon^{ijk}\partial_j\omega_k\right)=-\partial_i\left(\frac{\sqrt h}{f^2}\tau^i\right)=0
\ee
o\`u $\chi$ est appel\'e potentiel NUT.

Il existe donc une charge conserv\'ee appel\'ee charge NUT
\be
{\cal N}=-\frac{1}{8\pi}\int F_{\theta\varphi}d\theta d\varphi=\frac{1}{8\pi}\int\sqrt h f^{-2}\partial_r\chi d\theta d\varphi
\ee
o\`u $F_{ij}=\partial_i\omega_j-\partial_j\omega_i$.

D'autre part, on peut montrer que les \'equations \`a trois dimensions prennent une forme tr\`es similaire \`a celle des \'equations de Maxwell avec un champ ``\'electrique'' $\vec{E}_g$ proportionnel \`a la masse et un champ ``magn\'etique'' $\vec{B}_g$ proportionnel \`a la charge NUT. Pour cette raison la masse est parfois appel\'ee masse du type ``\'electrique'' et la charge NUT masse du type ``magn\'etique''.

\vspace{0.5cm}
$\bullet$
\underline{Isom\'etrie}: Une isom\'etrie est un changement de coordonn\'ees $x\rightarrow x'$ tel que 
\be \lb{iso}
g_{\mu\nu}(x)=\p_\mu x'^\rho \p_\nu x'^\sigma g'_{\rho\sigma}(x')=\p_\mu x'^\rho \p_\nu x'^\sigma g_{\rho\sigma}(x'),
\ee
c'est-\`a-dire $g'_{\rho\sigma}(x')=g_{\rho\sigma}(x')$.

Consid\'erons les isom\'etries infinit\'esimales
\be
x'^\mu=x^\mu+\varepsilon\xi^\mu(x),\qquad \varepsilon<<1.
\ee 
L'\'equation (\ref{iso}) est satisfaite au premier ordre en $\varepsilon$ si le vecteur $\xi$ appel\'e vecteur de Killing satisfait \`a l'\'equation
\be
\xi_{\sigma;\rho}+\xi_{\rho;\sigma}=0
\ee
appel\'ee \'equation de Killing.

\vspace{0.5cm}
$\bullet$  \underline{Stationnaire}: Un espace-temps est dit stationnaire lorsqu'il poss\`ede un vecteur de
Killing du genre temps. La m\'etrique d'un espace-temps stationnaire peut toujours \^etre mise sous la forme:
\be
ds^2=-f(dt-\omega_i dx^i)(dt-\omega_j dx^j)+f^{-1}h_{ij}dx^i dx^j.
\ee

\vspace{0.5cm}
$\bullet$  \underline{Statique}: Un espace-temps est dit statique lorsqu'il est stationnaire et qu'il existe une hypersurface du genre espace $\Sigma$, telle que les orbites du vecteur de Killing du genre temps soient perpendiculaire \`a $\Sigma$.  Il existe alors un syst\`eme de coordonn\'ees tel que la m\'etrique puisse se mettre sous la forme:
\be
ds^2=-fdt^2+f^{-1}h_{ij}dx^i dx^j.
\ee





\newpage
\thispagestyle{empty}
\null

\appendix
\newpage
\chapter*{Appendice A}
\addcontentsline{toc}{chapter}{{\bf {\tt {\bf Appendice A. ``Linear dilaton black holes''}}}}
\begin{center}
\vspace{3cm}

{\huge {\tt ``Linear dilaton black holes''}}
\vspace{2cm}

{\large {\sl publi\'e dans Physical Review D}}
\vspace{2cm}

{\large {\sl Phys Rev. {\bf D67} (2003) 024012}}

\vspace{2cm}

{\large {\sl arXiv: hep-th/0208225}}
\end{center}
\newpage
\chapter*{}
\thispagestyle{empty}
\newpage
\chapter*{Appendice B}
\addcontentsline{toc}{chapter}{{\bf {\tt {\bf Appendice B. ``The black holes of topologically massive gravity''}}}}
\begin{center}
\vspace{3cm}

{\huge {\tt ``The black holes of topologically massive gravity''}}
\vspace{2cm}

{\large {\sl publi\'e dans Classical and Quantum Gravity}}
\vspace{2cm}

{\large {\sl Class. Quantum Grav. {\bf 20} (2003) L277}}

\vspace{2cm}

{\large {\sl arXiv: gr-qc/0303042}}
\end{center}

\newpage
\chapter*{}
\thispagestyle{empty}
\chapter*{Appendice C}
\addcontentsline{toc}{chapter}{{\bf {\tt {\bf Appendice C. ``Non-asymptotically flat, non-AdS dilaton black holes''}}}}
\begin{center}
\vspace{3cm}

{\huge {\tt ``Non-asymptotically flat, non-AdS dilaton black holes''}}
\vspace{2cm}

{\large {\sl \`a para\^itre dans Physical Review D}}

\vspace{2cm}

{\large {\sl arXiv: gr-qc/0405034}}
\end{center}

\newpage
\chapter*{}
\thispagestyle{empty}
\newpage
\chapter*{}
\thispagestyle{empty}
\pagestyle{empty}
\newpage
\vspace{4cm}
\underline{R\'esum\'e}:

\vspace{.5cm}
Dans le cadre de th\'eories de la gravitation dilatonique inspir\'ees des th\'eories des cordes (de $4$ \`a $D$ dimensions d'espace-temps), nous construisons de nouvelles solutions trou noir ou branes noires non asymptotiquement plates. Pour certaines valeurs de la constante de couplage dilatonique, nous g\'en\'eralisons les trous noirs statiques \`a des trous noirs en rotation, en utilisant le groupe d'isom\'etrie de l'espace cible. Nous calculons leurs masses et leurs moments angulaires en utilisant l'approche moderne au calcul de l'\'energie en Relativit\'e G\'en\'erale, le formalisme quasilocal, et nous v\'erifions qu'ils satisfont \`a la premi\`ere loi de la thermodynamique des trous noirs. Enfin, nous \'etudions une famille de trous noirs en Gravitation Topologiquement Massive \`a $2+1$ dimensions.

\vspace{.5cm}
Mots-cl\'es: Relativit\'e G\'en\'erale, trous noirs, \'energie quasilocale.

\vspace{3cm}
\underline{Abstract}:

\vspace{.5cm}
In the framework of string-inspired dilatonic gravity theories (from $4$ to $D$ space-time dimensions), we construct new non-asymptotically flat black hole or black brane solutions. For particular values of the dilatonic coupling constant, we generalize static solutions to rotating ones, using the target space isometry group. We compute their masses and their angular momentum using the modern approach to the computation of energy in General Relativity, the quasilocal formalism, and we check the agreement of these solutions with the first law of black hole thermodynamics. Finally, we study a black hole family in the $2+1$ dimensional theory of Topologically Massive Gravity.

\vspace{.5cm}
Keywords: General Relativity, black holes, quasilocal energy.

\vspace{3cm}
\noindent {\bf L}aboratoire d'{\bf A}nnecy-le-Vieux de {\bf P}hysique {\bf Th}\'eorique (LAPTH)

\noindent UMR 5108

\noindent 9 chemin de Bellevue

\noindent B.P. 110

\noindent 74941 Annecy-le-Vieux, Cedex

\end{document}